\documentclass[a4paper,11pt]{article}
\pdfoutput=1 

\usepackage{jheppub} 

\usepackage[T1]{fontenc} 

\usepackage{amsfonts,amsmath,amssymb}
\usepackage{enumerate}
\usepackage{hyperref}
\usepackage{graphicx}
\usepackage{caption}
\usepackage{subcaption}
\usepackage{ytableau}
\usepackage{placeins}

\usepackage{tikz}
\usetikzlibrary{arrows.meta,calc}

\usepackage{enumitem}   
\usepackage{booktabs}
    
\newcommand{\be}{\begin{equation}}
\newcommand{\ee}{\end{equation}}

\newcommand{\bea}{\begin{eqnarray}}
\newcommand{\eea}{\end{eqnarray}}

\newcommand{\nn}{\nonumber}

\newcommand{\dd}{\textrm{d}}

\newcommand{\cB}{\mathcal{B}}
\newcommand{\cC}{\mathcal{C}}
\newcommand{\cD}{\mathcal{D}}

\newcommand{\cF}{\mathcal{F}}
\newcommand{\cG}{\mathcal{G}}

\newcommand{\cN}{\mathcal{N}}
\newcommand{\cO}{\mathcal{O}}

\newcommand{\cR}{\mathcal{R}}

\newcommand{\bN}{\mathbb{N}}

\newcommand{\bR}{\mathbb{R}}

\newcommand{\bV}{\mathbb{V}}

\newcommand{\bZ}{\mathbb{Z}}

\newcommand{\fkb}{\mathfrak{b}}

\newcommand{\fkm}{\mathfrak{m}}

\newcommand{\fkn}{\mathfrak{n}}

\newcommand{\fks}{\mathfrak{s}}

\DeclareMathOperator{\Li}{Li}
\newcommand{\ii}{\mathrm{i}}
\newcommand{\wt}{\widetilde}

\newcommand{\xmin}{x_{\text{min}}}
\newcommand{\xL}{x_{\text L}}
\newcommand{\xR}{x_{\text R}}
\newcommand{\xmax}{x_{\text{max}}}
\newcommand{\DeltaS}{\sum_I\Delta_I}
\newcommand{\diag}{\Theta}

\usepackage{soul}

\graphicspath{ {./figures/}}

\title{\boldmath  Large\,-$N$ limit of chiral quiver theories: free energy, supersymmetric indices and operator counting}

\author[a]{Edoardo Colombo,}
\author[b]{Hyojoong Kim,}
\author[a]{Nakwoo Kim}
\author[a]{and Hoseob Shin}

\affiliation[a]{Department of Physics and Research Institute of Basic Science, Kyung Hee University,
26, Kyungheedae-ro, Dongdaemun-gu, Seoul 02447, Republic of Korea}
\affiliation[b]{Department of Physics and Astronomy \& Center for Theoretical Physics, \\
Seoul National University, Seoul 08826, Korea}

\emailAdd{edoardo.colombo@khu.ac.kr}
\emailAdd{hjkim1996@gmail.com}
\emailAdd{nkim@khu.ac.kr}
\emailAdd{hoseob0610@khu.ac.kr}

\abstract{We find the first analytic solution to the large\,-$N$ saddle point equations for the $S^3$ free energy of a chiral $\mathcal N=2$ Chern-Simons-matter theory with an M-theory dual. Focusing on the chiral quiver dual to AdS$_4\times Q^{1,1,1}/\mathbb Z_{\rm k}$, we show how the long-range forces of the matrix model cancel at large $N$, resolving a long-standing tension with the expected $N^{3/2}$ scaling. The resulting free energy precisely matches the holographic prediction for all assignments of trial R-charges. Our results extend to the superconformal and topologically twisted indices on $S^2\times S^1$, whose large\,-$N$ and Cardy limit is determined by the same saddle. The solution agrees with the predictions of the operator counting method, giving strong evidence for this conjecture beyond non-chiral quivers. We support the analytic solution with a finite\,-$N$ numerical analysis for multiple assignments of R-charges, extending recent numerical results obtained for this quiver.}

\begin{document} 
\maketitle
\flushbottom
\section{Introduction}

Supersymmetric localization reduces several partition functions of three-dimensional $\cN=2$ theories to finite-dimensional matrix models, opening the way to exact tests of AdS$_4$/CFT$_3$ holography.
These include the partition function on $S^3$ \cite{Kapustin:2009kz,Jafferis:2010un,Hama:2010av}, and the superconformal \cite{Kim:2009wb,Imamura:2011su,Kapustin:2011jm} and topologically twisted \cite{Benini:2015noa} indices on $S^2\times S^1$.
The $S^3$ free energy $F_{S^3}=-\log Z_{S^3}$ is a measure of the degrees of freedom of the theory; for superconformal Chern-Simons-matter theories dual to a stack of $N$ M2 branes, holography predicts that the free energy should scale as $N^{3/2}$ \cite{Klebanov:1996un}, with the coefficient determined in terms of the volume of the Sasaki-Einstein internal manifold \cite{Herzog:2010hf}.
On the other hand, the $S^2\times S^1$ superconformal and topologically twisted indices of such theories count the microstates of supersymmetric black holes in AdS$_4$, and at large $N$ reproduce their Bekenstein-Hawking entropy \cite{Benini:2015eyy,Choi:2019zpz}.
In all these cases, the large\,-$N$ limit is captured by a saddle of the respective matrix model.
In the Cardy limit and at leading order in $N$, the saddle point equations of the matrix models of the $S^2\times S^1$ indices can be re-expressed in terms of (two copies of) the same equations that govern the saddles of the $S^3$ free energy \cite{Hosseini:2016tor,Choi:2019zpz, Choi:2019dfu, Hosseini:2022vho}, reducing the problem of computing the large\,-$N$ limit of the indices to the analogous problem for the $S^3$ free energy.

The large\,-$N$ limit of the supersymmetric partition functions is currently well understood only for a particular subclass of $\cN=2$ Chern-Simons-matter quiver gauge theories with an M-theory dual, the \emph{non-chiral} ones, for which every bifundamental chiral multiplet can be paired up with another chiral multiplet in the conjugate representation. For more general \emph{chiral} quiver theories, much less is known, and only recently a few results have been obtained in this direction \cite{Hosseini:2025jxb,Amariti:2026djf}.

The main difficulty lies in the so-called \emph{long range forces}: interactions between the eigenvalues of the matrix model that are far apart, and whose contribution would naively seem to scale as $N^2$, spoiling the expected $N^{3/2}$ behavior.
In the case of non-chiral theories, there is a simple ansatz for the eigenvalue distributions for which the long range forces cancel automatically, and the resulting large\,-$N$ free energy reproduces the holographic prediction \cite{Herzog:2010hf,Martelli:2011qj,Cheon:2011vi,Jafferis:2011zi,Gulotta:2011si,Gulotta:2011vp}.
Similar consideration also apply to the superconformal and topologically twisted indices.
On the other hand, for chiral quiver theories the long range forces do not vanish in the same way and the large\,-$N$ saddle point equations have not been solved analytically so far.
A proposed solution to this problem modifies the matrix model by symmetrizing its integrand \cite{Amariti:2011jp,Amariti:2011uw,Gang:2011jj};
however, this prescription does not descend from first principles  and the resulting saddle does not solve the saddle point equation of the original matrix model.
The operator counting method of \cite{Gulotta:2011si, Gulotta:2011aa} has also been applied to chiral quiver theories in order to make predictions about the eigenvalue distribution of the large\,-$N$ saddle \cite{Gulotta:2011aa,Kim:2012vza}. Even though this method is based on a conjecture, for the chiral quiver that we consider in this paper we find that the eigenvalue density and differences between eigenvalues that it predicts are accurate and match the solution to the saddle point equations that we find.

Recently, there has been some progress in the understanding of the large\,-$N$ behavior of chiral quiver theories.
First, in \cite{Hosseini:2025jxb} numerical finite\,-$N$ solutions to the saddle point equations have been found for two distinct chiral quivers, providing the first quantitative evidence that the free energy of these theories has the precise $N^{3/2}$ scaling predicted by holography.
The shape of these numerical saddles also provides a first important clue about which ansatz is appropriate when trying to solve the large\,-$N$ saddle point equations.
Then in \cite{Amariti:2026djf} it was shown that for certain chiral quivers
the $S^3$ matrix model may be rewritten using exact integral identities \cite{vandeBult:2007} related to Giveon-Kutasov dualities \cite{Giveon:2008zn}, mapping it to the matrix model of a non-chiral quiver whose large\,-$N$ limit is known.
This method applies to the theories dual to M-theory AdS$_4 \times Y_7$ whenever $Y_7$ is a toric Sasaki-Einstein manifold whose toric diagram does not have internal points.
However, this approach does not provide a saddle of the original chiral matrix model and does not explain how long range forces cancel.
We expect that addressing these gaps is the natural way forward towards a general understanding of chiral theories at large $N$, and finding explicit saddles for these models makes it possible to make direct contact with the supersymmetric indices on $S^2\times S^1$.

Understanding how the long range forces cancel in chiral models and finding an analytical solution to the large\,-$N$ saddle point equations are longstanding open problems, which we set out to solve, focusing on a particular chiral $\cN=2$ Chern-Simons-matter theory.
We consider the chiral quiver dual to M-theory on AdS$_4 \times Q^{1,1,1}/{\mathbb{Z}_{\rm k}}$ \cite{Franco:2008um,Davey:2009sr,Hanany:2009vx,Franco:2009sp} as our case study and present an analytic solution to its large\,-$N$ saddle point equations. Before that, in order to find a suitable ansatz for the analytic large\,-$N$ saddle we consider two other complementary approaches: a numerical study of the saddle point equations at finite $N$ that extends the results of \cite{Hosseini:2025jxb}, and the operator counting method of \cite{Gulotta:2011si, Gulotta:2011aa}.

For the numerical analysis, following the approach of \cite{Hosseini:2025jxb}, we begin by solving the saddle point equations of the three-sphere matrix model at finite $N$ for canonical R-charges and all values of $N$ up to $N=200$, pushing the algorithm beyond the $N=60$ stopping point of \cite{Hosseini:2025jxb}.
Going to larger values of $N$ not only improves the numerical precision, it also reveals an important trend in the shape of the eigenvalue distribution that is not very noticeable at smaller values of $N$.
We also extend the numerical analysis to some non-canonical assignments of R-charges, by gradually deforming the canonical solution along one-parameter families of R-charge assignments.
We extract the leading order in the large\,-$N$ expansion of the free energy by fitting its numerical finite-$N$ values, finding agreement with the holographic prediction both on-shell and off-shell. The numerical analysis also gives an important clue about the mechanism behind cancellation of long range forces: for the assignments of R-charges we consider initially the long range forces cancel primarily against single pairs of eigenvalues on adjacent nodes whose real parts are very close and produce a large contribution to the saddle point equation due to the proximity to a pole of the cotangent terms. For more general R-charge assignments there is another ingredient needed for the cancellation of long range forces to occur, as we discuss shortly. 

We next consider the operator counting method of \cite{Gulotta:2011si, Gulotta:2011aa}, which is a conjectural relation between the number of certain gauge invariant operators and  the large\,-$N$ distribution of eigenvalues of the matrix model.
The operator counting method has already been applied to the chiral quiver dual to M-theory on AdS$_4 \times Q^{1,1,1}/{\mathbb{Z}_{\rm k}}$ in \cite{Kim:2012vza}, but with a different assignment of Chern-Simons levels than the one we consider in this paper.
Following the same logic as \cite{Kim:2012vza}, we compute the operator counting prediction for the eigenvalue density and eigenvalue differences for general R-charges and compare them with the numerical solutions. The two match rather well, with a caveat.
The finite\,-$N$ numerical saddles contain a ``bulk region'' whose relative size progressively shrinks to zero as $N$ is increased; the operator counting prediction does not have such a bulk region: in its place there is only a jump discontinuity.
The fact that the bulk region shrinks to the point of disappearing at large $N$ resolves the apparent mismatch observed in \cite{Hosseini:2025jxb} between the operator counting prediction and the finite\,-$N$ saddles. At the smaller values of $N\leq60$ considered in \cite{Hosseini:2025jxb}, the size of the bulk region is much larger compared to $N=200$ and the fact that its relative size scales approximately as $1/\sqrt{N}$ is not very evident.
These observations suggest that the analytic solution to the large\,-$N$ saddle point equations should also have a jump discontinuity in place of the bulk region of the numerical solutions.

After gathering information about the behavior of the distribution of eigenvalues from the numerical finite\,-$N$ saddles and the operator counting method, we then derive an analytical solution to the large\,-$N$ saddle point equations.
The free energy computed from the analytic saddle agrees perfectly with the holographic prediction obtained from the Sasakian volume \cite{Herzog:2010hf,Martelli:2011qj,Jafferis:2011zi} off-shell.
Furthermore, the eigenvalue density and difference between eigenvalues of the solution perfectly match the operator counting prediction.
Special care must be taken in showing that long range forces cancel, which requires a few other ingredients beyond the operator counting results.
The cancellation of the long range forces of the matrix model happens in a novel way: rather than vanishing identically, as in the case of non-chiral quivers, they cancel due to the specific form of the saddle.
When the trial R-charge assignment matches the superconformal one, the long range forces (outside the bulk region) are completely cancelled by the proximity of the eigenvalues to a pole of the saddle point equations.
For the analytic large\,-$N$ saddle this means that subleading corrections of order $1/N$ to the eigenvalue distribution are responsible for this cancellation; we compute these corrections and successfully match them with the numerical saddles.
This mechanism is sufficient to cancel the long range forces whenever the trial R-charges obey a linear relation.
For the most generic R-charge assignment however there is another mechanism at play: the cancellation of long range forces is tied to the presence of a large $\cO(N)$ imaginary part in the saddles.
For the analytic large\,-$N$ saddle, this corresponds to the addition of a $\ii N w(x)$ term in the ansatz for the eigenvalues, with piecewise constant $w(x)$ that is equal for all nodes and cancels in eigenvalue differences.
Since the operator counting method only predicts eigenvalue differences and densities, it is blind to the addition of such $\ii N w(x)$ terms.
Instead, we test the correctness of this addition numerically, finding a numerical saddle that matches the analytic prediction well, outside the bulk region.
Unlike the numerical solutions that we described earlier, we use the expected form of the saddle from the analytic computation as an input in the search for a numerical solution that validates it.
Surprisingly, our numerical search also yields a large number of other saddles with large imaginary parts that compete with it, and seem to be subdominant at the values of parameters that we tested.

The paper is structured as follows.
In section \ref{sec:chiral-quiver}, after describing the chiral quiver model of interest and the saddle point equations of its matrix model, we present the finite\,-$N$ numerical saddles, first for the canonical R-charges in section \ref{sec:numerical_solution}, and then for a non-canonical assignment of trial R-charges in section \ref{sec:noncanonical}.
In section \ref{sec:operator_counting} we consider the operator counting method and present its analytic prediction, which we then compare with the numerical solutions in section \ref{sec:Numerics_OPC}.
In section \ref{sec:large N expansions} we also briefly review the superconformal and topologically twisted indices, and explain how the large\,-$N$ expansion of the free energy on the round three-sphere can be related to the large\,-$N$ and Cardy expansion of the indices.
Section \ref{sec: Analytic solution} is for the most part dedicated to solving the large\,-$N$ saddle point equations analytically.
In section \ref{sec: Saddle point of Q111} we present a first solution that is valid for subclass of R-charge assignments and does not require a major modification of the usual ansatz for the eigenvalues.
Then in sections \ref{sec:varying the discontinuity} and \ref{sec:subleading corrections} we check the stability of the solution at the discontinuity point and the cancellation of long range forces respectively.
In section \ref{sec: A proposal for general Delta} we present a solution valid for general R-charge assignments featuring the addition of large imaginary parts in the eigenvalue ansatz, which we then test numerically in section \ref{sec:Delta_1_numeric_analysis}.
In section \ref{Comparison with the Sasakian volume} we compare our results with the gravitational prediction determined by the Sasakian volume.
We conclude with a discussion of open problems and future prospects in section \ref{sec:Discussion}.

\section{Chiral quiver dual to AdS$_4 \times Q^{1,1,1}/{\mathbb{Z}_{\rm k}}$}
\label{sec:chiral-quiver}

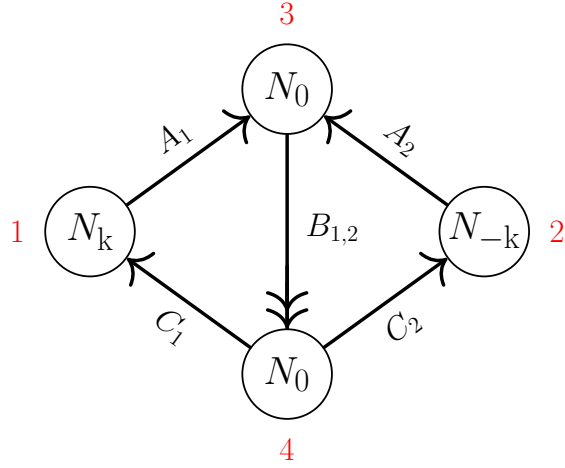
\begin{figure}[h]
  \centering 
\resizebox{0.5\textwidth}{!}{%
\begin{tikzpicture}[
    line cap=round,
    line join=round,
    quiver/.style={line width=2.2pt,-{>[scale=2.0]}},
    gauge/.style={
        circle,draw=black,fill=white,
        line width=1.15pt,
        minimum size=2.00cm,
        inner sep=0pt,
        font=\Huge
    },
    field/.style={font=\fontsize{20}{24}\selectfont},
    nodenumber/.style={font=\fontsize{20}{24}\selectfont,text=red}
]

\node[gauge] (T) at (0, 3.18) {$N_0$};
\node[gauge] (L) at (-4.40,0) {$N_{\rm k}$};
\node[gauge] (R) at ( 4.40,0) {$N_{-{\rm k}}$};
\node[gauge] (B) at (0,-3.18) {$N_0$};

\draw[quiver] (L) -- (T);
\draw[quiver] (R) -- (T);

\draw[line width=2.2pt] (T) -- (B);
\draw[quiver] ($(T)!0.62!(B)$) -- ($(T)!0.78!(B)$);
\draw[quiver] ($(T)!0.68!(B)$) -- ($(T)!0.84!(B)$);

\draw[quiver] (B) -- (L);
\draw[quiver] (B) -- (R);

\node[field,rotate= 36] at (-2.56, 2.11) {$A_1$};
\node[field,rotate=-36] at ( 2.56, 2.11) {$A_2$};
\node[field,rotate=-36] at (-2.58,-2.12) {$C_1$};
\node[field,rotate= 36] at ( 2.58,-2.12) {$C_2$};
\node[field,anchor=west] at (0.28,0.05) {$B_{1,2}$};

\node[nodenumber,anchor=east]  at ([xshift=-0.30cm]L.west)  {$1$};
\node[nodenumber,anchor=west]  at ([xshift= 0.30cm]R.east)  {$2$};
\node[nodenumber,anchor=south] at ([yshift= 0.30cm]T.north) {$3$};
\node[nodenumber,anchor=north] at ([yshift=-0.30cm]B.south) {$4$};

\end{tikzpicture}
}
\caption{The quiver diagram for the chiral Chern-Simons-matter theory dual to $\text{AdS}_4 \times Q^{1,1,1}/\mathbb{Z}_{\rm k}$. 
 In red we have indicated the ordering that we have chosen for the nodes.
  }
  \label{fig:Q111 quiver} 
\end{figure}

We focus our attention to a particular chiral $\cN=2$ Chern-Simons-matter theory, dual to M-theory on AdS$_4 \times Q^{1,1,1}/{\mathbb{Z}_{\rm k}}$ and described by the quiver diagram in Figure \ref{fig:Q111 quiver}. Its gauge group is U($N$)$^4$, with Chern-Simons levels $({\rm k},-{\rm k},0,0)$. It has six bifundamental chiral superfields, $A_{1,2}$, $B_{1,2}$ and $C_{1,2}$, each represented with an arrow in the quiver diagram with the convention that an arrow from the node $a$ to the node $b$ is a chiral multiplet in the $(\mathbf{N}_a,\overline{\mathbf{N}}_b)$ representation.
The fact that each arrow between two nodes is not accompanied by another with opposite direction makes the quiver \emph{chiral}.
The superpotential is given by 
\be
    W\,=\,h\:\text{Tr}\,\big(C_1\,A_1\,B_1\,C_2\,A_2\,B_2-C_1\,A_1\,B_2\,C_2\,A_2\,B_1\big)\:.
\ee
The global symmetry of the theory is SU(2)$^3\times$U(1)$_R\times$U(1)$^2_B$. Each one of the three pairs of chiral superfields $(A_1,A_2)$, $(B_1,B_2)$, $(C_1,C_2)$ transforms as a SU(2) doublet, with the  SU(2)$^3$ symmetry linked to the isometry of $Q^{1,1,1}=\:\,$SU(2)$^3/$U(1)$^2$. The superconformal U(1)$_R$ symmetry assigns R-charge equal to 1/3 to all chiral superfields. The remaining baryonic U(1)$^2_B$ symmetry may be parametrized as giving charges $(1,0,-1)$ and $(-1,1,0)$ to each SU(2) doublet. This chiral quiver theory describes the dynamics of a stack of $N$ M2-branes on the tip of the cone $\cC(Q^{1,1,1}/\bZ_{\rm k})$ over the Sasaki-Einstein manifold $Q^{1,1,1}/\bZ_{\rm k}$ \cite{Franco:2008um,Davey:2009sr,Hanany:2009vx,Franco:2009sp}.

In this section we focus on the free energy $F_{S^3}=-\log Z_{S^3}$ of this theory on the round three-sphere. The free energy of an $\cN=2$ Chern-Simons-matter quiver theory can be computed by means of supersymmetric localization, and the result is given by \cite{Kapustin:2009kz,Jafferis:2010un,Hama:2010av}
\begin{equation}
\label{matrix_model}
    Z_{S^3}(\Delta)\,=\int_{-\infty}^{+\infty}\bigg(\prod_{a=1}^{|\cG|}\prod_{i=0}^{N-1}\dd\lambda_{a,\,i}\bigg)\:\exp\Big(-\cF(\lambda,\Delta)\Big)\:,
\end{equation}
where we are specializing to the case where the gauge group $\cG$ is given by the product of $|\cG|$ U($N$) factors, and $\cF(\lambda,\Delta)$ is defined by
\begin{align}
\label{cF}
    \cF(\lambda,\Delta)={}&|\cG|\log((2\pi)^NN!)+\sum_{a=1}^{|\cG|}\sum_{i=0}^{N-1}
        \left(-\frac{\ii{\rm k}_a}{4\pi}\,\lambda^2_{a,\,i}+\Delta_{m;\, a}\,\lambda_{a,\,i}\right)-\\\nn
    {}&-2\sum_{a=1}^{|\cG|}\,\sum_{i>j\geq0}^{N-1}\log\left(2\,\sinh\,\frac{\lambda_{a,\,i}-\lambda_{a,\,j}}2\right)
        -\sum_{I=1}^{|\cR|}\sum_{\rho\,\in\,\cR_I}\ell\left(1-\Delta_I+\ii\,\frac{\rho(\lambda)}{2\pi}\right)\:.
\end{align}
The numeric factor $|\cG|\log((2\pi)^NN!)$ is the logarithm of the order of the Weyl group of $\cG$ times $(2\pi)^{N|\cG|}$, the latter factor coming from the fact that in our notation the $\lambda_{a,\,i}$ are $2\pi$ times the constant value of the vector multiplet scalars at the BPS locus.
The other terms in the first line of \eqref{cF} are the classical terms obtained by evaluating the action of the theory at the BPS locus, and they receive contributions from the Chern-Simons terms with levels ${\rm k}_a$ and topological symmetry terms; the latter depend on the classical R-charge $\Delta_{m;\,a}$ of a monopole operator with unit flux at node $a$.
The terms in the second line of \eqref{cF} are the one-loop determinants of the vector and chiral multiplets.
The index $I=1,\ldots,|\cR|$ runs over the chiral multiplets of the theory, each one transforming in the representation $\cR_I$ of the gauge group and carrying (trial) R-charge $\Delta_I$, and the function $\ell(z)$ is defined as
\be
\label{ell_def}
    \ell(z)=\frac{\ii}{2\pi}\Li_2(e^{2\pi\ii z})+z\Li_1(e^{2\pi\ii z})+\frac{\pi\ii}{2}\,z^2-\frac{\pi\ii}{12}\:.
\ee

For the chiral quiver theory dual to AdS$_4 \times Q^{1,1,1}/{\mathbb{Z}_{\rm k}}$ we have $|\cG|=4$, $|\cR|=6$, ${\rm k}_1=-\,{\rm k}_2\equiv{\rm k}$, $\:{\rm k}_3={\rm k}_4=0$, and the representations $\cR_I$ are all bifundamental with weights $(\rho_I^{ij})_{i,j\,=\,0,\ldots,N-1}\:\in\cR_I$ such that $\rho_I^{ij}(\lambda)=\lambda_{a,\,i}-\lambda_{b,\,j}$, with $a$ and $b$ being the start and end node of the $I$-th arrow in the quiver. For convenience we will mostly use $I\in\{A_1,A_2,B_1,B_2,C_1,C_2\}$ rather than $I\in\{1,\ldots,6\}$. The trial R-charges must satisfy the marginality constraint
\be
    \Delta_{A_1}+\Delta_{A_2}+\Delta_{B_1}+\Delta_{B_2}+\Delta_{C_1}+\Delta_{C_2}=2\:,
\ee
and generically differ from the superconformal R-charges, which are all equal to 1/3. By the principle of $F$-extremization \cite{Jafferis:2010un}, the latter should give the maximum \cite{Jafferis:2011zi} of $F_{S^3}(\Delta)$.

We are interested in finding explicit solutions to the saddle point equations for the matrix model \eqref{matrix_model}, namely finding (complex) values for the eigenvalues $\lambda_{a,\,i}$ that solve the following equations:
\be
    \partial_{\lambda_{a,\,i}}\cF(\lambda,\Delta)=0\:.
\ee
When computing the derivative of $\cF(\lambda,\Delta)$ it is useful to note that $\partial_z\ell(z)=-\pi z\cot(\pi z)$.

In this section, following \cite{Hosseini:2025jxb}, we will solve the saddle point equations for the finite-$N$ free energy $F_{S^3}$ at various values of $N$ and extrapolate the large\,-$N$ behavior numerically. In section \ref{sec:numerical_solution} we will focus on the superconformal assignment of R-charges $\Delta_I=1/3$, which is the only one considered in \cite{Hosseini:2025jxb} for this quiver, and we will push the numerical analysis to a higher value of $N$, $N=200$. The higher value of $N$ will make a certain trend in the behavior of the eigenvalues that went unnoticed in \cite{Hosseini:2025jxb} more apparent. Then in section \ref{sec:noncanonical} we will extend the numerical analysis to a non-canonical assignment of R-charges. Then in section \ref{sec:operator_counting} we will consider the operator counting conjecture \cite{Gulotta:2011si, Gulotta:2011aa}, and in section \ref{sec:Numerics_OPC} we will compare its prediction with the numerical analysis finding a good match.

\subsection{Numerical solution for canonical R-charges}
\label{sec:numerical_solution}
\begin{figure}[h]
     \centering
     \begin{subfigure}[b]{0.5\textwidth}
         \centering
         \includegraphics[width=\linewidth]{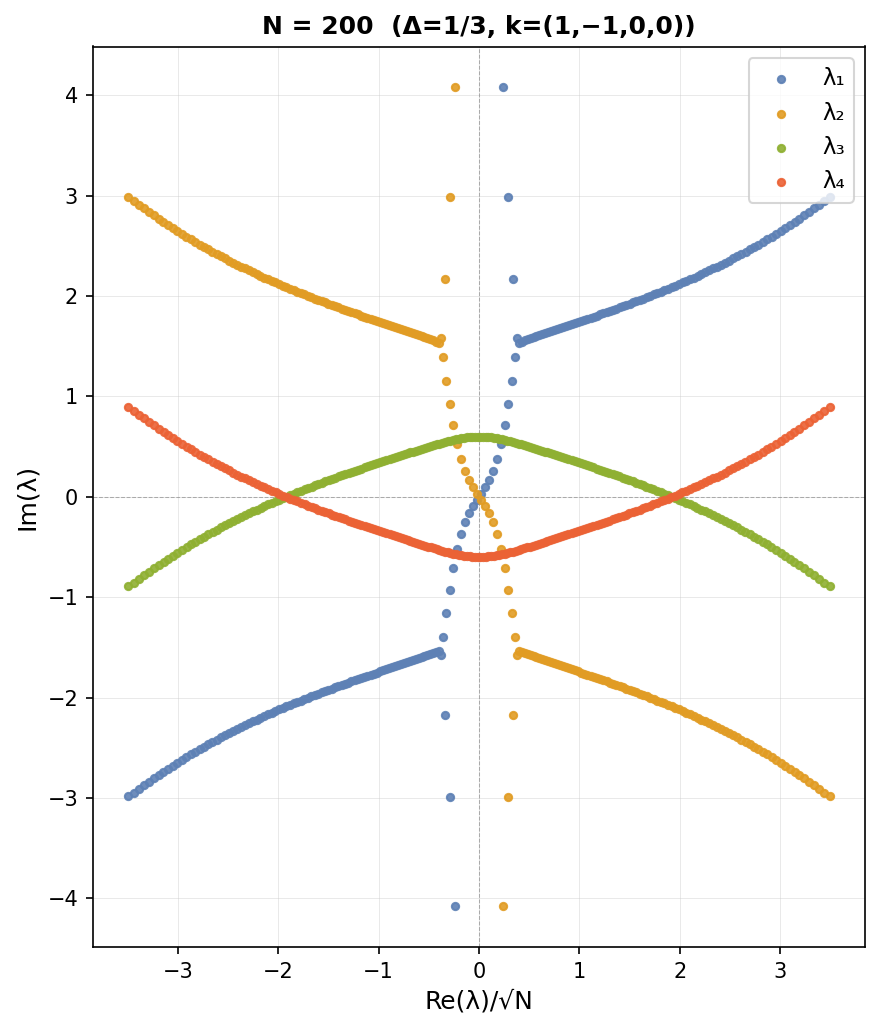}
         \caption{The eigenvalue distributions of all four nodes at $N=200$.}
         \label{fig:eigenvalue}
     \end{subfigure}
     \hfill 
     \begin{subfigure}[b]{0.45\textwidth}
         \centering
         \includegraphics[width=\linewidth]{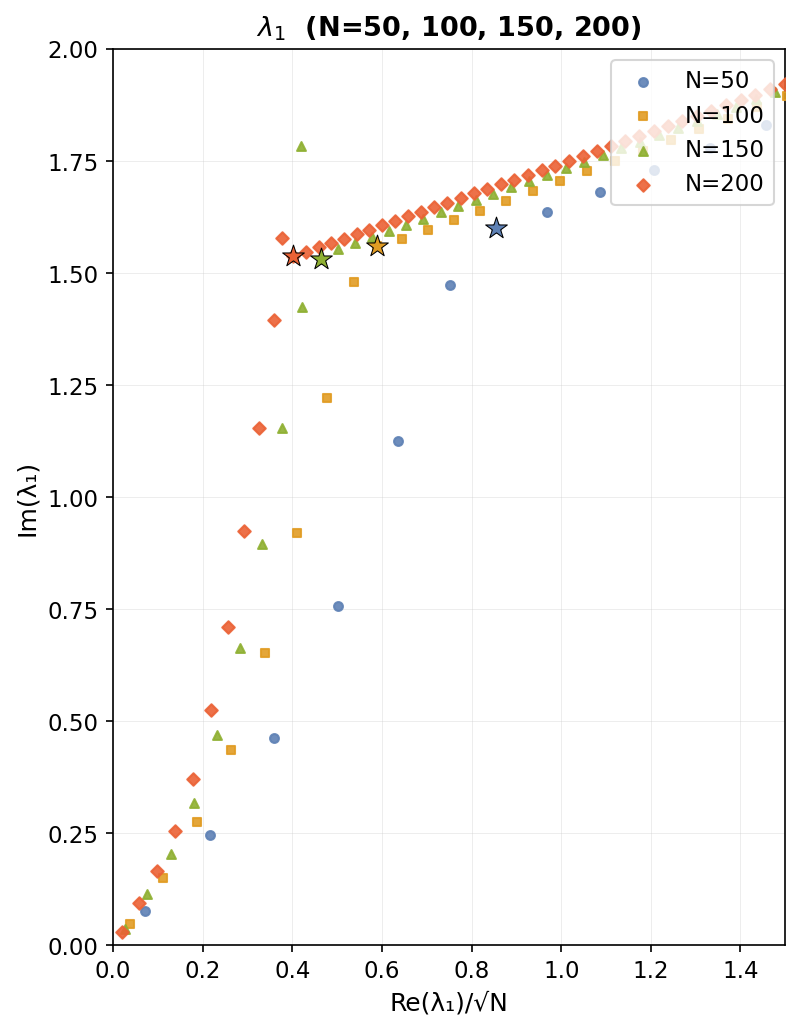}
         \caption{Zoomed-in views of the first quadrant for various values of $N$.}
         \label{fig:kink}
     \end{subfigure}
     
     \caption{The eigenvalue distribution for the chiral $Q^{1,1,1}/\mathbb{Z}_{\rm k}$ quiver.}
     \label{fig:nodes_comparison}
\end{figure}

In this section, we revisit the numerical solutions to the saddle point equations studied in \cite{Hosseini:2025jxb}\footnote{We solve the saddle point equations numerically using Python code developed with the assistance of Claude and Gemini.}. We find it crucial to choose the right initial values for the roots in numerical methods, as was alluded to in \cite{Hosseini:2025jxb}, for large\,-$N$.
We begin with a small value of $N$, for example $N=2$, and find the roots numerically. We then use the resulting solution to guess the best initial values for solving the equations at $N+1$, using the large\,-$N$ behavior of the roots we aim to find. By repeating this procedure, we obtain numerical solutions for large values of $N$.
In this way, we managed to secure numerical solutions up to $N=200$, significantly improving over the $N=60$ results reported in \cite{Hosseini:2025jxb}. The eigenvalue distributions at $N=200$ are shown in Figure \ref{fig:eigenvalue}. Throughout this section, we set ${\rm k}=1,\,\Delta_I=1/3$ and 
$\Delta_m=0$. As expected, the real parts of the eigenvalues scale as $N^{1/2}$, while the imaginary parts remain of order $\mathcal{O}(1)$ \cite{Herzog:2010hf,Martelli:2011qj,Cheon:2011vi,Jafferis:2011zi}. Their overall shapes agree well with the figures in \cite{Hosseini:2025jxb}, where the eigenvalue distributions of nodes $1$ and $2$ can be divided into three regions: let us call them left tail, the bulk, and the right tail. It is obvious that the nodes 1 and 2 constitute odd functions, while nodes 3 and 4 give even functions.

\begin{figure}[h]
    \centering

    \begin{minipage}[t]{0.4\textwidth}
        \centering

        \begin{minipage}[c][5.5cm][c]{\linewidth}
            \centering
            \begin{tabular}{cccc}
                \toprule
                \(N\)
                & \(\operatorname{Re}(\lambda_{\mathrm{kink}})/\sqrt{N}\)
                & \(n_1\)
                & \(n_2\) \\
                \midrule
                 50 & 0.85463 &  7 & 18 \\
                100 & 0.58943 & 10 & 40 \\
                150 & 0.46485 & 12 & 63 \\
                200 & 0.40229 & 14 & 86 \\
                \bottomrule
            \end{tabular}
        \end{minipage}

        \captionof{table}{The kink position and the numbers of data points in the $\text{Re}(\lambda)>0$ bulk and tail regions, respectively, for various values of $N$.}
        \label{tab:numerical-data}
    \end{minipage}
    \hfill
    \begin{minipage}[t]{0.59\textwidth}
        \centering

        \begin{minipage}[c][5.5cm][c]{\linewidth}
            \centering
            \includegraphics[
                width=\linewidth,
                height=5.5cm,
                keepaspectratio
            ]{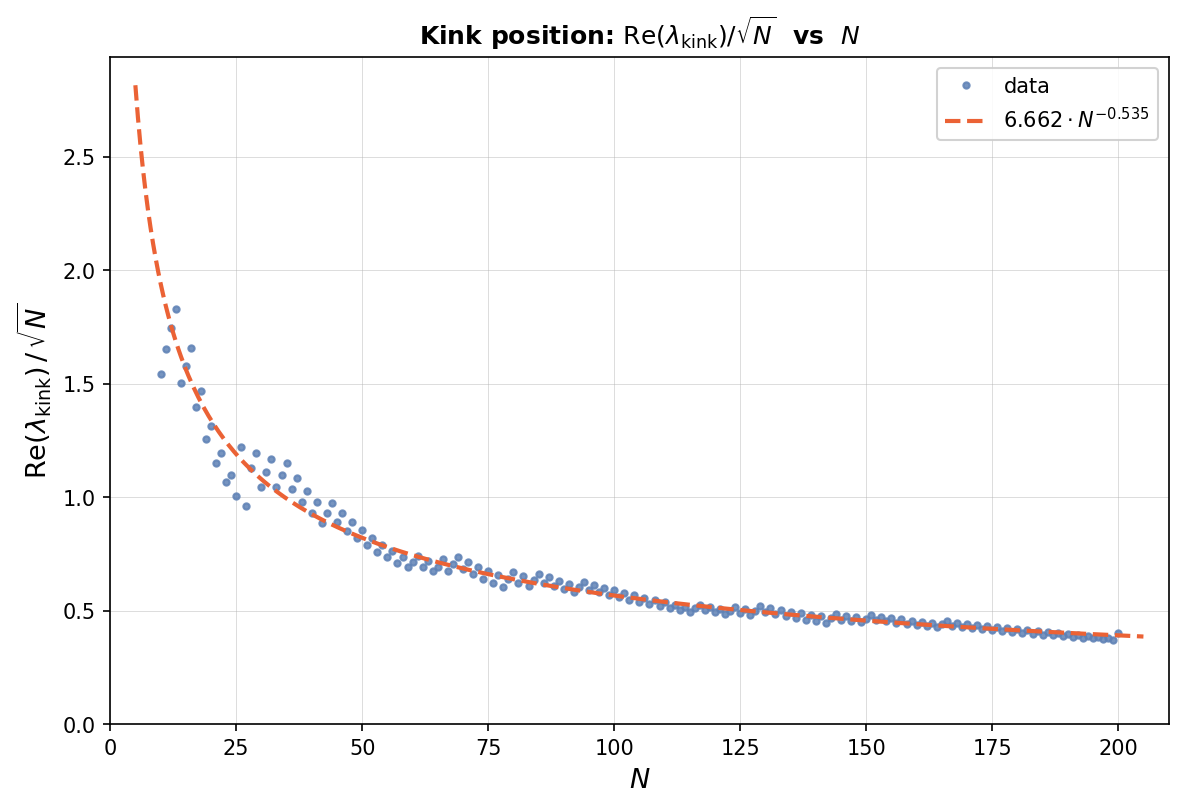}
        \end{minipage}

        \captionsetup{margin={1cm,0cm}}
        \captionof{figure}{The kink position as a function of \(N\), compared with a power law fit. We find that $\operatorname{Re}(\lambda_{\mathrm{kink}})/\sqrt{N}\,\sim\,N^{-0.535}$\:.}
        \label{fig:kink-position}
    \end{minipage}
\end{figure}

 Let us zoom in on the eigenvalue distributions of node $1$ for different values of $N$, as shown in more detail in Figure \ref{fig:kink}. The star-marked points denote the location of kink, i.e. the boundary between the bulk and tail. An interesting feature is that the scaled position of the kink approaches zero as $N$ increases, as shown in Table \ref{tab:numerical-data} and Figure \ref{fig:kink-position}. Moreover, the number of data points in the bulk region increases more slowly than that in the tail region. These observations indicate that the bulk region should be sub-dominant as $N\rightarrow\infty$, so we expect their contribution to the free energy to be suppressed.

\begin{figure}[h]
  \centering
  \includegraphics[width=0.75\textwidth]{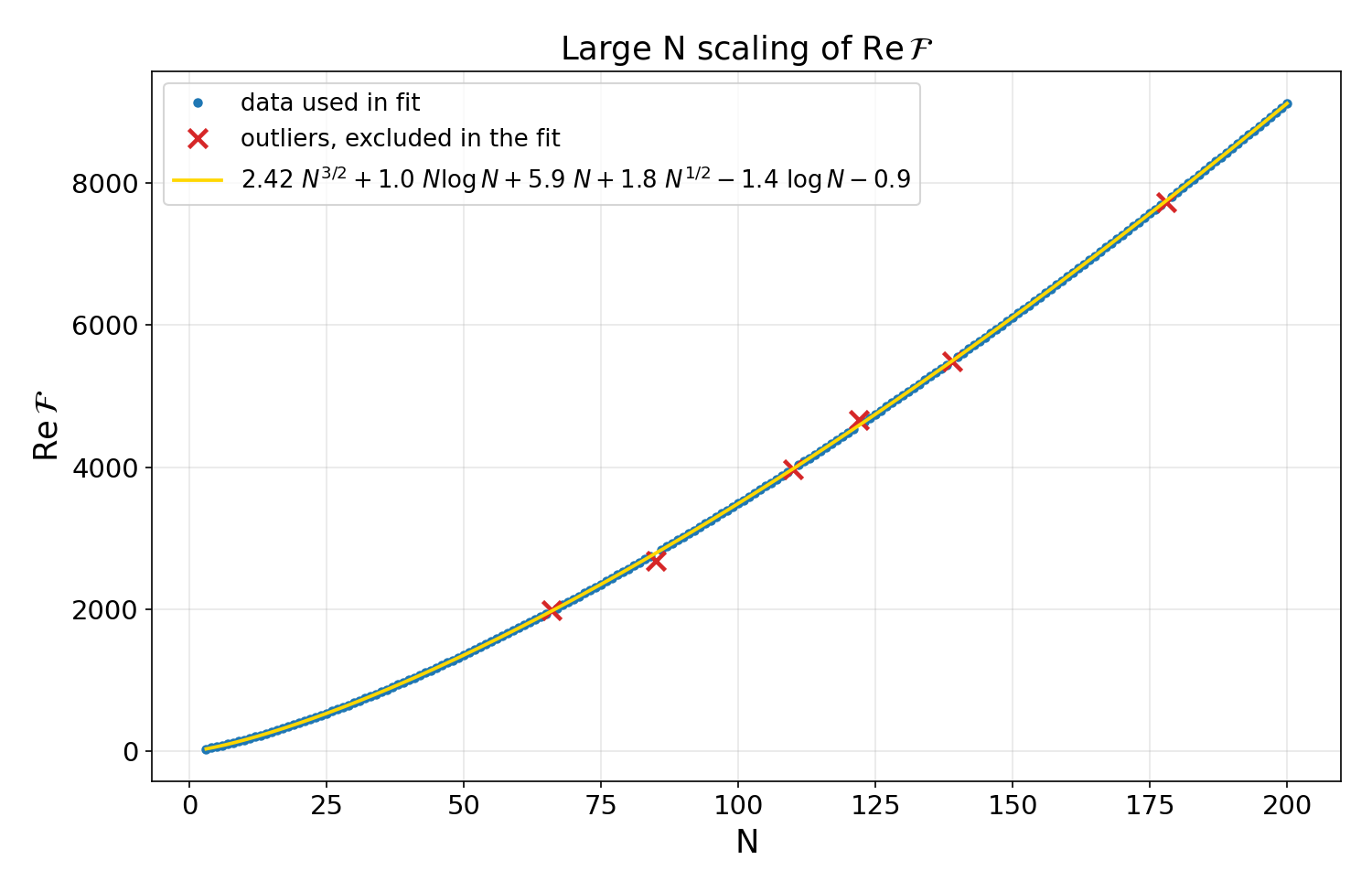}
  \caption{Numerical fit of the real part of $\cF$ as a function of $N$.}
  \label{fig:free energy fit}
\end{figure}

 We evaluate the free energy by substituting the numerical solutions in the function $\cF(\lambda,\Delta)$ given by \eqref{cF}, and fitting the result as a function of $N$ in the basis $\{N^{3/2}$,\, $N\, \textrm{log} N$,\, $N$,\, $N^{1/2}$,\, $\textrm{log} N$,\, $1\}$. 
 The plot of the fit is in Figure \ref{fig:free energy fit}, we obtain
 \footnote{The imaginary part is zero up to numerical accuracy, namely $\approx10^{-10}$.
    This requires some carefulness, since the polylogarithms in the definition \eqref{ell_def} of $\ell(z)$ have branch cut
    ambiguities. We fix these ambiguities by using
    \be
        \ell(z)=z\int_0^1\dd t\:\ell'(tz)=-\pi\,z^2\int_0^1\dd t\cdot t\:\cot(\pi tz)\:.
    \ee
    }
 \be
 \label{ReF0_numerical_fit}
    \cF_{\text{numerical}}(\Delta_I=1/3)\:=\:2.42\: N^{3/2} +1.0\: N\, \textrm{log}N+5.9\: N+\cdots.
 \ee
By saddle point approximation, the leading order of the free energy $F_{S^3}$ matches the leading order of $\cF$ evaluated
on the solution, and thus we find $F_{S^3}\,\approx\,2.42\:N^{3/2}$.
 The leading contribution to the free energy scales as $N^{3/2}$ and its numerical coefficient is in excellent agreement with the holographic result, $F_{\textrm{grav}}=\frac{4\pi}{3\sqrt{3}} \, N^{3/2}\approx 2.418\, N^{3/2}.$
We will discuss the matching with holography in section \ref{Comparison with the Sasakian volume}.
 The $ N\,\textrm{log}N$ term in the numerical fit is expected to be canceled by gaussian one-loop determinants around the saddle of the matrix model,
 along the lines of \cite{Liu:2019tuk,Hong:2021bsb,Geukens:2024zmt}, but a detailed analysis of such corrections is beyond the scope of the present paper.

\subsection{Numerical solution for a non-canonical assignment of R-charges}
\label{sec:noncanonical}

In the previous subsection we described a numerical solution for the canonical R-charge assignment $\Delta_I=1/3$.
For the chiral quiver of interest, dual to $Q^{1,1,1}/\bZ_{\rm k}$, the same algorithm struggles to converge to a solution
for non-canonical R-charge assignments, as already noticed in \cite{Hosseini:2025jxb}.
Nonetheless, we want to probe numerically the form of the saddle of the matrix model for more general R-charges.
Therefore, in this section, we will adopt a different approach: we will choose a one-parameter family of R-charge assignments
which includes the canonical R-charges $\Delta_I=1/3$, and probe numerically how the matrix model saddle deforms as the R-charges are gradually changed along the given one-parameter family.%
\footnote{The matrix model is invariant under the following flat directions \cite{Jafferis:2011zi}
\be
   \lambda_{a,\,i}\:\to\:\lambda_{a,\,i}-2\pi\ii\,\delta^{(a)}\:,\qquad
   \Delta_I\:\to\:\Delta_I+\delta^{(a)}-\delta^{(b)}\:,\qquad
   \Delta_{m;\,a}\:\to\:\Delta_{m;\,a}+{\rm k}_a\delta^{(a)}\:.
\ee
When considering non-canonical R-charges, we will avoid values of $\Delta_I$ that can be reached from the canonical ones by moving along a flat direction, as that would be trivial.}
With this approach, we will be able to converge to a numerical solution with non-canonical R-charges.
We keep ${\rm k}=1$ throughout this subsection.

A natural candidate for a simple one-parameter family of R-charge assignments would be
$\Delta_{A_i}=\Delta_{B_i}=\Delta_{C_i}\,\equiv\,\Delta_i\,$, with $\Delta_1+\Delta_2=2/3$.
This seemingly simple case however hides a very complicated structure: in section \ref{sec:Delta_1_numeric_analysis}
we will describe an algorithm that converges to a large number of different competing numerical saddles for these values of
R-charges. 
We postpone the discussion of this case to Section \ref{sec:Delta_1_numeric_analysis}. This is because the numerical algorithm most effective for these R-charge assignments relies on an observation regarding the expected large\,-$N$ analytic behavior of the eigenvalues, which we will introduce in Section \ref{sec: A proposal for general Delta}.

\begin{figure}[h]
  \centering
  \includegraphics[width=0.57\textwidth]{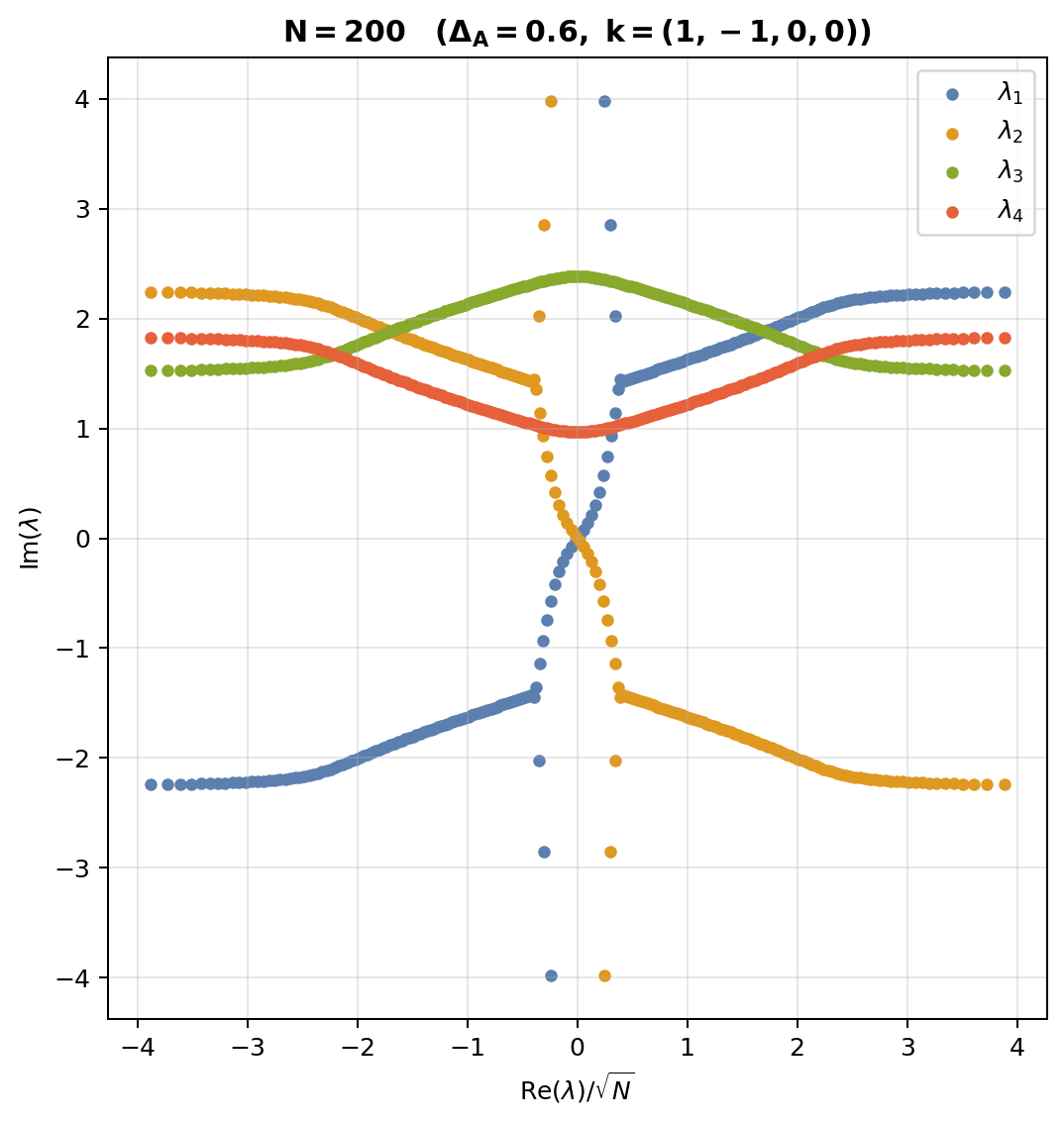}
  \caption{The eigenvalue distributions of all four nodes at $N=200$ for the solution with $\Delta_{A_i}=\Delta_{B_1}=0.6,\,\Delta_{C_i}=\Delta_{B_2}$.}
  \label{fig:N200_Delta_A}
\end{figure}

Instead, in this section we will consider the one-parameter family given by
\be
    \Delta_{A_1}=\Delta_{A_2}=\Delta_{B_1}\,\equiv\,\Delta_A\:,\qquad\Delta_{C_1}=\Delta_{C_2}=\Delta_{B_2}=2/3-\Delta_A\:,
\ee
which among various nontrivial one-parameter families that we have tested converged to a new numerical solution quickly and without issues for all values of $N$ up to $N=200$. For each $N$ we start from the solution described in the previous section and valid for $\Delta_A=1/3$, and we increase $\Delta_A$ in small steps. At each step the new eigenvalues are initially predicted by linear extrapolation from the solution at the previous value of $\Delta_A$ and the current rate of change of the eigenvalues with $\Delta_A$.%
\footnote{Namely after $\Delta_A^{\text{new}}=\Delta_A^{\text{old}}+\epsilon$, we set
$\lambda^{\text{new guess}}_{a,\,i}=\lambda^{\text{old solution}}_{a,\,i}+\epsilon\,v_{a,\,i}$, where $v$ is the solution to the following linear system: $\sum_{b,\,j}\partial^2\cF/(\partial\lambda_{a,\,i}\,\partial\lambda_{b,\,j})\:v_{b,\,j}=-\partial^2\cF/(\partial\lambda_{a,\,i}\,\partial\Delta_A)$.}
Then a damped Newton iteration is used to converge to a solution. We stop at the value $\Delta_A=0.6$.
The numerical solution for $N=200$ and $\Delta_A=0.6$ is plotted in Figure \ref{fig:N200_Delta_A}.

\begin{figure}[h]
     \centering
     \begin{subfigure}[t]{0.49\textwidth}
         \centering
         \includegraphics[width=\linewidth]{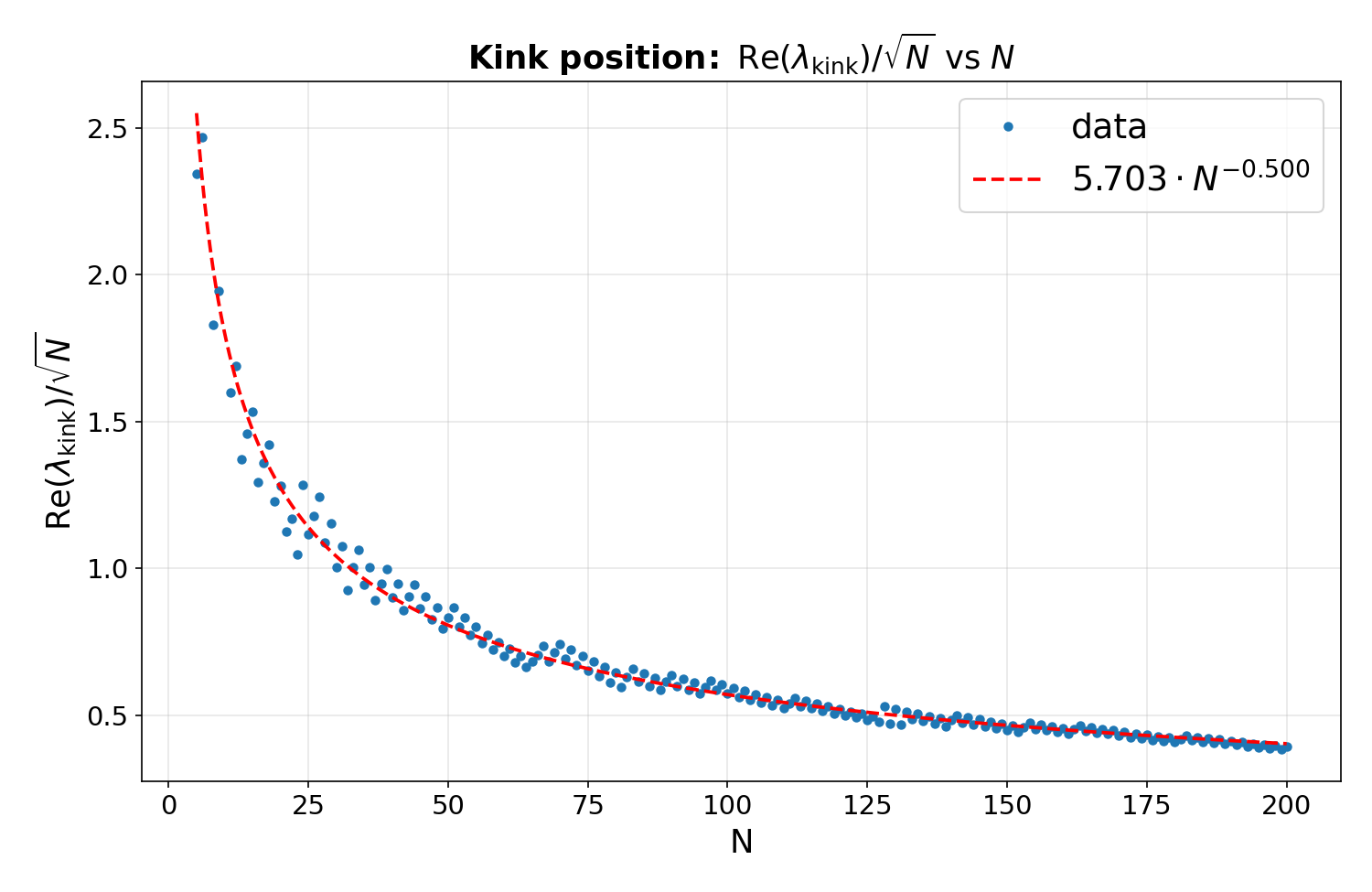}
         \caption{The kink position as a function of $N$ for $\Delta_{A_i}=\Delta_{B_1}=0.6,\,\Delta_{C_i}=\Delta_{B_2}$.
         We find that $\operatorname{Re}(\lambda_{\mathrm{kink}})/\sqrt{N}\,\sim\,N^{-0.500}$\:.}
         \label{fig:kink_vs_N_Delta_A}
     \end{subfigure}
     \hfill
     \begin{subfigure}[t]{0.49\textwidth}
         \centering
         \includegraphics[width=\linewidth]{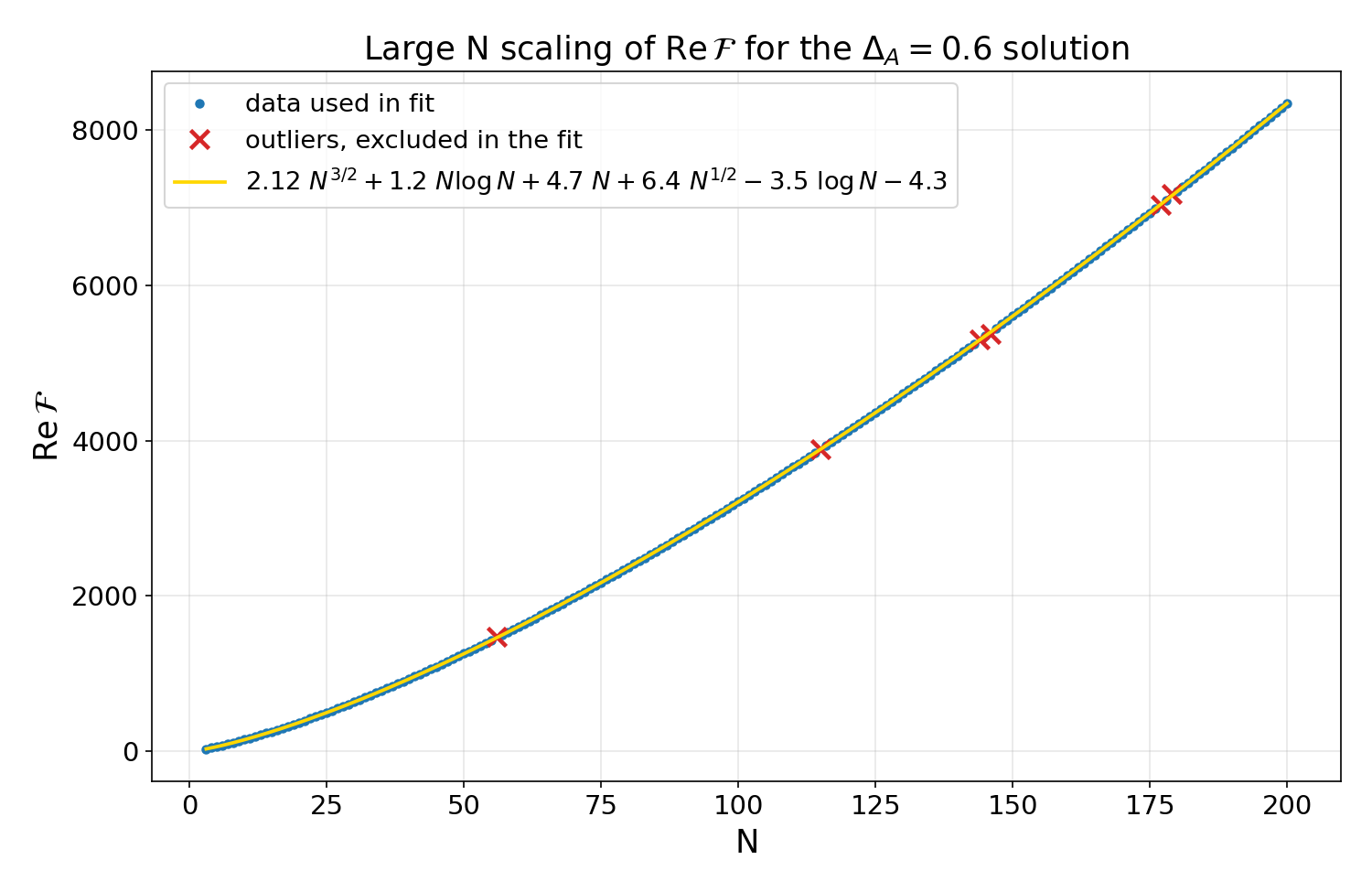}
          \captionsetup{margin={5mm,0cm}}
         \caption{Numerical fit of the real part of $\cF$ as a function of $N$ for $\Delta_{A_i}=\Delta_{B_1}=0.6,\,\Delta_{C_i}=\Delta_{B_2}$.}
         \label{fig:ReF_fit_Delta_A}
     \end{subfigure}
     \caption{}
     \label{fig:fits_for_Delta_A}
\end{figure}

Again, the real parts of the eigenvalues scale as $N^{1/2}$ while the imaginary parts remain of order $\cO(1)$.
In the plot, the $x$ axis has been rescaled to account for this, $x=\text{Re}(\lambda)/\sqrt N$.
The eigenvalue distributions follow an odd-function profile for the nodes 1 and 2, and an even-function profile for nodes 3 and 4. Near the origin, there is once again a bulk region where the eigenvalue distributions for the nodes 1 and 2 have a sharp change in behavior. As $N$ increases, the width of this bulk region in the rescaled plot shrinks to zero, as shown in Figure \ref{fig:kink_vs_N_Delta_A}.

We evaluate the numerical free energy with the same method that we described in the previous section, present the plot of the fit in Figure \ref{fig:ReF_fit_Delta_A}, and find the result as follows
 \be
 \label{ReF0_numerical_fit_Delta_A}
    \cF_{\text{numerical}}(\Delta_{A_i}=\Delta_{B_1}=0.6,\,\Delta_{C_i}=\Delta_{B_2})\:=\:2.12\: N^{3/2} +1.2\: N\, \textrm{log}N+4.7\: N+\cdots.
 \ee
 The leading order coefficient matches well with the holographic result, which for these values of R-charges is $F_{\text{grav}}\approx2.140\:N^{3/2}$.%
 \footnote{This value can be found from \eqref{free_energy_with_general_Delta}, after substituting the given assignment of R-charges.}
 Again, we expect the $N\log N$ term to be canceled by gaussian one-loop determinants around the saddle, but we will not perform such an analysis.

\subsection{Analytic solution from the operator counting method}
 \label{sec:operator_counting}
We next address the operator counting method \cite{Gulotta:2011si, Gulotta:2011aa}, which relates the number of certain gauge invariant operators in the chiral ring of the field theory to matrix model quantities in the large\,-$N$ limit. This method has passed several nontrivial tests for non-chiral quiver gauge theories \cite{Gulotta:2011si, Gulotta:2011aa, Kim:2012vza}.  It has also been used to provide prescriptions for obtaining matrix model quantities in chiral quiver theories \cite{Gulotta:2011aa, Kim:2012vza}, to which the conventional large\,-$N$ matrix model approach had long been considered inapplicable. 

Let $\psi(r,m)$ be the number of gauge invariant operators with R-charge and monopole charge less than $r$, $m$ respectively, and let $\psi_{X_{ab}}(r,m)$ be the number of such operators which do not contain the bifundamental field $X_{ab}$.\footnote{We use the convention where the bifundamental field $X_{ab}$ is in the $(\mathbf{N}_a,\overline{\mathbf{N}}_b)$ representations. For the chiral quiver dual to $Q^{1,1,1}/\bZ_{\rm k}$, $X_{ab}$ corresponds to the fields $A_1$, $A_2$, $B_1$, $B_2$, $C_1$ and $C_2$.} 
Then it was proposed in \cite{Gulotta:2011si, Gulotta:2011aa} that the 
large\,-$N$ saddle of the matrix model be related to $\psi$ and $\psi_{X_{ab}}$ as follows:
\footnote{Note that $(x)_{\text{here}}=2\pi(x)_{\text{in \cite{Gulotta:2011si, Gulotta:2011aa,Kim:2012vza}}}$, $(\rho)_{\text{here}}=\frac{1}{2\pi}(\rho)_{\text{in \cite{Gulotta:2011si, Gulotta:2011aa,Kim:2012vza}}}$, $(y_a)_{\text{here}}=2\pi (y_a)_{\text{in \cite{Gulotta:2011si, Gulotta:2011aa,Kim:2012vza}}}$, $(\mu)_{\text{here}} = 4 \pi^2 (\mu)_{\text{in \cite{Gulotta:2011si, Gulotta:2011aa,Kim:2012vza}}}$.
Also $X_{ab}$ in \cite{Gulotta:2011si, Gulotta:2011aa,Kim:2012vza} is in the $(\overline{\mathbf{N}}_a ,\mathbf{N}_b)$ representation, while here it is in $(\mathbf{N}_a,\overline{\mathbf{N}}_b)$.}
\begin{align}
    \frac1{(2\pi)^3}\:\frac{\partial^3 \psi}{\partial r^2 \partial m}\bigg |_{m\,=\,2\pi r\,x/\mu}&= \frac{r}{\mu}\,\rho(x),\label{opcrho} \\
    \frac1{(2\pi)^2}\:\frac{\partial^2\psi_{X_{ab}}}{\partial r\partial m}\bigg |_{m\,=\,2\pi r\,x/\mu} &= \frac{r}{\mu}\,\rho(x) \Big(y_a (x)- y_b(x)+2\pi\Delta_{X_{ab}}\Big),\label{opccy}
\end{align}
where the functions $\rho$ and $y_a$ are determined from the eigenvalues $\lambda_{a,\,i}$ by
\begin{equation}
    \lambda_{a,\,i}=N^{1/2}\,x_i+\ii\,y_a(x_i)\:,\qquad x_i,\,y_a(x_i)\in\mathbb R\:,\qquad\rho(x_i)=\frac1{N(x_{i+1}-x_i)}\:,
\end{equation}
and at large\,-$N$ the discrete $x_i$ can be replaced by a continuous variable $x$.
Lastly, $\mu$ is a Lagrange multiplier whose value is fixed by imposing the appropriate normalization of the eigenvalue density $\rho$,
namely $\int dx\,\rho(x)=1$.

We may consider the abelian case $N=1$ and count the gauge invariant operators for this simple case without
losing generality. In the non-abelian case many more operators may be written, but they are not expected to be relevant
at large\,-$N$.
The gauge-invariant operators whose counting encodes the eigenvalue data of the chiral quiver dual to $Q^{1,1,1}/\bZ_{\rm k}$ can be constructed as
\be
\begin{aligned}
{}&T^{(m)}\:A_1^{n_1}\:A_2^{m{\rm k}+n_2}\:B_1^{n_3}\:B_2^{m{\rm k}+n_1+n_2-n_3}\:C_1^{m{\rm k}+n_1}\:C_2^{n_2}\:,\qquad&&\text{if }\:m>0, \\
{}&T^{(m)}\:A_1^{|m|{\rm k}+n_1}\:A_2^{n_2}\:B_1^{n_3}\:B_2^{|m|{\rm k}+n_1+n_2-n_3}\:C_1^{n_1}\:C_2^{|m|{\rm k}+n_2}\:,\qquad&&\text{if }\:m<0,
\label{GIops}
\end{aligned}
\ee
where $n_1,\,n_2,\,n_3$ are non-negative and $n_3<|m|\,{\rm k}+n_1+n_2$, while $T^{(m)}$ is a diagonal monopole operator with monopole number $m$ and R-charge $m \Delta _m$,%
\footnote{When $m<0$, $T^{(m)}$ represents the anti-monopole operator with R-charge $-|m| \Delta_m$.}
it has charge $m\,({\rm k},-{\rm k},0,0)$ under each one of the four gauge nodes.

The explicit computation proceeds along the same lines as in \cite{Kim:2012vza}, where the same quiver with a different Chern-Simons level assignment $({\rm k},{\rm k},-{\rm k},-{\rm k})$ was analyzed. 
We also allowed generic values of the R-charges here. When $m>0$, in the limit where $n_1,\,n_2,\,n_3$ may be treated
as continuous variables, we have
\be
\begin{aligned}
    \frac{\partial^2\psi}{\partial r\partial m}=\int_0^\infty\dd n_1\int_0^\infty\dd n_2\int_0^{m\,{\rm k}+n_1+n_2}\dd n_3\:
        \delta\big({}&m \Delta_m +n_1 \Delta_{121}+n_2 \Delta_{222}\,+\\
    {}&+ n_3 (\Delta_{B_1}-\Delta_{B_2})+m{\rm k}\Delta_{221} - r\big)\:,
\end{aligned}
\ee
where we used a Dirac delta to fix the R-charge of the operators being counted,
and we introduced the following shorthand notation for linear combinations of $\Delta_I$:
\begin{equation}
	\Delta_{ijk}\,\equiv\,\Delta_{A_i}+\Delta_{B_j}+\Delta_{C_k}\:,\qquad i,j,k\in\{1,2\}\:.
\end{equation}
An analogous expression is also valid for $\psi_{X_{ab}}$ in place of $\psi$ if a further Dirac delta is inserted
in the integrand to impose the absence of the $X_{ab}$ field, e.g.\ $\delta(n_1)$ for $X_{ab}=A_1$ and so on.
The $m<0$ case works in a similar way.
In the following, we omit further details and only record the final result.

\begin{figure}[htbp]
     \centering
     \begin{subfigure}[b]{0.48\textwidth}
         \centering
         \includegraphics[width=\linewidth]{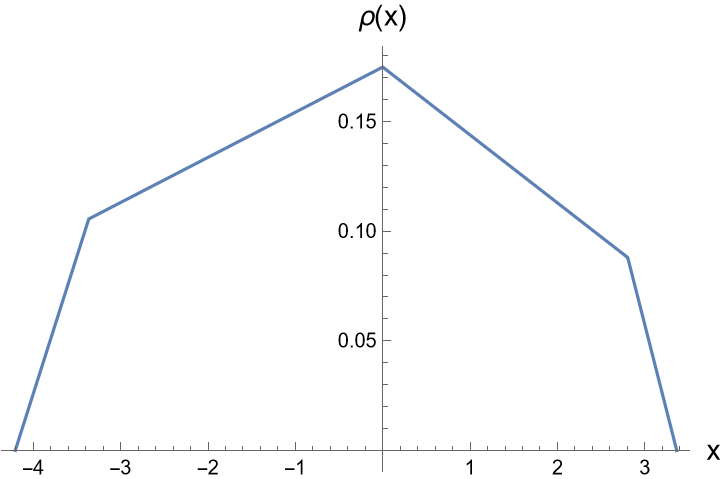}
         \caption{Eigenvalue density $\rho(x)$}
         \label{fig:operator_counting_rho}
     \end{subfigure}
     \hfill
     \begin{subfigure}[b]{0.48\textwidth}
         \centering
         \includegraphics[width=\linewidth]{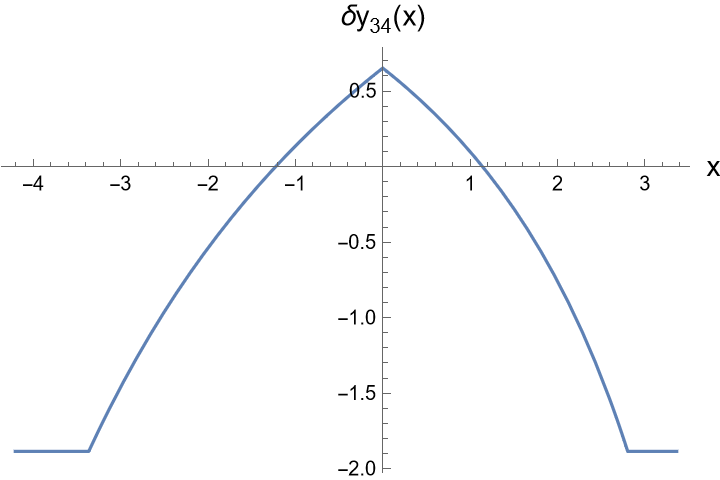}
         \caption{Eigenvalue difference $y_3(x)-y_4(x)$}
         \label{fig:operator_counting_y34}
     \end{subfigure}
     
     \vspace{5mm}
     \begin{subfigure}[b]{0.48\textwidth}
         \centering
         \includegraphics[width=\linewidth]{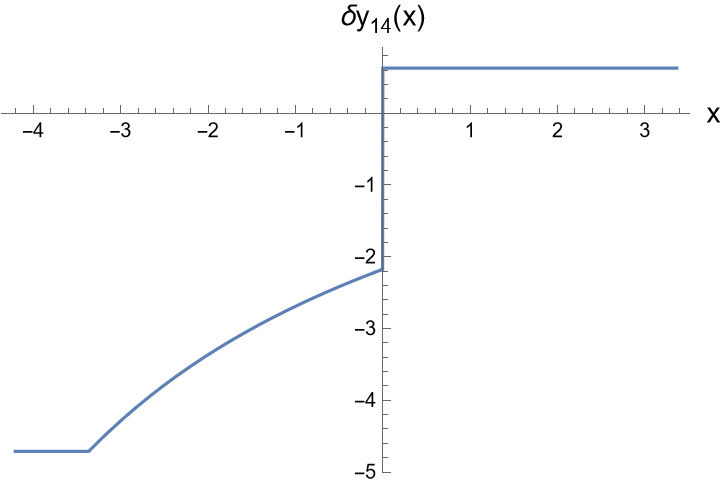}
         \caption{Eigenvalue difference $y_1(x)-y_4(x)$}
         \label{fig:operator_counting_y14}
     \end{subfigure}
     \hfill
     \begin{subfigure}[b]{0.48\textwidth}
         \centering
         \includegraphics[width=\linewidth]{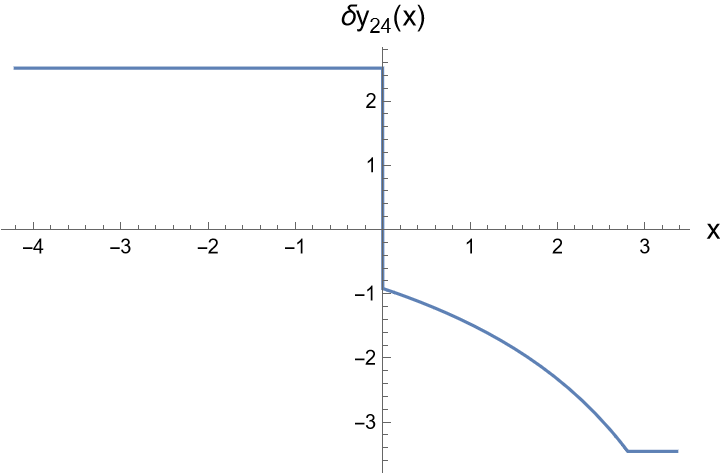}
         \caption{Eigenvalue difference $y_2(x)-y_4(x)$}
         \label{fig:operator_counting_y24}
     \end{subfigure}     
     \caption{Plot of the operator counting result for the following assignment of values: ${\rm k}=1$, $\Delta_{A_1}=0.45$, \,$\Delta_{A_2}=0.25$, \,$\Delta_{B_1}=0.5$, \,$\Delta_{B_2}=0.3$, \,$\Delta_{C_1}=0.1$, \,$\Delta_{C_2}=0.4$, \,$\Delta_m=0.35$.}
     \label{fig:operator_counting_plot}
\end{figure}

Without loss of generality, we can assume that
\begin{equation}
    \Delta_{B_1} \ge \Delta_{B_2}\:,\qquad{\rm k}\,\Delta_{122}-\Delta_m>0\:,\qquad{\rm k}\,\Delta_{221}+\Delta_m>0\:.
\end{equation}
Then the result for the eigenvalue density that we obtain from \eqref{opcrho} is%
\footnote{For $\Delta_{B_1}=\Delta_{B_2}$ it would seem that $\rho(x)$ is ill-defined in the regions $\xmin<x<\xL$ and
        $\xR<x<\xmax$. However, as evident from \eqref{x_bounds_oc}, when $\Delta_{B_1}=\Delta_{B_2}$
        we have $\xmin=\xL$ and $\xR=\xmax$, so the $\xmin<x<\xL$ and $\xR<x<\xmax$ regions are not present
        and $\rho(x)$ is well defined.}
\begin{equation}
\label{rho_operator_counting}
	\rho(x)=\begin{cases}
		\displaystyle\frac{2\pi\,{\rm k}\,x\,\Delta_{122}+\mu-2\pi x \Delta_{m}}{(2\pi)^3(\Delta_{B_1}-\Delta_{B_2})\,\Delta_{121}\,\Delta_{222}}\:,
			&\makebox[15mm][r]{$\xmin<\:$}x<\xL\:,\\[7mm]
		\displaystyle\frac{2\pi\,{\rm k}\,x\left(\Delta_{122}\DeltaS-\Delta_{121}\,\Delta_{222}\right)+(\mu-2\pi x \Delta_{m})\,\DeltaS}
			{(2\pi)^3\,\Delta_{111}\,\Delta_{121}\,\Delta_{212}\,\Delta_{222}}\:,&\makebox[15mm][r]{$\xL<\:$}x<0\:,\\[7mm]
		\displaystyle\frac{-2\pi\,{\rm k}\,x\left(\Delta_{221}\DeltaS-\Delta_{121}\Delta_{222}\right)+(\mu-2\pi x \Delta_{m})\,\DeltaS}
			{(2\pi)^3\,\Delta_{111}\,\Delta_{121}\,\Delta_{212}\,\Delta_{222}}\:,&\makebox[15mm][r]{$0<\:$}x<\xR\:,\\[7mm]
		\displaystyle\frac{-2\pi\,{\rm k}\,x\,\Delta_{221}+\mu-2\pi x \Delta_{m}}{(2\pi)^3(\Delta_{B_1}-\Delta_{B_2})\,\Delta_{121}\,\Delta_{222}}\:,
			&\makebox[15mm][r]{$\xR<\:$}x<\xmax\:,\end{cases}
\end{equation}
which is a piecewise linear function with compact support on $[\xmin,\,\xmax]$, whose slope changes at the
$x$-values $\xL$, $0$, $\xR$, where $\xmin,\:\xmax,\:\xL,\:\xR$ are defined by
\begin{equation}
\label{x_bounds_oc}
\begin{aligned}
	\xmin&=-\frac{\mu}{2\pi\,({\rm k}\,\Delta_{122}-\Delta_{m})}\:,\qquad\xL=-\frac{\mu}{2\pi\,({\rm k}\,\Delta_{112}-\Delta_{m})}\:,\qquad\\
		\xR&=\frac{\mu}{2\pi\,({\rm k}\,\Delta_{211}+\Delta_{m})}\:,\qquad\xmax=\frac{\mu}{2\pi\,({\rm k}\,\Delta_{221}+\Delta_{m})}\:.
\end{aligned}
\end{equation}
From \eqref{opccy} we derive the eigenvalue differences $y_a(x)-y_b(x)$.
For the nodes $a=3$, $b=4$ we obtain the following:
\begin{align}
	{}&\frac{y_{3}(x)-y_4(x)}{2\pi}=\\[2mm]\nonumber
    {}&=\begin{cases}
		-\Delta_{B_2}\:,&\makebox[7mm][r]{$\xmin<\:$}x<\xL\:,\\[5mm]
		\Delta_{A_2}+\Delta_{C_2}+\displaystyle\frac{\left(2\pi\,{\rm k}\,x\,(\Delta_{C_1}-\Delta_{C_2})-\mu+2\pi x \Delta_{m}\right)\,\Delta_{212}\,\Delta_{222}}
			{2\pi\,{\rm k}\,x\left(\Delta_{122}\DeltaS-\Delta_{121}\,\Delta_{222}\right)+(\mu-2\pi x \Delta_{m})\,\DeltaS}\:,&\makebox[7mm][r]{$\xL<\:$}x<0\:,\\[7mm]
		\Delta_{A_1}+\Delta_{C_1}+\displaystyle\frac{\left(2\pi\,{\rm k}\,x\,(\Delta_{C_1}-\Delta_{C_2})-\mu+2\pi x \Delta_{m}\right)\,\Delta_{111}\,\Delta_{121}}
			{-2\pi\,{\rm k}\,x\left(\Delta_{221}\DeltaS-\Delta_{121}\,\Delta_{222}\right)+(\mu-2\pi x \Delta_{m})\,\DeltaS}\:,&\makebox[7mm][r]{$0<\:$}x<\xR\:,\\[7mm]
		-\Delta_{B_2}\:,&\makebox[7mm][r]{$\xR<\:$}x<\xmax\:.\end{cases}
\end{align}
All other possible eigenvalue differences may be obtained from the above expression together with the relations
\begin{equation}
\begin{cases}
	\,0\,=\,\dfrac{y_1(x)-y_3(x)}{2\pi}+\Delta_{A_1}\,=\,\dfrac{y_4(x)-y_2(x)}{2\pi}+\Delta_{C_2}\:,&\makebox[15mm][r]{$x<0$}\:,\\[4mm]
	\,0\,=\,\dfrac{y_2(x)-y_3(x)}{2\pi}+\Delta_{A_2}\,=\,\dfrac{y_4(x)-y_1(x)}{2\pi}+\Delta_{C_1}\:,&\makebox[15mm][r]{$x>0$}\:,
\end{cases}
\end{equation}
which are also derived from \eqref{opccy}. In Figure \ref{fig:operator_counting_plot} we have plotted the eigenvalue differences and eigenvalue density for a choice of parameters.

The Lagrange multiplier $\mu$ is then fixed by imposing the normalization $\int dx\,\rho(x)=1$, and its value is expected to be proportional
to the large\,-$N$ free energy. The value that we obtain this way turns out to be exactly the same as the result from the free energy computation \eqref{free_energy_with_general_Delta} that we will discuss in section \ref{sec: Analytic solution}.

\FloatBarrier

\subsection{Numerics vs. Operator counting results}
 \label{sec:Numerics_OPC}
Now let us compare the numerical solutions to the saddle point equations discussed in sections \ref{sec:numerical_solution} and \ref{sec:noncanonical} with the results from the operator counting method in section \ref{sec:operator_counting}. We first substitute the assignments $\Delta_{A_i}=\Delta_{B_1}=\Delta_A$, \:$\Delta_{C_i}=\Delta_{B_2}=2/3-\Delta_A$, \:${\rm k=1}$, \:$\Delta_m=0$ in \eqref{rho_operator_counting} and subsequent formulas. These formulas only determine differences between eigenvalues, but if we further impose that $y_1(x)$ and $y_2(x)$ are odd functions while
$y_3(x)$ and $y_4(x)$ are even functions we find a complete prediction that we can compare with the numerical solution.
For the nodes 1 and 2, when $x>0$ we find
\begin{equation}
y_1(x) =-y_2(x)=
\begin{cases}
\displaystyle\frac{-\pi\mu\left(\Delta_A+\frac23\right)\left(\Delta_A-\frac43\right)}
{2\pi\,x\left(\Delta_A+\frac23\right)\left(\Delta_A-\frac43\right)+2\mu},
&\makebox[15mm][r]{$0<\:$}x<\xR\:,\\[5mm]
-\pi\left(\Delta_A-\tfrac43\right),
&\makebox[15mm][r]{$\xR<\:$}x<\xmax\:,
\end{cases}
\end{equation}
with $\xR=\mu/(2\pi(\Delta_A+\frac23))$ and $\xmax=\mu/(2\pi(\frac43-\Delta_A))$, and the $x<0$ behavior
is fixed by the fact that $y_1(x)$ and $y_2(x)$ are odd.
On the other hand, for the nodes 3 and 4 we have
\begin{equation}
y_3(x) =4\pi(\Delta_A-\tfrac13)-y_4(x)=
\begin{cases}
2\pi\Delta_A+y_1(x)\:, &\quad x<0, \\[3.5mm]
2\pi\Delta_A-y_1(x)\:, &\quad x>0.
\end{cases}
\end{equation}
The eigenvalue density $\rho(x)$ is an even function, and for $x>0$ it is given by
\begin{equation}
\rho(x)=
\begin{cases}
\displaystyle\frac{\pi x\left(\Delta_A+\frac23\right)\left(\Delta_A-\frac43\right)+\mu}
{4\pi^3\left(\Delta_A+\frac23\right)^2\left(\Delta_A-\frac43\right)^2}\:,
&\makebox[15mm][r]{$0<\:$}x<\xR\:,\\[5mm]
\displaystyle\frac{2\pi\,x\left(\Delta_A-\frac43\right)+\mu}
{16\pi^3\left(\Delta_A-\frac43\right)^2\left(\Delta_A-\frac13\right)}\:,
&\makebox[15mm][r]{$\xR<\:$}x<\xmax\:.
\end{cases}
\end{equation}
Imposing the normalization $\int dx\,\rho(x)=1$ then fixes $\mu$ to
\begin{equation}
\label{mu_Delta_A}
    \mu=\frac{(2\pi)^2}9\,\frac{(3\Delta_A+2)^{3/2}(4-3\Delta_A)^{3/2}}{(9\Delta_A^2-6\Delta_A+28)^{1/2}}\:.
\end{equation}
The operator counting prediction for the canonical R-charge assignment $\Delta_I=1/3$ may be found by further setting $\Delta_A=1/3$,
obtaining $\mu=(2\pi)^2/\sqrt3$ and
\be
\begin{aligned}
    & y_1(x) =-y_2(x)=\frac{2\pi^2\:\text{sign}(x)}{4\pi-\sqrt3\,|x|}\:,\qquad&& y_3(x) =-y_4(x)=\frac{2\pi}{3}-\frac{2\pi^2}{4\pi-\sqrt3\,|x|}\:,\\
    &\rho(x)=\dfrac1{\pi\sqrt3}-\dfrac{|x|}{(2\pi)^2}\:,&&-\dfrac{2\pi}{\sqrt{3}} <x<\dfrac{2\pi}{\sqrt{3}}\:.
\end{aligned}
\ee

\begin{figure}[h]
     \centering
     \begin{subfigure}[b]{0.49\textwidth}
         \centering
         \includegraphics[width=\linewidth]{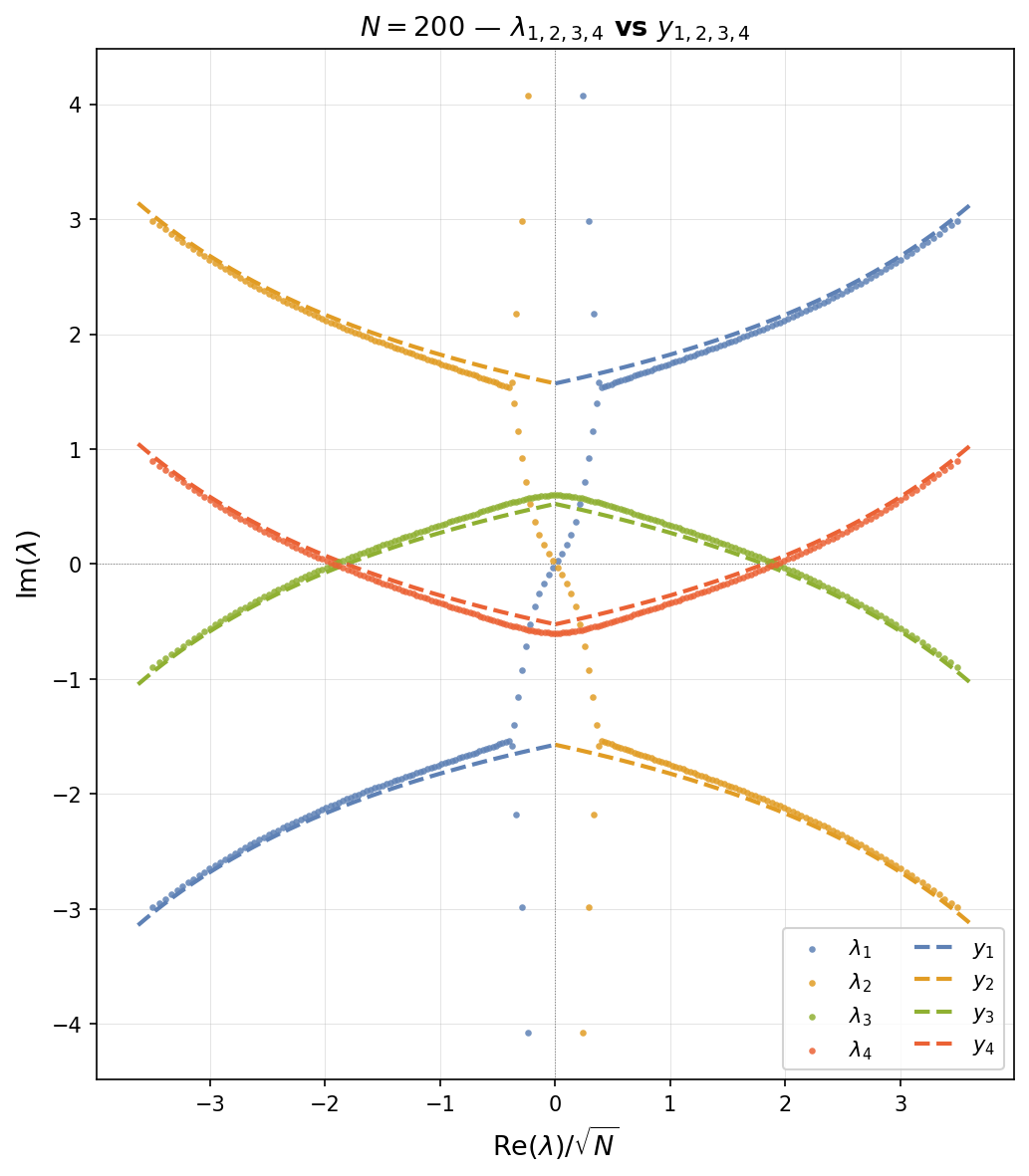}
         \caption{$\Delta_I=1/3\:$ solution.}
         \label{fig:operator counting_canonical}
     \end{subfigure}
     \hfill
     \begin{subfigure}[b]{0.49\textwidth}
         \centering
         \includegraphics[width=\linewidth]{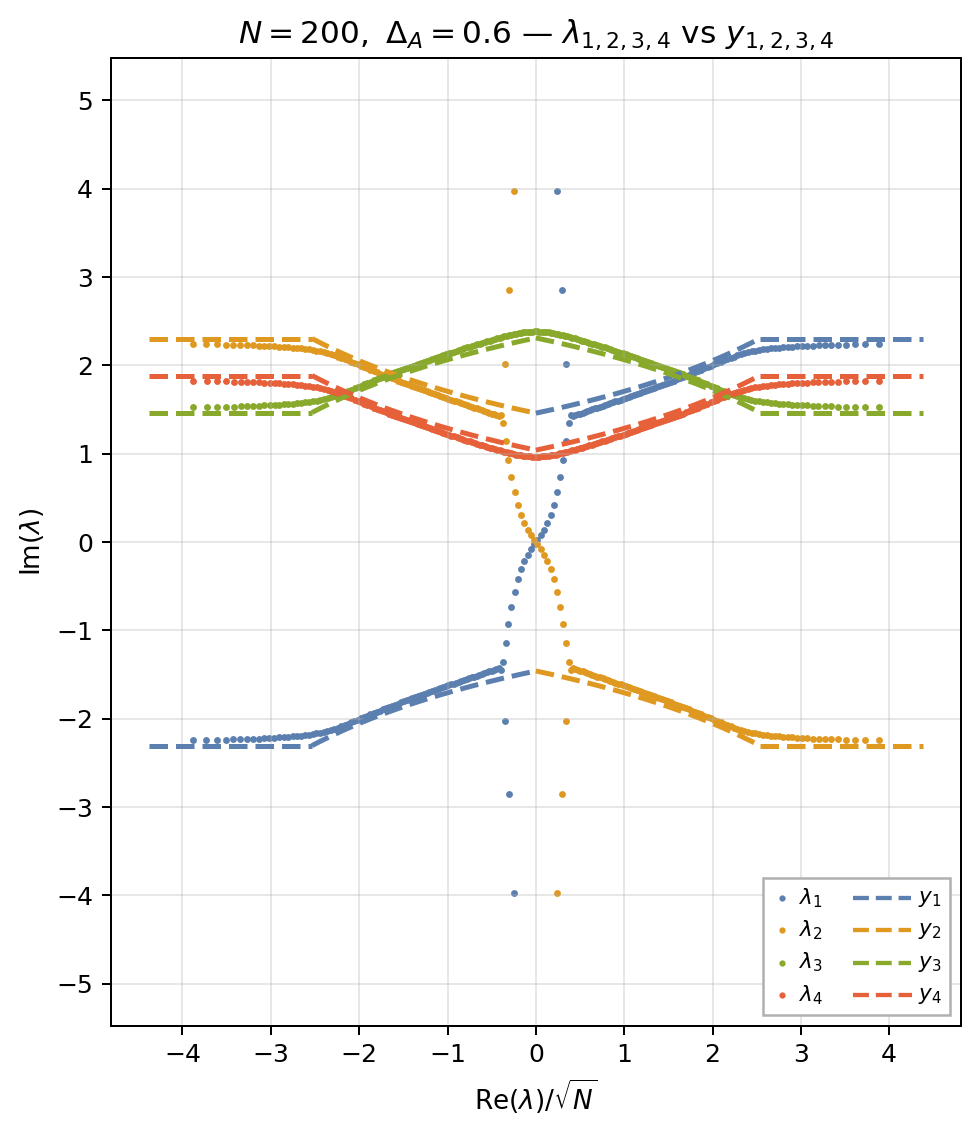}
         \caption{$\Delta_{A_i}=\Delta_{B_1}=0.6,\,\Delta_{C_i}=\Delta_{B_2}\:$ solution.}
         \label{fig:operator counting_non_canonical}
     \end{subfigure}
     
     \caption{Comparison of the numerical solutions and the $y_a(x)$ functions from operator counting.}
     \label{fig:operator counting}
\end{figure}

\begin{figure}[h]
     \centering
     \begin{subfigure}[b]{0.49\textwidth}
         \centering
         \includegraphics[width=\linewidth]{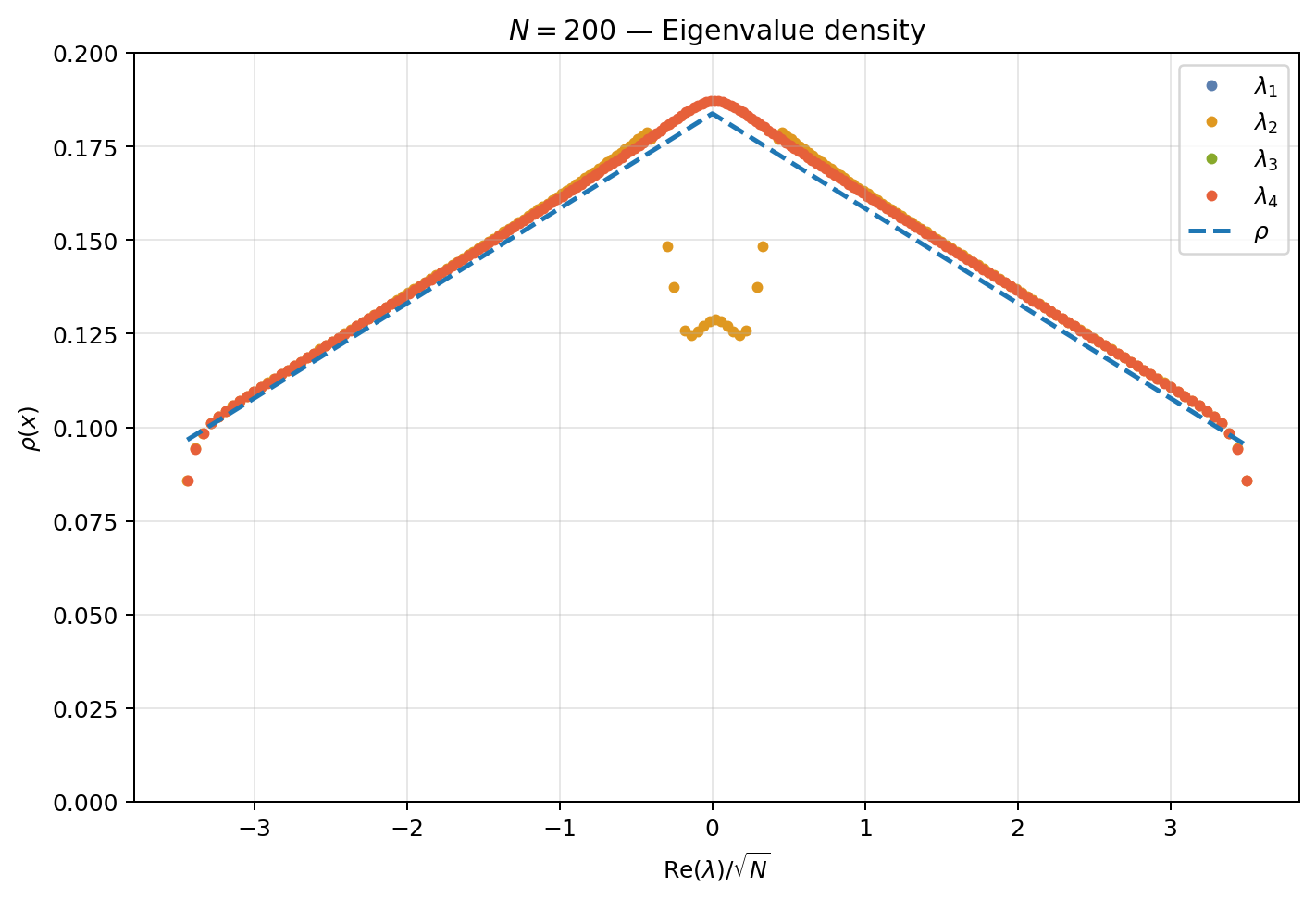}
         \caption{$\Delta_I=1/3\:$ solution.}
         \label{fig:eigenvalue density_canonical}
     \end{subfigure}
     \hfill
     \begin{subfigure}[b]{0.49\textwidth}
         \centering
         \includegraphics[width=\linewidth]{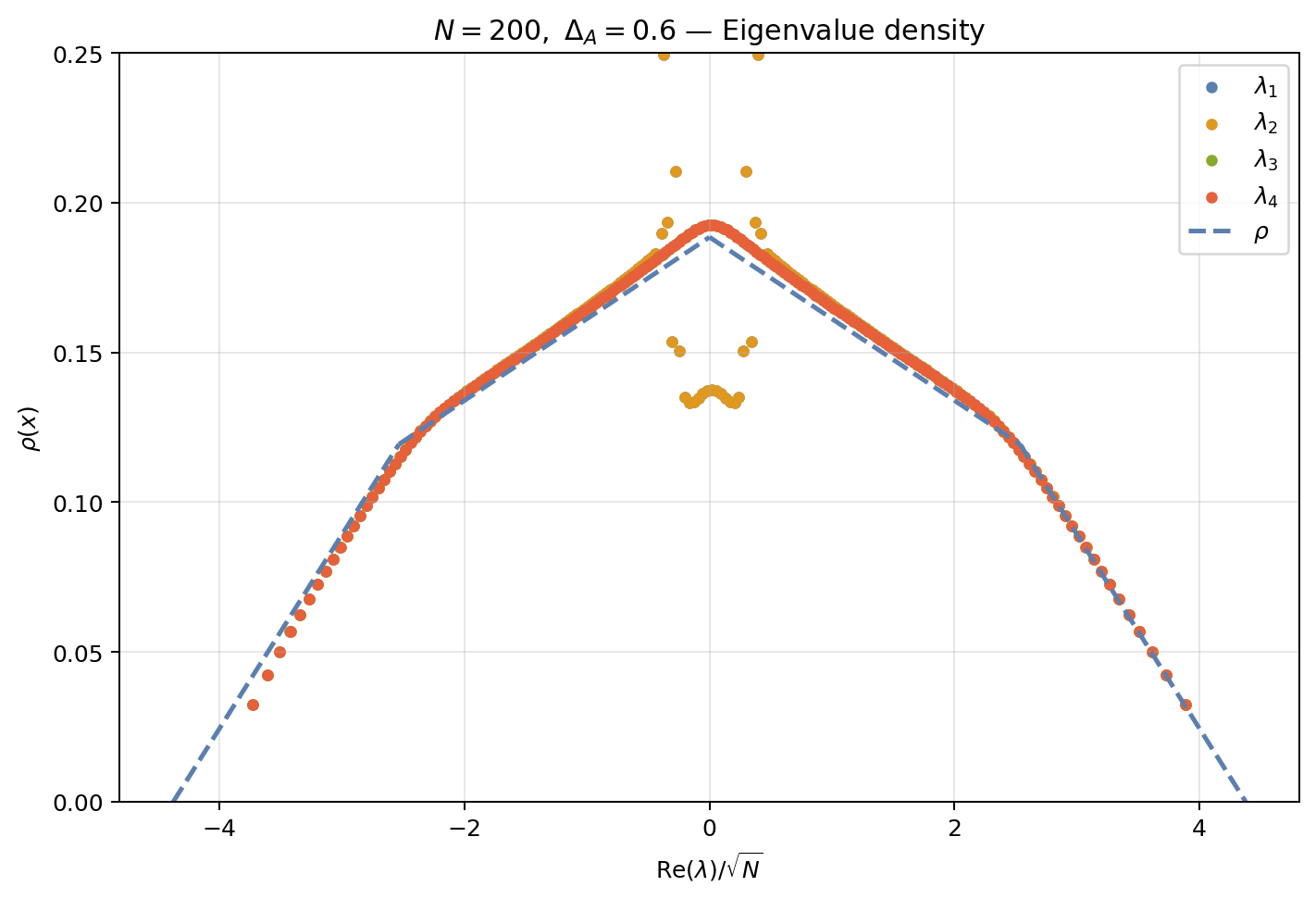}
         \caption{$\Delta_{A_i}=\Delta_{B_1}=0.6,\,\Delta_{C_i}=\Delta_{B_2}\:$ solution.}
         \label{fig:eigenvalue density_non_canonical}
     \end{subfigure}
     
     \caption{Comparison of the eigenvalue density of the numerical solutions and the $\rho(x)$ function from operator counting.}
     \label{fig:eigenvalue density}
\end{figure}

Figures \ref{fig:operator counting} and \ref{fig:eigenvalue density} show the numerical solutions overlaid with the results obtained from the operator counting method. The dotted lines represent the numerical solutions at finite $N$, with $N=200$, while the dashed lines correspond to the large\,-$N$ eigenvalues derived from the operator counting method. They agree well in the tail regions, but not in the bulk region,
where the eigenvalues of the nodes 1 and 2 in the numeric solutions interpolate between the values in the left and right tail regions,
whereas the operator counting method predicts a sudden jump at $x=0$. 
 Together with the observation that the width of the bulk region decreases as $N$ increases, this comparison suggests that the bulk region observed in the numerical solution may be a finite $N$ artifact that disappears in the large\,-$N$ limit.

\FloatBarrier


\section{The large\,-$N$ limit of the free energy and supersymmetric indices}
\label{sec:large N expansions}

Having discussed the numerical solution to the saddle point equations in section \ref{sec:numerical_solution},
in this section we switch gears and lay the groundwork for deriving an analytic solution at large\,-$N$.
First, we will discuss the large\,-$N$ expansion of the free energy of a general chiral quiver by following the same approach as
\cite{Herzog:2010hf,Martelli:2011qj,Cheon:2011vi,Jafferis:2011zi} while also keeping track of all the long range forces
that in a non-chiral quiver would otherwise automatically cancel. We postpone the full derivation of the large\,-$N$
saddle of the chiral quiver dual to $Q^{1,1,1}/\bZ_{\rm k}$ to section \ref{sec: Analytic solution}.
Before that, in this section we also consider the superconformal \cite{Imamura:2011su,Kapustin:2011jm} and topologically twisted \cite{Benini:2015noa} indices on $S^2\times S^1$,
and compute their expansion at leading order in $N$ and all perturbative orders in the Cardy limit. 
The result of this expansion for a generic chiral quiver can be expressed in terms of the large\,-$N$ free energy,
in agreement with previous results in the literature \cite{Hosseini:2016tor,Choi:2019zpz,Choi:2019dfu,Hosseini:2022vho},
allowing us to extend the results of this paper to the $S^2\times S^1$ indices as well.

\subsection{The free energy at large\,-$N$}
\label{sec:Free energy at large N}

We consider the usual Ansatz for the large\,-$N$ distribution of eigenvalues \cite{Herzog:2010hf},
\be
\label{eigenvalue_Ansatz}
    \lambda_{a,\,i}=N^{1/2}\,x_i+\ii\,y_a(x_i)\:,
\ee
and replace the discrete variables $x_i$ with a single continuous variable $x$ and an eigenvalue density
$\rho(x)$ defined by $\rho(x_i)=N^{-1}(x_{i+1}-x_i)^{-1}$.
Then it is possible to approximate sums over the index $i=0,\ldots,N-1$ with integrals over $x$ as follows:
\begin{equation}
\label{sums_to_integrals}
	\sum_{i=0}^{N-1}f(\lambda_{a,\,i})\:=\:N\int\dd x\,\rho(x)\,f\big(N^{1/2}\,x+\ii\,y_a(x)\big)\:+\:\cO(N^0)\:,\qquad\int\dd x\,\rho(x)=1\:.
\end{equation}
The function $\cF(\Delta,\lambda_{a,\,i})$ \eqref{cF} in this limit becomes a functional of $\rho(x)$ and $y_a(x)$
whose extremal value  gives the large\,-$N$ free energy.
For non-chiral quivers the ansatz \eqref{eigenvalue_Ansatz} is known to reproduce the large\,-$N$ behavior of the eigenvalues
\cite{Herzog:2010hf,Martelli:2011qj,Cheon:2011vi,Jafferis:2011zi}. 
For the chiral quiver dual to $Q^{1,1,1}/\bZ_{\rm k}$, we will argue that this same ansatz also captures the large\,-$N$
    saddle point solution, at least when a condition on the R-charges is imposed.
    In section \ref{sec: A proposal for general Delta} we will discuss a possible modification of the ansatz
    \eqref{eigenvalue_Ansatz} for general R-charges.

Let us start from the Chern-Simons and topological symmetry terms
in $\cF$, which at large\,-$N$ with the Ansatz \eqref{eigenvalue_Ansatz} become
\be
\label{F_classical_terms}
    \sum_{a=1}^{|\cG|}\sum_{i=0}^{N-1}\left(-\frac{\ii{\rm k}_a}{4\pi}\,\lambda^2_{a,\,i}+
        \Delta_{m;\, a}\,\lambda_{a,\,i}\right)=N^{3/2}\int\dd x\,\rho(x)\cdot x\cdot
        \sum_{a=1}^{|\cG|}\left(\frac{{\rm k}_a}{2\pi}\,y_a(x)+\Delta_{m;\, a}\right)+\cO(N)\:,
\ee
where we have used that for quiver theories with an M-theory dual the sum of Chern-Simons levels vanishes:
\be
\label{CS_level_sum}
    \sum_{a=1}^{|\cG|}{\rm k}_a=0\:.
\ee

Let us now consider the one-loop determinant of a chiral multiplet
corresponding to an arrow in the quiver diagram from the node $a$ to the node $b$.
If we assume that $y_a(x)-y_b(x)$ takes values in the range
\be
\label{free_energy_determination}
    0<2\pi\Delta_I\,+\,y_a(x)-y_b(x)<4\pi\:,
\ee
which is consistent with the numerical result, we may write the chiral multiplet contribution as
\be
\begin{aligned}
\label{chiral_contribution_to_F}
    {}&-\sum_{i,j=0}^{N-1}\ell\Big(1-\Delta_I+\ii\,\tfrac{\lambda_{a,\,i}\,-\,\lambda_{b,\,j}}{2\pi}\Big)
        =-N^2\int\dd x\,\dd\wt x\:\rho(x)\rho(\,\wt x\,)\:
        \ell\big(z_I(x,\wt x)\big)+\cO(N)=\\
    {}&\qquad=N^{3/2}\int\dd x\,\rho(x)^2\left(\frac{4\pi^2}{3}\,B_3\big(z_I(x,x)\big)
        -2\pi^2\,z_I(x,x)\,B_2\big(z_I(x,x)\big)\right)+\\
    {}&\qquad\quad+\frac{\pi\ii}2\,N^2\!\!\int\!\dd x\,\dd\wt x\:\rho(x)\rho(\,\wt x\,)\:\text{sign}(\wt x-x)\!
        \left(z_I(x,\wt x)^2-\frac16\right)+\cO(N)\:,
\end{aligned}
\ee
where we have defined
\be
    z_I(x,\wt x)\,\equiv\,1-\Delta_I-\tfrac{y_a(x)-y_b(\,\wt x\,)}{2\pi}+\tfrac{\ii N^{1/2}(x-\wt x)}{2\pi}\:.
\ee
In the second step of \eqref{chiral_contribution_to_F} we have used that $\ell(z)$ is defined as a sum of
polylogarithms and a polynomial by \eqref{ell_def}, and we applied the following asymptotic formulas:
\begin{align}
\label{Li_asymptotic}
   	{}&\int_0^{\bar t>0}\dd t\,f(t)\Li_s\left(e^{-N^\alpha t\,+\,\ii y(t)}\right)=
		N^{-\alpha}f(0)\Li_{s+1}\left(e^{\ii y(0)}\right)+\cO(N^{-2\alpha})\:,\\
\label{Li_asymptotic2}
    {}&\int_0^{\bar t>0}\dd t\,f(t)\cdot t\cdot\Li_s\left(e^{-N^\alpha t\,+\,\ii y(t)}\right)=
		N^{-2\alpha}f(0)\Li_{s+2}\left(e^{\ii y(0)}\right)+\cO(N^{-3\alpha})\:,
\end{align}
which can be found using the series expansion for $\Li_s$ and integration by parts,
see e.g.\ equation (4.18) of \cite{Choi:2019zpz} for the first one.
The formulas \eqref{Li_asymptotic} and \eqref{Li_asymptotic2} are used to integrate $\wt x$ after
changing variable to $t=|x-\wt x|$; in the $x>\wt x$ region of the domain these formulas can be applied directly,
whereas in the $x<\wt x$ region we first replace $\Li_s(e^{2\pi\ii\,z})$ with $\Li_s(e^{-2\pi\ii\,z})$ plus a polynomial
by using the polylogarithm inversion property,
\be
\label{polylog_inversion}
    \Li_s(e^{\ii y})+(-1)^s\Li_s(e^{-\ii y})=-\frac{(2\pi\ii)^s}{s!}\,
        B_s\left(\frac y{2\pi}-\left\lfloor\frac{\text{Re}\,y}{2\pi}\right\rfloor\right)\:,\qquad s\in\bN.
\ee
The terms in the last line of \eqref{chiral_contribution_to_F}
are the combination of the polynomial terms in the definition of $\ell(z)$ and the polynomial terms from the application of
the above inversion identity in the $x<\wt x$ region.
The value of the floor function only shifts these terms by integer multiples of $2\pi\ii$, which may be reabsorbed in the choice of branch for $\ell(z)$.
Lastly, the polylogarithms coming from the right hand side of \eqref{Li_asymptotic}
and \eqref{Li_asymptotic2} combine nicely using \eqref{polylog_inversion} to give the $B_3$ and $B_2$ polynomials
in the second line of \eqref{chiral_contribution_to_F}, and the domain \eqref{free_energy_determination} is used to fix the floor function to be either $0$ or $-1$; both values lead to the same result, since $B_k(z+1)=B_k(z)+k\,z^{k-1}$.

For the one-loop determinant of a vector multiplet we can write 
\begin{align}
\label{vector_contribution_to_F}\nn
    -2\!\sum_{i>j\geq0}^{N-1}\log\left(2\sinh\tfrac{\lambda_{a,\,i}-\lambda_{a,\,j}}2\right)
        =N^2\!\int_{x>\wt x}\!\!\!\!\!\!\!{}&\dd x\,\dd\wt x\:\rho(x)\rho(\,\wt x\,)
        \bigg(2\Li_1\Big(e^{N^{1/2}(\wt x-x)+\ii\,(y_a(\,\wt x\,)-y_a(x))}\Big)+\\
    {}&+N^{1/2}(\wt x-x)+\ii\big(y_a(\,\wt x\,)-y_a(x)\big)\bigg)\:,
\end{align}
where we have used \eqref{sums_to_integrals} and the definitions $\:\sinh z=\frac12(e^z-e^{-z})$, $\:\Li_1(z)=-\log(1-z)$.
We can apply \eqref{Li_asymptotic} to perform the integral in $\wt x$ after changing variable to $t=x-\wt x$, obtaining%
\footnote{$\Li_s(z)$ has a branch cut going from $z=1$ to $z=+\infty$,
    which may seem problematic for applying \eqref{Li_asymptotic}, as we would get $\Li_2(1)$. 
    However, the derivation of \eqref{Li_asymptotic} uses the series expansion valid in $|z\,|\leq1$, and
    $\lim_{\,\epsilon\to0+}\Li_2(e^{-\epsilon})=\pi^2/6$.}
\be
    2N^2\!\int_{x>\wt x}\!\!\!\!\!\!\!\dd x\,\dd\wt x\:\rho(x)\rho(\,\wt x\,)
        \Li_1\Big(e^{N^{1/2}(\wt x-x)+\ii\,(y_a(\,\wt x\,)-y_a(x))}\Big)=\frac{\pi^2}3\,N^{3/2}\int\dd x\,\rho(x)^2\:,
\ee
which matches the terms in the second line of \eqref{chiral_contribution_to_F}
with $\Delta_I\to2$ and $b\equiv a$, as expected.
The same is true for the terms in the second line of \eqref{vector_contribution_to_F} compared to the terms
in the last line of \eqref{chiral_contribution_to_F}, after terms in the integrand that are antisymmetric in the
exchange of $x$ and $\wt x$ are canceled.

In order to write the results in a more compact manner and to later compare with the supersymmetric indices, it is convenient to
rewrite some expressions by imposing the marginality constraint
\be
\label{marginality_general}
    \sum_{I\in W}\Delta_I\:=\:2\:,
\ee
where $I\in W$ denotes the set of chiral multiplets appearing in the same superpotential term $W$.
In the following we will make use of \eqref{marginality_general}, and if the theory has multiple superpotential terms
involving different chiral multiplets $W$ could be freely chosen among any of them.
Following a similar approach as \cite{Hosseini:2016tor}, we will use \eqref{marginality_general} to make the coefficients
of $\cF$ as a functional of $\rho$ and $y_a$ to be homogeneous functions in $\Delta$.
We can focus on the expression \eqref{chiral_contribution_to_F} for the chiral multiplet contribution,
and in the case of vector multiplets use the same expression with $\Delta_I\to\sum_{I\in W}\Delta_I$ and $b\equiv a$.

First, let us define the following quantities: 
\begin{equation}
\label{cDa_nab_def}
	\cD_a=\sum_{J\in W}\Delta_J+\sum_{I\in(a)}\Big(\Delta_I-\frac12\sum_{J\in W}\Delta_J\Big)\:,\qquad n_{a,b}=
		\underset{I\,:\,a\to b}{\sum_{I=1}^{|\cR|}}1-\underset{I\,:\,b\to a}{\sum_{I=1}^{|\cR|}}1\:,
\end{equation}
where $I\in(a)$ is a summation over all the arrows in the quiver with a leg on the node $a$, counting twice for adjoint multiplets.
The numbers $n_{a,b}$ count the difference of arrows in the quiver from the node $a$ to the node $b$,
whereas $\cD_a$ would compute ABJ anomalies if the quiver theory was four-dimensional rather than three-dimensional.
Analytic solutions to the saddle point equations in the M-theory limit with fixed ${\rm k}_a$ and large\,-$N$ have been described in the literature only for the quiver theories for which all the quantities defined in \eqref{cDa_nab_def} vanish for all $a,b$.
By definition, chiral theories are the ones for which some of the $n_{a,b}$ are nonzero;
for the chiral quivers of interest, generically only the sum over $a$ of these quantities vanishes:
\begin{equation}
\label{quiver_conditions0}
	\sum_{a=1}^{|\cG|}\cD_a=0\:,\qquad\sum_{a=1}^{|\cG|}n_{a,b}=0\:.
\end{equation}
Then the long range force terms in the last line of \eqref{chiral_contribution_to_F}, after summing over all
chiral and vector multiplets, can be written as
\be
    \pi\ii\,N^2\int\dd x\,\dd\wt x\:\rho(x)\rho(\,\wt x\,)\,\text{sign}(\wt x-x)\sum_{a=1}^{|\cG|}\bigg[\frac1{2\pi}\,\cD_a\,y_a(x)-\frac1{8\pi^2}\sum_{b=1}^{|\cG|}n_{a,b}\:y_a(x)\,y_b(\,\wt x\,)\bigg]\:,
\ee
where the $\cO(N^3)$ terms have vanished due to anti-symmetry of the integrand under the exchange of $x$ and $\wt x$,
while the $\cO(N^{5/2})$ terms have vanished as a consequence of \eqref{quiver_conditions0}.

In order to simplify the $\cO(N^{3/2})$ terms in \eqref{chiral_contribution_to_F}, we first substitute the marginality condition
\eqref{marginality_general} as follows:
\be
\label{marginality_substitution}
    B_n(z)\,=\,1^n\cdot B_n(1^{-1}\cdot z)\:,\qquad 1\:\equiv\:\frac12\sum_{I\in W}\Delta_I\:.
\ee
The resulting expression can then be written as
\begin{align}
\label{introducing_G3_polynomial}
    N^{3/2}\int\dd x\,\rho(x)^2\bigg[\frac4\pi\,G_3\Big(\pi\Delta_I+\frac12\delta y_I(x)\Big)
        +\frac{\pi^2}{12}\Big(\Delta_I-\frac12\sum_{J\in W}\Delta_J+\frac1{2\pi}\delta y_I(x)\Big)
        \Big(\sum_{J\in W}\Delta_J\Big)^2\bigg]\:,
\end{align}
where we have introduced the shorthand $\delta y_I(x)\equiv y_a(x)-y_b(x)$ and defined the cubic polynomial $G_3(z)$ as
\begin{equation}
\label{G3_def}
	G_3(z)=\frac z6\Big(z-\frac\pi2\sum_{I\in W}\Delta_I\Big)\Big(z-\pi\sum_{I\in W}\Delta_I\Big)\:.
\end{equation}
Notice that the $G_3$ polynomials in \eqref{introducing_G3_polynomial} vanish for vector multiplets after the replacements
$\Delta_I\to\sum_{I\in W}\Delta_I$ and $b\equiv a$.
After summing over all chiral and vector multiplets, the terms in \eqref{introducing_G3_polynomial} other than the $G_3$
polynomial vanish due to \eqref{quiver_conditions0}.

After combining all terms, the large\,-$N$ free energy functional is given by the following expression:
\begin{align}
\label{free_energy_functional}
	{}&N^{-3/2}\:\cF(\Delta,\rho,y,\nu)=\\\nn
	{}&\quad=\int\dd x\,\rho(x)\cdot x\cdot\sum_{a=1}^{|\cG|}\left(\frac{{\rm k}_a}{2\pi}\,y_a(x)+\Delta_{m;\,a}\right)
		+\int\dd x\,\rho(x)^2\sum_{I=1}^{|\cR|}\frac4\pi\,G_3\Big(\pi\Delta_I+\frac12\delta y_I(x)\Big)+\\\nn
	{}&\quad\quad+\nu\int\dd x\,\dd\wt x\,\rho(x)\,\rho(\,\wt x\,)\,\text{sign}(\wt x-x)
		\sum_{a=1}^{|\cG|}\bigg[\frac1{2\pi}\,\cD_a\,y_a(x)
        -\frac1{8\pi^2}\sum_{b=1}^{|\cG|}n_{a,b}\:y_a(x)\,y_b(\,\wt x\,)\bigg]\:,
\end{align}
where $\nu$ is a parameter that we have introduced for later convenience; here it takes the value $\nu=\pi\ii\,N^{1/2}$.
The terms proportional to $\nu$ are called long range forces,
since they involve a double integration in $x$ and $\wt x$ which makes them interaction terms between distant eigenvalues.

The saddles of the partition function $F_{S^3}$ are found by extremizing \eqref{free_energy_functional},
which amounts to solving the following equations:
\be
\label{saddle_point_equations}
    \frac{\delta\cF}{\delta\rho(x)}=\frac{\mu\,N^{3/2}}{2\pi}\:,\qquad\frac{\delta\cF}{\delta y_a(x)}=0\:,
\ee
where $\mu$ is a Lagrange multiplier that is introduced to impose that the eigenvalue density $\rho$ is normalized to one,
\be
   \frac{\cF}{N^{3/2}}\:\longrightarrow\:\frac{\cF}{N^{3/2}}-\frac{\mu}{2\pi}\,\bigg[\int\dd x\,\rho(x)-1\bigg]\:.
\ee
There is an important caveat to consider: when one of the inequalities in \eqref{free_energy_determination} is saturated,
a modification to the above extremization equations must be made. It is easy to verify that when these inequalities
are saturated the function $\partial_z\,\ell(z)=-\pi z\cot(\pi z)$ that appears in the derivatives of
$\cF(\Delta,\lambda_{a,\,i})$ \eqref{cF} diverges.
As a consequence, the eigenvalues of a saddle of the free energy cannot escape from the bounds
of \eqref{free_energy_determination}; when they get close to saturating one such bound, the divergence generates
an insurmountable force. Many eigenvalues remain stuck very close to these bounds, creating ``tail regions'' in the graph of
the eigenvalue distribution.
Since this divergence is not visible in the large\,-$N$ free energy \eqref{free_energy_functional},
its effects must be accounted for separately by extremizing $\cF$ only along directions that do not
violate the bounds of \eqref{free_energy_determination}. In practice, in a tail region where the bound
$\delta y_I(x)+2\pi\Delta_I>0$ is saturated,
the saddle point solution at leading order in $N$ may be found by imposing the constraint $\delta y_I(x)+2\pi\Delta_I=0$ while extremizing the large\,-$N$
free energy functional \eqref{free_energy_functional}, which also reduces the number of extremization equations 
\eqref{saddle_point_equations} to be imposed by one.
After a solution to the extremization equations \eqref{saddle_point_equations} is found, by saddle point approximation
the large\,-$N$ limit of the free energy is given by
\begin{equation}
\label{crtical_value_free_energy}
	F_{S^3}(\Delta)\:\sim\:\cF_{\text{crit.}}(\Delta)\:\equiv\:\cF(\Delta,\rho,y,\nu)\,\big|_{\text{ solution to }\eqref{saddle_point_equations}}\:.
\end{equation}

So far, analytic solutions to the extremization equations of the large\,-$N$ free energy functional \eqref{free_energy_functional} have been found only for non-chiral
quivers that satisfy $\cD_a = 0$ and $n_{a,b}=0$, which makes the long range forces vanish identically.
In this paper we find an analytic saddle point for the chiral quiver dual to $Q^{1,1,1}/\bZ_{\rm k}$,
for which the long range forces are not identically zero.
The key idea, inspired from the numerical analysis and by comparison with the operator counting method,
is that the large\,-$N$ saddle point should only have tail regions, since the central ``bulk'' region 
where none of the \eqref{free_energy_determination} bounds are saturated shrinks to zero as $N$ increases.
We find that the long range forces between two points within the same tail region sum to zero in the free energy, after imposing the saturation of the bound 
\eqref{free_energy_determination}. Long range forces between different tail regions also vanish, if a linear condition
among the $\Delta_I$ is assumed, namely $\Delta_{A_1}+\Delta_{C_1}=\Delta_{A_2}+\Delta_{C_2}$. Therefore, for this subset of values of $\Delta_I$ we can immediately provide an analytic solution to
the saddle point equations, as we will describe in detail in section \ref{sec: Saddle point of Q111}.
We also verify that the long range forces cancel from the saddle point equations, and we compute the subleading corrections
required for this cancellation to happen in section \ref{sec:subleading corrections};
when $\Delta_I=1/3$ we can also compare these subleading corrections with the numerical solution and we find a perfect match.
At last, we also find an analytic solution that is valid for the most general assignment of R-charges in section \ref{sec: A proposal for general Delta}, although that requires a small modification to the ansatz for the eigenvalues $\lambda_{a,\,i}$.

Lastly, let us point out that the free energy functional \eqref{free_energy_functional} has many flat directions \cite{Jafferis:2011zi}.
Even at finite\,-$N$ the partition function is invariant under
\be
\label{flat_directions}
\begin{aligned}
    {}&\lambda_{a,\,i}\:\to\:\lambda_{a,\,i}-2\pi\ii\,\delta^{(a)}\:,\\
    {}&\Delta_I\:\to\:\Delta_I+\delta^{(a)}-\delta^{(b)}\:,\\
    {}&\Delta_{m;\,a}\:\to\:\Delta_{m;\,a}+{\rm k}_a\delta^{(a)}\:,
\end{aligned}
\ee
which gives a number of flat directions, parametrized by the $\delta^{(a)}$, equal to the number of U($N$) gauge nodes $|\cG|$.
At large\,-$N$, the functional \eqref{free_energy_functional}
is also invariant under $y_a(x)\to y_a(x)+\wt y(x)$ for any function
$\wt y(x)$, and it only depends on the sum of all the $\Delta_{m;\,a}$, which we denote as
\be
    \Delta_{m}\,\equiv\,\sum_{a=1}^{|\cG|}\Delta_{m;\,a}\:.
\ee
The parameters $\Delta_{m}$, $\{\Delta_I\}_{I=1}^{|\cR|}$ transform under $|\cG|-1$ of the flat directions \eqref{flat_directions},
therefore the large\,-$N$ functional \eqref{free_energy_functional} effectively only depends on $|\cR|-|\cG|+2$ such parameters,
and if the marginality conditions \eqref{marginality_general} are imposed this number can be further reduced.

\subsection{The superconformal and topologically twisted indices on $S^2\times S^1$}

On $S^2\times S^1$ it is possible to define two distinct supersymmetric indices, the superconformal index \cite{Imamura:2011su,Kapustin:2011jm}
and the topologically twisted index \cite{Benini:2015noa}. The two indices correspond to two different supersymmetry-preserving backgrounds on
$S^2\times S^1$, which can be distinguished by the value of the flux of the R-symmetry connection on the sphere $S^2$:
\begin{equation}
	\int_{S^2}F=\pi\,(1+\sigma)\:,
\end{equation}
where we have defined
\begin{equation}
	\sigma=\begin{cases}+1\qquad&\text{for the topologically twisted index,}\\-1\qquad&\text{for the superconformal index.}\end{cases}
\end{equation}
The two indices can be written in a unified manner in terms of expressions that depend on the sign $\sigma$ following the approach of
\cite{Inglese:2023wky,Inglese:2023tyc,Colombo:2024mts}, which more generally deal with the spindle index but whose formulas recover the topologically twisted
and superconformal indices when $n_\pm=1$.
For a $\cN=2$ Chern-Simons-matter quiver theory with gauge group $\cG=\prod_{a=1}^{|\cG|}$U$(N)_{{\rm k}_a}$ the indices take the form
\begin{equation}
	Z_{S^2\times S^1,\,\sigma}(\Delta^\pm_I,\,\Delta^{\pm}_{m;a},\,\epsilon)=
		\sum_{\{\fkm_i^a\in\bZ\}}\oint_\cC\frac{\prod_{a,i}\dd u^a_i}{|W_\cG|}\:Z_{\text{classical}}\:Z_{\text{1-loop}}\:,
\end{equation}
where $\cC$ is a suitable integration contour, $|W_\cG|\sim\cO(e^{N\log N})$ is the order of the Weyl group of $\cG$, the index $i$ runs from 0 to $N-1$,
$Z_{\text{classical}}$ is the action evaluated at the BPS locus and $Z_{\text{1-loop}}$ are the one-loop determinants.
The variables $u^a_i$ and the integers $\fkm^a_i$ that are respectively integrated and summed over are the holonomies along $S^1$ and the fluxes over
$S^2$ of the gauge fields.
On the other hand, the variables $\Delta$ that the index depends on are linear combinations of holonomies and fluxes for the background
gauge connections that couple to the global symmetries. The index $I=1,\ldots,|\cR|$ runs over all the chiral multiplets of the theory and we have
\begin{equation}
\label{Delta^pm_def}
	\Delta^\pm_I=(2\,u_I+n\,r_I)+\epsilon\,(\pm\,\fkn_I+\sigma_\pm\,r_I)\:,
\end{equation}
where $u_I$ and $\fkn_I$ are the holonomy and fluxes associated to the flavour symmetry of the $I$-th chiral multiplet, $r_I$ is its R-charge,
and we can take $n=1$ without loss of generality, if we demand that the spinors are anti-periodic along the $S^1$.
The sign $\sigma_\pm$ is defined as
\begin{equation}
	\sigma_+=\sigma\:,\qquad\sigma_-=-1\:.
\end{equation}
The $\Delta^\pm_{m;\,a}$ are defined as
\begin{equation}
\label{Delta_m_def}
	\Delta^\pm_{m;\,a}=2\,u_{m;\,a}\pm\epsilon\,\fkn_{m;\,a}\:,
\end{equation}
where $u_{m;\,a}$ and $\fkn_{m;\,a}$ are the holonomy and flux of the background field coupled to the topological symmetry of the node
U$(N)_{{\rm k}_a}$.
Lastly $\epsilon$ is the chemical potential for the angular momentum, it enters the metric as the parameter that fibers $S^1$ over $S^2$.

The $\Delta^\pm_I$ are not all independent: for every superpotential term $W$ in the Lagrangian they obey the constraint
\begin{equation}
\label{Delta_constraint}
	\frac12\sum_{I\in W}\Delta^\pm_I=n+\sigma_\pm\epsilon\:.
\end{equation}
From now on we will take $n=1$.

The integrand of the index is more conveniently written in terms of the following linear combinations of the integration variables $u^a_i$
and gauge fluxes $\fkm^a_i$:
\begin{equation}
	u^\pm_{a;\,i}=\pm\,u^a_i-\frac\epsilon2\,\fkm^a_i\:.
\end{equation}
The classical part of the integrand, which takes contributions from the Chern-Simons terms and the topological symmetries, can be written as
\begin{equation}
\label{Z_classical}
	Z_{\text{classical}}=\prod_\pm\prod_{a=1}^{|\cG|}\prod_{i=0}^{N-1}e^{\frac{\pi\ii}\epsilon\left[\mp{\rm k}_a(u^\pm_{a;\,i})^2
		\,+\,\Delta^\pm_{m;\,a}\,u^\pm_{a;\,i}\right]}\:,
\end{equation}
while the one-loop determinants can be written as \cite{Colombo:2024mts}
\begin{equation}
	Z_{\text{1-loop}}=\prod_\pm\Bigg[\:\prod_{I=1}^{|\cR|}\Psi^\pm_I\:\cB^\pm_I\:\diag^\pm_I\:\cdot\:\prod_{a=1}^{|\cG|}\Psi^\pm_a\:\cB^\pm_a\:\Bigg]\:,
\end{equation}
where the contributions from a chiral multiplet in the representation $\textbf{N}_a\otimes\overline{\textbf{N}}_b$ are%
\footnote{The $\diag^\pm_I$ terms were neglected in \cite{Colombo:2024mts} due to their subleading nature at large\,-$N$.
        In the notations of \cite{Colombo:2024mts}, they are found as
        $\diag^+_I\cdot\diag^-_I=\prod_{i=0}^{N-1}\zeta^\sigma_q(y^I_{ii},\fkb^I_{ii})=\prod_{i=0}^{N-1}
        \zeta^\sigma_q(y^I_{ii},\fkb^I_{ii})^{1/2}\zeta^\sigma_q((y^I_{ii})^{-\sigma},1-\sigma-\fkb^I_{ii})^{-\sigma/2}$.}
\begin{equation}
\label{blocks}
\begin{aligned}
	\Psi^\pm_I={}&\prod_{i,j=0}^{N-1}e^{\frac{\pi\ii}{2\epsilon}\:\text{sign}(j-i)\left[
		(\Delta^\pm_I\,-\,\sigma_\pm\epsilon-1)(u^\pm_{a;\,i}-\,u^\pm_{b;\,j})\,\mp\,(u^\pm_{a;\,i}-\,u^\pm_{b;\,j})^2\right]}\:,\\
	\mathcal B^\pm_I={}&\underset{\pm(i-j)>0}{\prod_{i,j=0}^{N-1}}\frac
		{\left(\mathrm{e}^{\pm2\pi\ii\sigma_\pm(u^\pm_{a;\,i}-u^\pm_{b;\,j})-\pi\ii\sigma_\pm\Delta^\pm_I+2\pi\ii\epsilon}\,
		;\,e^{2\pi\ii\epsilon}\right)_\infty}
		{\left(\mathrm{e}^{\mp2\pi\ii\sigma_\pm(u^\pm_{a;\,j}-u^\pm_{b;\,i})+\pi\ii\sigma_\pm\Delta^\pm_I}\,;\,e^{2\pi\ii\epsilon}\right)_\infty}\:,\\
    \diag^\pm_I={}&\prod_{i=0}^{N-1}\frac
		{\left(\mathrm{e}^{\pm2\pi\ii\sigma_\pm(u^\pm_{a;\,i}-u^\pm_{b;\,i})-\pi\ii\sigma_\pm\Delta^\pm_I+2\pi\ii\epsilon}\,
		;\,e^{2\pi\ii\epsilon}\right)^{1/2}_\infty}
		{\left(\mathrm{e}^{\mp2\pi\ii\sigma_\pm(u^\pm_{a;\,i}-u^\pm_{b;\,i})+\pi\ii\sigma_\pm\Delta^\pm_I}\,;\,e^{2\pi\ii\epsilon}\right)^{1/2}_\infty}\:,
\end{aligned}
\end{equation}
and the vector multiplet contributions $\Psi^\pm_a$, $\cB^\pm_a$ are given by the same expressions
with $b=a$ and $\Delta^\pm_I$ being replaced by $2(1+\sigma_\pm\epsilon)$.
Here $(z\,;\,q)_\infty=\prod_{k=0}^\infty(1-zq^k)$ is the $q$-Pochhammer symbol.
At large\,-$N$, the $\Psi^\pm$ terms will give rise to the $\cO(N^2)$ long range forces,
the $\cB^\pm$ terms will provide the $\cO(N^{3/2})$ contributions, whereas $\diag_I^\pm$ only gives a subleading contribution
but plays a role in correcting the saddle point equations in the tail regions.

Note that, compared to \cite{Colombo:2024mts} (with $n_\pm=1$) we are using slightly different conventions,
including a different sign convention for ${\rm k}_a$ and the following change of notations:
\begin{equation}
	\left(\Delta^\pm_I\right)_{\text{here}}=\frac1\pi\left(\Delta^\pm_I\right)_{\text{in \cite{Colombo:2024mts}}}\:,\quad
		\left(\epsilon\,\right)_{\text{here}}=\frac1{2\pi}\left(\epsilon\,\right)_{\text{in \cite{Colombo:2024mts}}}\:,\quad
		\left(e^{2\pi \ii\,u^\pm_{a;\,i}}\right)_{\text{here}}=\left(z^\pm_{a;\,i}\right)_{\text{in \cite{Colombo:2024mts}}}\:.
\end{equation}
After some algebra, one can also verify that the product $\Psi^+_I\:\Psi^-_I$ here matches with
$\Psi_I$ in \cite{Colombo:2024mts} up to a phase.%
\footnote{This phase was discarded in \cite{Colombo:2024mts} on the grounds that phases are expected to be subleading at large\,-$N$. Here we will be more precise and keep it.}

\subsection{The Cardy and large\,-$N$ limit of the indices}
In this section we extend the large\,-$N$ computation of the $S^2\times S^1$ indices to the chiral quivers of interest,
using the approach of \cite{Hosseini:2022vho,Colombo:2024mts}, while carefully keeping track of long range forces.
In this approach, the logarithm of the one-loop determinants is expanded in powers of $\epsilon$;
this expansion is valid up to non-perturbative corrections that are exponentially suppressed in the Cardy limit $\epsilon\to0$.%
\footnote{More generically, one could also consider a generalized Cardy limit where $\epsilon$ approaches a rational number rather than zero, see \cite{BenettiGenolini:2023rkq}.}

The ansatz for the large\,-$N$ distribution of $u^\pm_{a;\,i}$ is reminiscent of \eqref{eigenvalue_Ansatz},
up to factors and signs:
\begin{equation}
	u^\pm_{a;\,i}\,=\,\frac1{4\pi\ii}\left(-\sigma_\pm\,N^{1/2}\,x_i^\pm\,\mp\,\ii\,y_a^\pm(x^\pm_i)\right)\:.
\end{equation}
In the same spirit of \eqref{sums_to_integrals},
at large\,-$N$ the $x_i^\pm$ is replaced with a continuous variable $x$, and the summations over $i=0,\ldots,N-1$ can be replaced with
integrals in $x$ by introducing eigenvalue densities $\rho^\pm(x)$ such that
\begin{equation}
	\sum_{i=0}^{N-1}\bullet\:=\:N\int\dd x\,\rho^\pm(x)\,\bullet\:+\:\cO(N^0)\:,\qquad\int\dd x\,\rho^\pm(x)=1\:.
\end{equation}
The classical terms \eqref{Z_classical} in the integrand then become 
\begin{equation}
	\log Z_{\text{classical}}=N^{3/2}\sum_\pm\bigg[-\frac{\sigma_\pm}{4\,\epsilon}
		\int\dd x\,\rho^\pm(x)\cdot x\cdot\sum_{a=1}^{|\cG|}\left(\frac{{\rm k}_a}{2\pi}\,y^\pm_a(x)+\Delta^\pm_{m;\,a}\right)\bigg]+\cO(N)\:,
\end{equation}
where the $\cO(N^2)$ terms are canceled by the fact that the sum of Chern-Simons levels vanishes \eqref{CS_level_sum}
for the quiver theories of interest.
Similarly, the prefactors $\Psi^\pm_I$ in \eqref{blocks} become
\begin{align}
\nonumber
	\log\Psi^\pm_I={}&\int\dd x\,\dd\wt x\,\rho^\pm(x)\,\rho^\pm(\,\wt x\,)\,\text{sign}(\wt x-x)\bigg[\pm N^3\frac\ii{32\pi\epsilon}\,(x-\wt x)^2\,-\\
	{}&-N^{5/2}\,\frac{\sigma_\pm}{8\,\epsilon}\,(x-\wt x)\,
		\Big(\tfrac{y^\pm_a(x)-y^\pm_b(\,\wt x\,)}{2\pi}+(\Delta^\pm_I-\sigma_\pm\epsilon-1)\Big)\,\mp\\\nonumber
	{}&\mp N^2\,\frac{\pi\ii}{4\,\epsilon}\Big(\tfrac{y^\pm_a(x)-y^\pm_b(\,\wt x\,)}{2\pi}\,(\Delta^\pm_I-\sigma_\pm\epsilon-1)
		+\tfrac12\big(\tfrac{y^\pm_a(x)-y^\pm_b(\,\wt x\,)}{2\pi}\big)^2\Big)\bigg]+\cO(N)\:.
\end{align}
It is convenient to rewrite the above in terms of quantities that are analogous to  \eqref{cDa_nab_def}:
\begin{equation}
	\cD^\pm_a=\sum_{J\in W}\Delta^\pm_J+\sum_{I\in(a)}\Big(\Delta^\pm_I-\frac12\sum_{J\in W}\Delta^\pm_J\Big)\:,\qquad n_{a,b}=
		\underset{I\,:\,a\to b}{\sum_{I=1}^{|\cR|}}1-\underset{I\,:\,b\to a}{\sum_{I=1}^{|\cR|}}1\:,
\end{equation}
Again, for the chiral quivers of interest only the sum over $a$ of the above vanishes generically:
\begin{equation}
\label{quiver_conditions}
	\sum_{a=1}^{|\cG|}\cD^\pm_a=0\:,\qquad\sum_{a=1}^{|\cG|}n_{a,b}=0\:,
\end{equation}
whereas for a non-chiral theory all the $\cD_a$ and $n_{a,b}$ vanish identically.
Using $\sigma_\pm\epsilon+1=\frac12\sum_{J\in W}\Delta^\pm_J$, the $\Psi^\pm_I$ prefactors can be written as
\begin{align}
	{}&\sum_{I=1}^{|\cR|}\log\Psi^\pm_I+\sum_{a=1}^{|\cG|}\log\Psi^\pm_a\,=\\\nonumber
	{}&=\mp\,\frac{\pi\ii\,N^2}{4\,\epsilon}\int\dd x\,\dd\wt x\,\rho^\pm(x)\,\rho^\pm(\,\wt x\,)\,\text{sign}(\wt x-x)
	\sum_{a=1}^{|\cG|}\bigg[\frac1{2\pi}\,\cD^\pm_a\,y^\pm_a(x)-\frac1{8\pi^2}\sum_{b=1}^{|\cG|}n_{a,b}\:y^\pm_a(x)\,y^\pm_b(\,\wt x\,)\bigg]\:,
\end{align}
where the $\cO(N^3)$ terms have vanished because the integrand is anti-symmetric in the exchange of $x$ and $\wt x$, while the $\cO(N^{5/2})$ terms vanish
after imposing \eqref{quiver_conditions}.

Let us now expand the $\cB^\pm$ terms in \eqref{blocks}.
The computation is the same as \cite{Hosseini:2022vho,Colombo:2024mts}, let us review it for completeness.
For simplicity, let us assume that $\Delta^\pm_I$ is real; we can extend the result to complex $\Delta^\pm_I$ by analytic continuation at the end.%
\footnote{One has to be careful about Stokes lines, namely for complex $\Delta_I^\pm$ the inequality \eqref{index_bounds}
        should be corrected to $0<\text{Im}\Big(\frac1\epsilon\Big(\pi\Delta^\pm_I+\frac{y^\pm_a(x)-y^\pm_b(x)}2\Big)\Big)
        <\text{Im}\left(\frac{2\pi}\epsilon\right)$, see \cite{Colombo:2024mts}.}
We can make use of the asymptotic formula
\begin{equation}
\label{Pochhammer_asymptotic1}
	\log\left(z\,;\,e^{2\pi\ii\epsilon}\right)_\infty\:\sim\:\sum_{k=0}^\infty\frac{\left(2\pi\ii\epsilon\right)^{k-1}}{k!}\,B_k^-\,\Li_{2-k}(z)\:,
\end{equation}
which is valid in the domain $|z|<1$ up to corrections that are exponentially suppressed for $\epsilon\to0$.
Here $B_k^\pm$ are the Bernoulli numbers with the convention $B^\pm_1=\pm\frac12$.%
\footnote{Namely, $B_k^-=B_k(0)$ and $B^+_k=B_k(1)$. $B^+_k=B^-_k$ when $k\ne1$.}
Using $(zq\,;\,q)_\infty=(z\,;\,q)_\infty/(1-z)$
and $\Li_1(z)=-\log(1-z)$ we also have that
\begin{equation}
\label{Pochhammer_asymptotic2}
	\log\left(z\,e^{2\pi\ii\epsilon}\,;\,e^{2\pi\ii\epsilon}\right)_\infty\:\sim\:
		\sum_{k=0}^\infty\frac{\left(2\pi\ii\epsilon\right)^{k-1}}{k!}\,B_k^+\,\Li_{2-k}(z)\:.
\end{equation}
Using \eqref{Pochhammer_asymptotic1} and \eqref{Pochhammer_asymptotic2} we can expand the blocks $\cB^\pm$ in \eqref{blocks} as
\begin{align}
\nonumber
	\log\cB_I^\pm\sim N^2\int_{\pm(x-\wt x)>0}{}&\dd x\,\dd\wt x\,\rho^\pm(x)\,\rho^\pm(\,\wt x\,)
		\sum_{k=0}^\infty\frac{(2\pi\ii\epsilon)^{k-1}}{k!}\:\cdot\\
	&\bigg[B_k^+\,\Li_{2-k}\left(e^{\mp\frac12N^{1/2}(x-\wt x\,)
		-\frac{\ii}2\sigma_\pm\big(2\pi\Delta^\pm_I+y_a^\pm(x)-y_b^\pm(\,\widetilde x\,)\big)}\right)-\\\nonumber
	&\quad-B_k^-\,\Li_{2-k}\left(e^{\mp\frac12N^{1/2}(x-\wt x\,)
		+\frac{\ii}2\sigma_\pm\big(2\pi\Delta^\pm_I+y_a^\pm(\,\wt x\,)-y_b^\pm(x)\big)}\right)\bigg]\:.
\end{align}
We can then perform the integral in $\wt x$ by using the polylogarithm asymptotic formula \eqref{Li_asymptotic}
with $t=\pm\frac12(x-\wt x)$, obtaining
\begin{align}
\nn
	\log\cB_I^\pm\sim 2N^{3/2}\int\dd x\,\rho^\pm(x)^2
		\sum_{k=0}^\infty{}&\frac{(2\pi\ii\epsilon)^{k-1}}{k!}\bigg[B_k^+\,\Li_{3-k}\left(e^{
		-\frac{\ii}2\sigma_\pm\big(2\pi\Delta^\pm_I+y_a^\pm(x)-y_b^\pm(x)\big)}\right)-\\
	&-B_k^-\,\Li_{3-k}\left(e^{\frac{\ii}2\sigma_\pm\big(2\pi\Delta^\pm_I+y_a^\pm(x)-y_b^\pm(x)\big)}\right)\bigg]\:.
\end{align}
The polylogarithm inversion formula \eqref{polylog_inversion}
and the fact that $B_k^\pm$ vanishes when $k$ is an odd number greater than one imply that
\begin{equation}
	B_k^+\,\Li_{3-k}\left(e^{\ii\,y}\right)-B_k^-\,\Li_{3-k}\left(e^{-\ii\,y}\right)=
		\begin{cases}-B_k^+\frac{(2\pi\ii)^{3-k}}{(3-k)!}B_{3-k}\left(\frac y{2\pi}-\lfloor\frac y{2\pi}\rfloor\right)\qquad&\text{for }\:k\leq2\:,\\
		0&\text{for }\:k>2\:.\end{cases}
\end{equation}
Similarly to \eqref{marginality_substitution}, 
we substitute the constraint \eqref{Delta_constraint} in the above expressions as follows:
\begin{equation}
	B_{3-k}\left(\frac y{2\pi}-\left\lfloor\frac y{2\pi}\right\rfloor\right)=1^{3-k}\,B_{3-k}\left(\frac y{2\pi\cdot1}-\left\lfloor\frac y{2\pi}\right\rfloor\right)\:,
		\qquad1\:\equiv\:\frac12\sum_{I\in W}\Delta_I^\pm-\sigma_\pm\epsilon\:.
\end{equation}
Then, if the functions $y^\pm_a(x)$ are such that
\begin{equation}
\label{index_bounds}
	0<\pi\Delta^\pm_I+\frac12\big(y^\pm_a(x)-y^\pm_b(x)\big)<2\pi\:,
\end{equation}
the asymptotic expansion for $\cB^\pm$ takes the form
\begin{align}
\label{blocks_last}
	\log\cB_I^\pm\sim N^{3/2}\int\dd x{}&\,\rho^\pm(x)^2\bigg(-\frac{\sigma_\pm}{\pi\epsilon}\,
		G_3^\pm\Big(\pi\Delta^\pm_I+\frac12\big(y^\pm_a(x)-y^\pm_b(x)\big)\Big)+\\\nonumber
	&+\frac{\pi^2}3\Big(\Delta_I^\pm-\frac12\sum_{J\in W}\Delta^\pm_J+\tfrac{y^\pm_a(x)-y^\pm_b(x)}{2\pi}\Big)
		\Big(\sigma_\pm\epsilon-\frac12\sum_{J\in W}\Delta^\pm_J\Big)\bigg)\:,
\end{align}
where the cubic polynomials $G_3^\pm$ are defined in a similar fashion as \eqref{G3_def},
\begin{equation}
\label{G3_def_pm}
	G_3^\pm(x)=\frac x6\Big(x-\frac\pi2\sum_{I\in W}\Delta^\pm_I\Big)\Big(x-\pi\sum_{I\in W}\Delta^\pm_I\Big)\:.
\end{equation}
Using \eqref{quiver_conditions}, the terms in the bottom line of \eqref{blocks_last} vanish when taking the sum over all multiplets.

It is easy to see that the $\diag_I^\pm$ terms in \eqref{blocks} only give a $\cO(N)$ contribution and can be neglected.
After putting everything together, the large\,-$N$, small\,-\,$\epsilon$ expansion of the integrand of the indices can be written as
\begin{equation}
	\log(Z_{\text{classical}}\:Z_{\text{1-loop}})\sim\frac{1}{4\,\epsilon}\Big(\cF\big(\Delta^-,\rho^-,y^-,\nu^-\,\big)
		-\sigma\,\cF\big(\Delta^+,\rho^+,y^+,\nu^+\,\big)\Big)\:,
\end{equation}
where $\cF(\Delta,\rho,y,\nu)$ is the free energy functional given by \eqref{free_energy_functional}
and for convenience we have defined $\nu^\pm=\pm\pi\ii\sigma_\pm N^{1/2}$.
Provided that one can solve the saddle point equations \eqref{saddle_point_equations} for the functional $\cF(\Delta,\rho,y,\nu)$,
it would then be possible to perform an independent saddle point approximation with respect to the functions $\rho^-,y^-$
and the functions $\rho^+,y^+$, leading to the following result for the limit of the indices:
\begin{equation}
\label{indices_vs_F}
	\log Z_{S^2\times S^1,\,\sigma}(\Delta^\pm,\epsilon)\sim\frac{1}{4\,\epsilon}\Big(\cF_{\text{crit.}}(\Delta^-)
		-\sigma\,\cF_{\text{crit.}}(\Delta^+)\Big)\:,
\end{equation}
where $\cF_{\text{crit.}}$ is defined as in \eqref{crtical_value_free_energy}, it matches the large\,-$N$ limit
of the $S^3$ free energy $F_{S^3}$.
These results are consistent with the analysis of \cite{Choi:2019zpz,Choi:2019dfu},
which considers the leading order in the Cardy limit,
and also extend the results of \cite{Hosseini:2022vho} to both values of $\sigma$ and chiral quivers.

For the topologically twisted index, the static limit $\epsilon=0$ is well defined.
Indeed, after setting $\sigma=+1$, we can rewrite \eqref{Delta^pm_def} as
\be
    \Delta^\pm_I=\Delta_I\pm\,\epsilon\,\fks_I\:,\qquad\Delta_I\,\equiv\,2\,u_I+r_I\:,\qquad\fks_I\,\equiv\,\fkn_I+r_I\:,
\ee
where $\Delta_I$ and $\fks_I$ satisfy
\be
   \sum_{I\in W}\Delta_I=2\:,\qquad\sum_{I\in W}\fks_I=2
\ee
for all superpotential terms $W$.
Similarly we can write $\Delta^\pm_{m}=\Delta_{m}\pm\,\epsilon\,\fks_{m}$, where
$\Delta_{m}\,\equiv\,2{\sum}_{a=1}^{|\cG|}u_{m;\,a}\:$ and
$\:\fks_{m}\,\equiv\,{\sum}_{a=1}^{|\cG|}\fkn_{m;\,a}$.
Then the divergent $\epsilon^{-1}$ terms in \eqref{indices_vs_F} cancel and
the topologically twisted index at $\epsilon=0$ becomes
\be
\label{indices_vs_F_unrefined}
    \log Z_{S^2\times S^1,\,\sigma\,=\,+1}\,(\Delta^\pm,\epsilon=0)\,\sim\,-\,\frac12\sum_{I=1}^{|\cR|}\fks_I\,
        \frac\partial{\partial\Delta_I}\cF_{\text{crit.}}(\Delta)
        \,-\,\frac12\,\fks_{m}\,\frac\partial{\partial\Delta_{m}}\cF_{\text{crit.}}(\Delta)\:,
\ee
which is consistent with \cite{Hosseini:2016tor}.

Using \eqref{indices_vs_F} and \eqref{indices_vs_F_unrefined}, the analytic saddle point solution for the chiral quiver dual to
$Q^{1,1,1}/\bZ_{\rm k}$ that we will describe in section \ref{sec: Analytic solution}
will characterize the large\,-$N$ leading order behavior of the free energy $F_{S^3}$ and the indices
$Z_{S^2\times S^1,\,\sigma}$ at the same time.%
\footnote{The inequalities \eqref{free_energy_determination} and \eqref{index_bounds},
    which generate the tail regions when saturated,
    can also be mapped into each other.}
Even if the rest of this paper will focus exclusively on the chiral quiver dual to $Q^{1,1,1}/\bZ_{\rm k}$, the derivation of this section applies more broadly to any chiral quiver with bifundamental matter, linking the large\,-$N$ behaviour of $F_{S^3}$ with the one of the $S^2\times S^1$ indices.
As recently observed in \cite{Bobev:2026lvl}, this correspondence between $F_{S^3}$ and the indices is expected to also hold for all the subleading orders in $N$, provided that the $S^3$ is taken to be \emph{squashed}.
In this paper we only consider the round $S^3$ free energy, which is sufficient to derive the leading order in $N$ of the indices.

\section{Analytic solution to the saddle point equations}
\label{sec: Analytic solution}
In this section we solve analytically the large\,-$N$ saddle point equations for the chiral quiver dual to AdS$_4\times Q^{1,1,1}/\bZ_{\rm k}$, giving us the large\,-$N$ behavior of both the free energy and the Cardy limit of the superconformal and topologically twisted indices of the chiral theory. In section \ref{sec: Saddle point of Q111} we first provide a simple preliminary solution valid when $\Delta_I$ obey a linear relation, namely $\Delta_{A_1}+\Delta_{C_1}=\Delta_{A_2}+\Delta_{C_2}$. The solution that we find is in agreement with both the numerical analysis and the operator counting results of section \ref{sec:chiral-quiver}. Then in section \ref{sec:varying the discontinuity} we verify that this solution is stable under variation of the position of the discontinuity, while in section \ref{sec:subleading corrections} we check that the cancellation of long range forces is caused by subleading corrections to the solution in the tail region, and we compare the analytic prediction for such corrections with the numerical saddles of section \ref{sec:chiral-quiver}, finding a good match. In section \ref{sec: A proposal for general Delta} we find a solution valid for general $\Delta_I$, by adding a large $\cO(N)$ imaginary component to the ansatz for the eigenvalues. In section \ref{sec:Delta_1_numeric_analysis} we verify the correctness of the general analytical solution with a numerical analysis. Lastly, in section \ref{Comparison with the Sasakian volume} we compare our results with the gravitational prediction governed by the Sasakian volume of $Q^{1,1,1}/\bZ_{\rm k}$.

\subsection{A large\,-$N$ saddle for a subclass of $\Delta$ values}
\label{sec: Saddle point of Q111}

For the chiral quiver dual to $Q^{1,1,1}/\bZ_{\rm k}$, the functional \eqref{free_energy_functional} that we need to extremize
takes the following form:
\begin{align}
\label{Q111_functional}
\nonumber
	{}&\frac{\cF}{N^{3/2}}=\int\dd x\,\rho(x)\cdot x\cdot\Big({\rm k}\,\tfrac{y_1(x)-y_2(x)}{2\pi}+\Delta_{m}\Big)
		+\frac{4}{\pi}\int\dd x\,\rho(x)^2\bigg[G_3\Big(\pi\Delta_{A_1}+\tfrac{y_1(x)-y_3(x)}2\Big)+\\\nonumber
	&\quad+G_3\Big(\pi\Delta_{A_2}+\tfrac{y_2(x)-y_3(x)}2\Big)+G_3\Big(\pi\Delta_{B_1}+\tfrac{y_3(x)-y_4(x)}2\Big)
		+G_3\Big(\pi\Delta_{B_2}+\tfrac{y_3(x)-y_4(x)}2\Big)+\\\nonumber
	&\quad+G_3\Big(\pi\Delta_{C_1}+\tfrac{y_4(x)-y_1(x)}2\Big)+G_3\Big(\pi\Delta_{C_2}+\tfrac{y_4(x)-y_2(x)}2\Big)\bigg]+
		\nu\int\dd x\,\dd\wt x\,\rho(x)\,\rho(\,\wt x\,)\,\text{sign}(\wt x-x)\cdot\\\nonumber
	&\quad\cdot\bigg[\tfrac{y_1(x)}{2\pi}\Big(\Delta_{A_1}+\Delta_{C_1}-\tfrac{y_3(\,\wt x\,)-y_4(\,\wt x\,)}{4\pi}\Big)
		+\tfrac{y_2(x)}{2\pi}\Big(\Delta_{A_2}+\Delta_{C_2}-\tfrac{y_3(\,\wt x\,)-y_4(\,\wt x\,)}{4\pi}\Big)-\\\nonumber
	&\quad\quad-\tfrac{y_3(x)}{2\pi}\Big(\Delta_{C_1}+\Delta_{C_2}+\tfrac{2y_4(\,\wt x\,)-y_1(\,\wt x\,)-y_2(\,\wt x\,)}{4\pi}\Big)
		-\tfrac{y_4(x)}{2\pi}\Big(\Delta_{A_1}+\Delta_{A_2}+\tfrac{y_1(\,\wt x\,)+y_2(\,\wt x\,)-2y_3(\,\wt x\,)}{4\pi}\Big)\bigg]-\\
	&\quad-\frac{\mu}{2\pi}\,\bigg[\int\dd x\,\rho(x)-1\bigg]\:,
\end{align}
where $\mu$ is a Lagrange multiplier, $\nu\sim N^{1/2}\,$ is a large parameter, and the cubic polynomial $G_3$ is defined as in \eqref{G3_def} with
\begin{equation}
	{\sum}_{I\in W}\,\Delta_I\:=\:\Delta_{A_1}+\Delta_{A_2}+\Delta_{B_1}+\Delta_{B_2}+\Delta_{C_1}+\Delta_{C_2}\:.
\end{equation}
When considering the free energy $F_{S^3}$, the above quantity is equal to 2 using \eqref{marginality_general}, whereas in the case of the superconformal and topologically twisted indices it is fixed by \eqref{Delta_constraint}; throughout the rest of this section we will not fix its value so that this discussion applies to both the free energy and the indices.
Without loss of generality we can assume that ${\rm k}>0$ and $\Delta_{m}=0$. 
The case $\Delta_{m}\ne0$ can then be easily derived by exploiting the flat directions \eqref{flat_directions} of $\cF$.%
\footnote{\label{Delta_m_reintroduction}Explicitly, it is possible to reintroduce $\Delta_{m}$ by taking any of the equations in this section and making the
replacements $\Delta_{A_1}\longrightarrow\Delta_{A_1}-\frac1{\rm k}\Delta_{m}\:$,
$\:\Delta_{C_1}\longrightarrow\Delta_{C_1}+\frac1{\rm k}\Delta_{m}\:$
and $\:y_1(x)\longrightarrow y_1(x)+\frac{2\pi}{\rm k}\Delta_{m}\:$.}

Motivated by the numerical analysis and the operator counting results,
we take an ansatz so that the solution at large\,-$N$ is subdivided in four tail regions,
with each region being characterized by the saturation of the left inequality \eqref{free_energy_determination}
for a subset of the chiral multiplets as follows:
\begin{equation}
\label{Q111_solution_ansatz}
\begin{cases}
	0=\tfrac{y_1(x)-y_3(x)}{2\pi}+\Delta_{A_1}=\tfrac{y_3(x)-y_4(x)}{2\pi}+\Delta_{B_2}=
		\tfrac{y_4(x)-y_2(x)}{2\pi}+\Delta_{C_2}\:,&\makebox[15mm][r]{$\xmin<\:$}x<\xL\:,\\[4mm]
	0=\tfrac{y_1(x)-y_3(x)}{2\pi}+\Delta_{A_1}=\tfrac{y_4(x)-y_2(x)}{2\pi}+\Delta_{C_2}\:,&\makebox[15mm][r]{$\xL<\:$}x<0\:,\\[4mm]
	0=\tfrac{y_2(x)-y_3(x)}{2\pi}+\Delta_{A_2}=\tfrac{y_4(x)-y_1(x)}{2\pi}+\Delta_{C_1}\:,&\makebox[15mm][r]{$0<\:$}x<\xR\:,\\[4mm]
	0=\tfrac{y_2(x)-y_3(x)}{2\pi}+\Delta_{A_2}=\tfrac{y_3(x)-y_4(x)}{2\pi}+\Delta_{B_2}=
		\tfrac{y_4(x)-y_1(x)}{2\pi}+\Delta_{C_1}\:,&\makebox[15mm][r]{$\xR<\:$}x<\xmax\:.
\end{cases}
\end{equation}
The presence of tail regions where \eqref{free_energy_determination} is saturated is a well-established feature
of the large\,-$N$ saddles of non-chiral quivers;
what distinguishes the above ansatz for the chiral quiver in consideration from the known saddles of non-chiral quivers
is the lack of a ``central region'' where \eqref{free_energy_determination} is not saturated, and the fact that
we allow for discontinuities at $x=0$ in the functions $y_a(x)$. For finite $N$, such a central region is
present in the numerical analysis of section \ref{sec:numerical_solution} and in the one of \cite{Hosseini:2025jxb};
however, at large\,-$N$ this central region shrinks to zero, as it is evident from Figure \ref{fig:kink-position}, leaving behind
only the $x=0$ discontinuity for large enough $N$.
This discontinuity is also a key feature of the operator counting results. 

The value of $\xL$ and $\xR$ will be determined by imposing that the solution
is continuous at each of these points, while the value of $\xmin$ and $\xmax$ is determined by imposing 
$\rho(\xmin)=0=\rho(\xmax)$, and it will not be possible to extend the solution beyond these points without violating
the condition $\rho(x)\geq0$.
After imposing the ansatz \eqref{Q111_solution_ansatz}, the free energy functional effectively depends only on two functions,
namely $\rho(x)$ and $\delta y_{34}(x)\,\equiv\,y_3(x)-y_4(x)$, the latter being further constrained to the value
$\delta y_{34}(x)=-2\pi\Delta_{B_2}$ when $x\in(\xmin,\xL)\cup(\xR,\xmax)$.

If we substitute the ansatz \eqref{Q111_solution_ansatz} in the free energy functional, we find that the long range forces take the following form:
\begin{align}
\label{Q111_long_range_forces}
\nonumber
	\frac{\cF-\cF\,\big|_{\nu\,=\,0}}{\nu N^{3/2}}\,=\,\big(\Delta_{A_1}\!+\!\Delta_{C_1}\!-\!\Delta_{A_2}\!-\!\Delta_{C_2}\big)
		{}&\int_{\xmin}^0\dd x\int_0^{\xmax}\dd\wt x\:\rho(x)\rho(\,\wt x\,)\Big[\tfrac{\delta y_{34}(x)+\delta y_{34}(\,\wt x\,)}{2\pi}\,-\\
	{}&\quad-\big(\Delta_{A_1}+\Delta_{C_1}+\Delta_{A_2}+\Delta_{C_2}\big)\Big]\:.
\end{align}
In particular, the long range forces vanish when the following condition is satisfied:
\begin{equation}
\label{Delta_restriction}
	\Delta_{A_1}+\Delta_{C_1}=\Delta_{A_2}+\Delta_{C_2}\:.
\end{equation}
In the rest of this section we will restrict to values of $\Delta_I$ that satisfy the above constraint.
We will describe a solution for general $\Delta_I$ in section \ref{sec: A proposal for general Delta}.

\begin{figure}[htbp]
     \centering
     \begin{subfigure}[b]{0.5\textwidth}
         \centering
         \includegraphics[width=\linewidth]{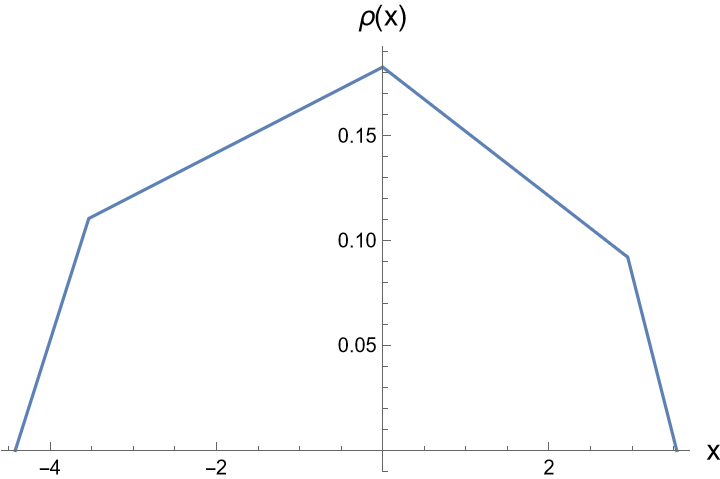}
         \caption{Eigenvalue density $\rho(x)$}
         \label{fig:analytic_rho}
     \end{subfigure}
     \hfill
     \begin{subfigure}[b]{0.45\textwidth}
         \centering
         \includegraphics[width=\linewidth]{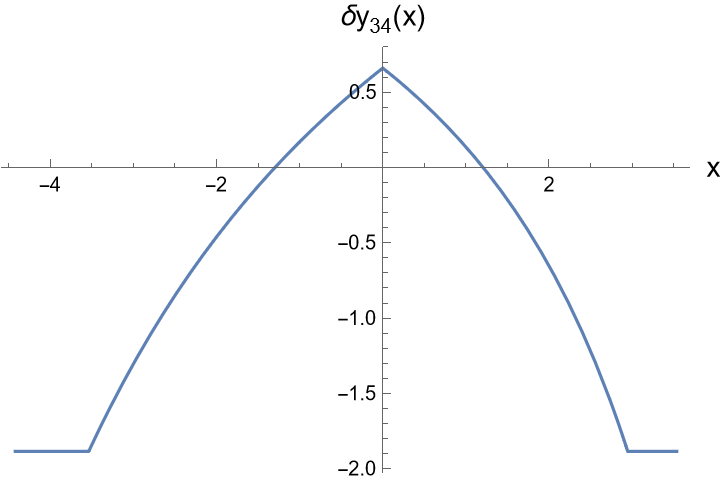}
         \caption{Eigenvalue difference $y_3(x)-y_4(x)$}
         \label{fig:analytic_y34}
     \end{subfigure}
     
     \vspace{3mm}
     \begin{subfigure}[b]{0.5\textwidth}
         \centering
         \includegraphics[width=\linewidth]{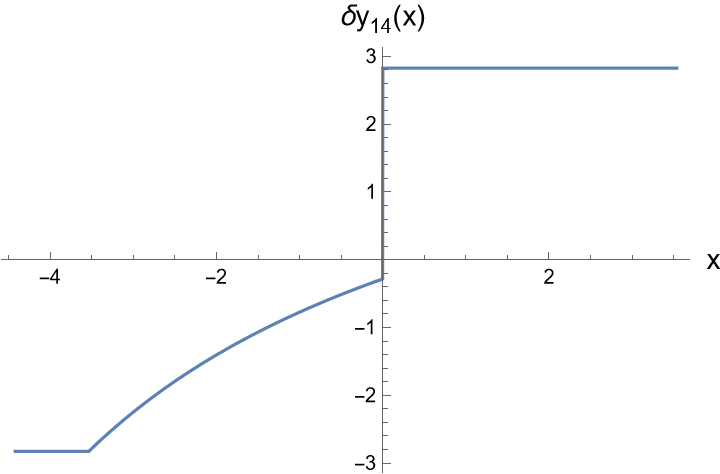}
         \caption{Eigenvalue difference $y_1(x)-y_4(x)$}
         \label{fig:analytic_y14}
     \end{subfigure}
     \hfill
     \begin{subfigure}[b]{0.45\textwidth}
         \centering
         \includegraphics[width=\linewidth]{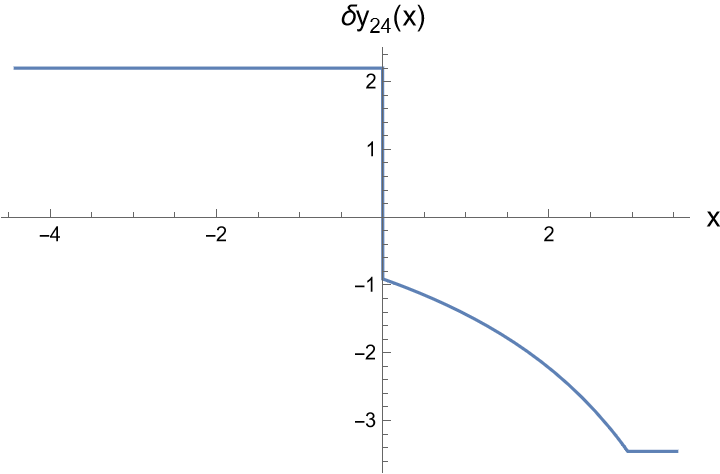}
         \caption{Eigenvalue difference $y_2(x)-y_4(x)$}
         \label{fig:analytic_y24}
     \end{subfigure}     
     \caption{Plot of the analytic solution for the following assignment of values: ${\rm k}=1$, $\Delta_{A_1}=0.15$, \,$\Delta_{A_2}=0.25$, \,$\Delta_{B_1}=0.5$, \,$\Delta_{B_2}=0.3$, \,$\Delta_{C_1}=0.45$, \,$\Delta_{C_2}=0.35$. These values obey the constraint $\Delta_{A_1}+\Delta_{C_1}=\Delta_{A_2}+\Delta_{C_2}$.}
     \label{fig:analytic_plot}
\end{figure}

After imposing \eqref{Delta_restriction}, the extremization equations for $\cF$ in each of the four regions become algebraic and are easily solved.
In order to write the results in a concise manner, we use the following notation for particular linear combinations of the $\Delta_I$:
\begin{equation}
	\Delta_{ijk}\,\equiv\,\Delta_{A_i}+\Delta_{B_j}+\Delta_{C_k}\:,\qquad i,j,k\in\{1,2\}\:.
\end{equation}
Then the result of the extremization of $\cF$ can be written as
\begin{equation}
\label{rho_sol}
	\rho(x)=\begin{cases}
		\displaystyle\frac{2\pi\,{\rm k}\,x\,\Delta_{122}+\mu}{(2\pi)^3(\Delta_{B_1}-\Delta_{B_2})\,\Delta_{121}^2}\:,
			&\makebox[15mm][r]{$\xmin<\:$}x<\xL\:,\\[7mm]
		\displaystyle\frac{2\pi\,{\rm k}\,x\left(\Delta_{122}\DeltaS-\Delta_{121}^2\right)+\mu\,\DeltaS}
			{(2\pi)^3\,\left(\Delta_{111}\,\Delta_{121}\right)^2}\:,&\makebox[15mm][r]{$\xL<\:$}x<0\:,\\[7mm]
		\displaystyle\frac{-2\pi\,{\rm k}\,x\left(\Delta_{221}\DeltaS-\Delta_{121}^2\right)+\mu\,\DeltaS}
			{(2\pi)^3\,\left(\Delta_{111}\,\Delta_{121}\right)^2}\:,&\makebox[15mm][r]{$0<\:$}x<\xR\:,\\[7mm]
		\displaystyle\frac{-2\pi\,{\rm k}\,x\,\Delta_{221}+\mu}{(2\pi)^3(\Delta_{B_1}-\Delta_{B_2})\,\Delta_{121}^2}\:,
			&\makebox[15mm][r]{$\xR<\:$}x<\xmax\:,\end{cases}
\end{equation}
and also
\begin{equation}
\label{y34_sol}
	\frac{\delta y_{34}(x)}{2\pi}=\begin{cases}
		-\Delta_{B_2}\:,&\makebox[15mm][r]{$\xmin<\:$}x<\xL\:,\\[5mm]
		\Delta_{A_2}+\Delta_{C_2}+\displaystyle\frac{\left(2\pi\,{\rm k}\,x\,(\Delta_{C_1}-\Delta_{C_2})-\mu\right)\,\Delta_{111}\,\Delta_{121}}
			{2\pi\,{\rm k}\,x\left(\Delta_{122}\DeltaS-\Delta_{121}^2\right)+\mu\,\DeltaS}\:,&\makebox[15mm][r]{$\xL<\:$}x<0\:,\\[7mm]
		\Delta_{A_1}+\Delta_{C_1}+\displaystyle\frac{\left(2\pi\,{\rm k}\,x\,(\Delta_{C_1}-\Delta_{C_2})-\mu\right)\,\Delta_{111}\,\Delta_{121}}
			{-2\pi\,{\rm k}\,x\left(\Delta_{221}\DeltaS-\Delta_{121}^2\right)+\mu\,\DeltaS}\:,&\makebox[15mm][r]{$0<\:$}x<\xR\:,\\[7mm]
		-\Delta_{B_2}\:,&\makebox[15mm][r]{$\xR<\:$}x<\xmax\:.\end{cases}
\end{equation}
The values of $\xL$ and $\xR$ are determined by imposing continuity of $\rho(x)$ and $\delta y_{34}(x)$, whereas $\xmin$ and $\xmax$ are determined
as the points where $\rho(x)$ vanishes. The result is
\begin{equation}
\label{x_bounds}
	\xmin=-\frac{\mu}{2\pi\,{\rm k}\,\Delta_{122}}\:,\qquad\xL=-\frac{\mu}{2\pi\,{\rm k}\,\Delta_{112}}\:,\qquad
		\xR=\frac{\mu}{2\pi\,{\rm k}\,\Delta_{211}}\:,\qquad\xmax=\frac{\mu}{2\pi\,{\rm k}\,\Delta_{221}}\:,
\end{equation}
which is consistent with the ordering $\xmin<\xL<0<\xR<\xmax$ if
\be
\label{x_bounds_inequalities}
	\Delta_{B_1}-\Delta_{B_2}>0\:,\qquad\Delta_{122}>0\:,\qquad\Delta_{221}>0\:.
\ee
The case $\Delta_{B_1}=\Delta_{B_2}$ may be obtained as a well defined limit, in this case $\xmin=\xL$ and $\xR=\xmax$.
Furthermore, in order to satisfy the inequalities in \eqref{free_energy_determination}, we also demand that
\be
    0<\Delta_{121}<2\:,\qquad\Delta_{111}^2<2\textstyle\sum_I\Delta_I\:.
\ee
In regions of the parameter space that do not satisfy the inequalities above, a modification of the ansatz \eqref{Q111_solution_ansatz} would be required;
for our purposes it will be sufficient to restrict ourselves to the above region.
Let us observe that the function $\delta y_{34}(x)$ as given by \eqref{y34_sol} is continuous in the entirety of its domain.
Any other difference of eigenvalues $\delta y_{ab}(x)\,\equiv\,y_a(x)-y_b(x)$ can be determined from $\delta y_{34}(x)$
and the ansatz \eqref{Q111_solution_ansatz}, and generically these will have a discontinuity at $x=0$,
see Figure \ref{fig:analytic_plot}. Also the above analytic solution to the saddle point equation matches perfectly
with the operator counting prediction of section \ref{sec:operator_counting}, if $\Delta_I$ in the latter are restricted to the values under consideration in this section.

Many expressions that we have written so far contain the Lagrange multiplier $\mu$, whose value is fixed by imposing that
\begin{equation}
	\int\dd x\,\rho(x)=1\:,\qquad\mu>0\:.
\end{equation}
After fixing $\mu$ as above, the saddle point solution described by \eqref{rho_sol}, \eqref{y34_sol} and \eqref{x_bounds}
is fully determined, and we can finally compute the value of $\cF$ evaluated at the solution ($\cF_{\text{crit.}}$)
by performing the integrals in \eqref{Q111_functional} using the explicit expressions for $\rho$ and $\delta y_{ab}$.
The results that we find for $\mu$ and $\cF_{\text{crit.}}$ are
\begin{align}
\label{free_energy_with_Delta_constraint}
\nn
	\frac{\cF_{\text{crit.}}(\Delta)}{N^{3/2}}=\frac{\mu}{3\pi}=\frac{4\pi}3\,\sqrt{2{\rm k}\,\Delta_{121}}\,\Big[{}&
		\Delta_{111}^{-2}\,\big(\Delta_{112}^{-1}+\Delta_{211}^{-1}\big)
		+\Delta_{112}^{-1}\,\Delta_{121}^{-1}\,\big(\Delta_{111}^{-1}+\Delta_{122}^{-1}\big)\,+\\
	{}&+\Delta_{211}^{-1}\,\Delta_{121}^{-1}\,\big(\Delta_{111}^{-1}+\Delta_{221}^{-1}\big)\Big]^{-1/2}\:.
\end{align}
The proportionality between $\cF_{\text{crit.}}$ and $\mu$ is no coincidence, rather it is the consequence of a fairly general
argument outlined in appendix B of \cite{Gulotta:2011si}.%
\footnote{We remark that we computed $\mu$ and $\cF_{\text{crit.}}$ independently, and then checked that they are indeed
proportional, rather than just computing $\mu$ and then deriving $\cF_{\text{crit.}}$ by imposing the proportionality.}
The free energy of the theory on $S^3$ is then
\begin{equation}
	F_{S^3}(\Delta)=\cF_{\text{crit.}}(\Delta)+o(N^{3/2})\:,
\end{equation}
whereas the large\,-$N$ and small\,-$\epsilon$ limits of the superconformal and topologically twisted indices on $S^2\times S^1$ are given by
\begin{equation}
	\log Z_{S^2\times S^1,\,\sigma}(\Delta^\pm,\epsilon)\sim\frac{1}{4\,\epsilon}\Big(\cF_{\text{crit.}}(\Delta^-)
		-\sigma\,\cF_{\text{crit.}}(\Delta^+)\Big)\:.
\end{equation}

\subsection{Varying the position of the discontinuity is impossible}
\label{sec:varying the discontinuity}

It is possible to generalize the ansatz \eqref{Q111_solution_ansatz} by relaxing the condition that the point of discontinuity
of the solution between the second and third tail regions is at $x=0$. If we require instead that the discontinuity point
is at $x=x_0$, with $\xL<x_0<\xR$, then we could in principle find a different solution to the saddle point equations.
Indeed, if the condition \eqref{Delta_restriction} on the $\Delta_I$ is still assumed,
the long range forces would cancel in the same manner.
Then the solution would still be described by the expression \eqref{rho_sol} and \eqref{y34_sol},
with the only differences being that the conditions $\xL<x<0$, $\:0<x<\xR$ should be replaced with
$\xL<x<x_0$, $\:x_0<x<\xR$ and that the value of the Lagrange multiplier is a function of the new parameter, $\mu(x_0)$.
The value of $\mu(x_0)$ is fixed by $\mu(x_0)>0$ and
\be
\label{mu_x_0}
    1=\int\dd x\:\rho(x)=\frac{\mu(x_0)^2}{\mu(x_0=0)^2}+\frac{{\rm k}\,x_0^2}{4\pi^2\Delta_{111}\Delta_{121}}\:,
\ee
where $\mu(x_0=0)$ is given by \eqref{free_energy_with_Delta_constraint}.
Evaluating the functional $\cF$ on this putative solution leads to the following critical value as a function of $x_0$:%
\footnote{Notice that using \eqref{mu_x_0} this expression for $\cF_{\text{crit.}}(\Delta,x_0)$ can also be written as
\be
\label{F_x_0_v2}
    \frac{\cF_{\text{crit.}}(\Delta,x_0)}{N^{3/2}}=\frac{\mu(x_0)}{3\pi}
        +\frac{{\rm k}\,x_0^2}{12\pi^3\Delta_{111}\Delta_{121}}(2\pi{\rm k}\,
        \big(\Delta_{C_1}-\Delta_{C_2})\,x_0-\mu(x_0)\big)\:,
\ee
meaning that there is no proportionality between $\cF_{\text{crit.}}$ and $\mu$ for general values of $x_0$.
This is possibly related to the fact that for general $x_0$ the functions $\rho(x)$ or $\delta y_{34}(x)$ are
no longer continuous.
}
\be
\label{F_x_0}
    \frac{\cF_{\text{crit.}}(\Delta,x_0)}{N^{3/2}}=\frac{1}{3\pi}\,\frac{\mu(x_0)^3}{\mu(x_0=0)^2}
            +\frac{{\rm k}^2(\Delta_{C_1}-\Delta_{C_2})\,x_0^3}{6\pi^2\Delta_{111}\Delta_{121}}\:.
\ee
However, since $x_0$ is a continuous parameter, small variations in $x_0$ would correspond to small variations of the saddle
point configuration which by definition should extremize $\cF$. This suggests that we should only consider values of
$x_0$ that extremize $\cF_{\text{crit.}}(\Delta,x_0)$, or in other words we should impose
\be
\label{x_0_extremization}
    0=\pi N^{-3/2}\:\partial_{x_0}\cF_{\text{crit.}}(\Delta,x_0)=
            \frac{\mu(x_0)^2\,\partial_{x_0}\mu(x_0)}{\mu(x_0=0)^2}
            +\frac{{\rm k}^2(\Delta_{C_1}-\Delta_{C_2})\,x_0^2}{2\pi\,\Delta_{111}\Delta_{121}}\:.
\ee
From \eqref{mu_x_0} we can derive 
\be
    \mu(x_0)\,\partial_{x_0}\mu(x_0)=\frac12\,\partial_{x_0}\big(\mu(x_0)^2\big)=
        -\frac{{\rm k}\:x_0\:\mu(x_0=0)^2}{4\pi^2\Delta_{111}\Delta_{121}}\:,
\ee
which can be used to rewrite \eqref{x_0_extremization} as
\be
    0=\frac{2\pi(\Delta_{C_1}-\Delta_{C_2})({\rm k}\,x_0)^2-\mu(x_0)({\rm k}\,x_0)}{4\pi^2\Delta_{111}\Delta_{121}}\:.
\ee
This equation only has two solutions, one of them being
\be
    x_0=0\:,
\ee
which corresponds to the saddle point solution that we have already discussed,
confirming that it is indeed an extremum of $\cF$
even under variations of the discontinuity point.
The other solution corresponds to the value of $x_0$ that satisfies the equation%
\footnote{We observe that both solutions to $\partial_{x_0}\,\cF_{\text{crit.}}(\Delta,x_0)=0$
when plugged into equation \eqref{F_x_0_v2} restore the proportionality between $\cF_{\text{crit.}}$ and $\mu$.}
\be
    x_0=\frac{\mu(x_0)}{2\pi\,{\rm k}\,(\Delta_{C_1}-\Delta_{C_2})}\:,
\ee
which is unique and can be easily found by substituting $\mu(x_0)=2\pi{\rm k}\,(\Delta_{C_1}-\Delta_{C_2})\,x_0$
in equation \eqref{mu_x_0}.
However this solution must be discarded, since it cannot satisfy both $\rho(x)\geq0$ and the inequalities
$\xmin<\xL<x_0<\xR<\xmax$. Indeed, the latter inequalities now read
\be
	-\frac{1}{{\rm k}\,\Delta_{122}}<-\frac{1}{{\rm k}\,\Delta_{112}}<\frac1{{\rm k}\,(\Delta_{C_1}-\Delta_{C_2})}<\frac1{{\rm k}\,\Delta_{211}}<\frac1{{\rm k}\,\Delta_{221}}\:,
\ee
and it is easy to check that if $\Delta_{C_1}>\Delta_{C_2}$ the above inequalities imply $\rho(\xmin+\varepsilon)<0$
for a small $\varepsilon>0$ regardless of the sign of ${\rm k}$,
whereas if $\Delta_{C_1}<\Delta_{C_2}$ we would find $\rho(\xmax-\varepsilon)<0$.
Therefore $x_0=0$ is the only possible solution.

\subsection{Subleading corrections to the saddle in the tail regions}
\label{sec:subleading corrections}

In this section we show precisely how the cancellation of long range forces in the saddle point equation
is tied to subleading corrections to the leading order ansatz \eqref{Q111_solution_ansatz}, and provide an
explicit expression for such corrections.

Let us focus on the $S^3$ free energy for the moment, and discuss the $S^2\times S^1$ indices later.
The finite\,-$N$ saddle point equations are
\be
\label{saddle_point_eq_finite_N}
\begin{aligned}
    0=\partial_{\lambda_{a,\,i}}\,\cF(\lambda,\Delta)={}&-\frac{\ii{\rm k}_a}{2\pi}\,\lambda_{a,\,i}+\Delta_{m;\,a}
        +\frac\ii2\sum_{I=1}^{|\cR|}\sum_{j,k=0}^{N-1}(\delta_a^{b_I}\delta_i^j-\delta_a^{c_I}\delta_i^k)\,z_{jk}^I\,\cot(\pi z_{jk}^I)\,-\\
        {}&-\ii\sum_{j\ne i}\cot\left(\ii\:\tfrac{\lambda_{a,\,i}\,-\,\lambda_{a,\,j}}2\right)\:,
\end{aligned}
\ee
where we have denoted with $b_I$ and $c_I$ the start and the end nodes of the $I$-th arrow in the quiver
and we also defined $z^I_{jk}$ as
\be
    z_{jk}^I=1-\Delta_I+\frac\ii{2\pi}(\lambda_{b_I,\,j}-\lambda_{c_I,\,k})\:.
\ee
The large\,-$N$ limit of equation \eqref{saddle_point_eq_finite_N} can be taken using the same method we reviewed in
section \ref{sec:Free energy at large N}, after noticing that $\cot(\pi z)=\ii+2\ii\Li_0(e^{-2\pi\ii z})$
and thus equations \eqref{Li_asymptotic}, \eqref{Li_asymptotic2} and \eqref{polylog_inversion}
can be applied in the same manner. In a region where the bounds \eqref{free_energy_determination} are all
satisfied with a $\sim N^0$ margin we would obtain
\be
\label{naive_large_N_saddle_point_eq}
    0\,=\,\frac1{\ii N\rho(x_i)}\,\frac{\delta\cF(\Delta,\rho,y,\nu)}{\delta y_a(x_i)}+\cO(N^0)\:,
\ee
which is equivalent to the statement that $\cF(\Delta,\rho,y,\nu)$ at the saddle point is extremized with
respect to small variation in all the $y_a(x)$.%
\footnote{Deriving a similar statement for small variations in $\rho(x)$ would require a more careful analysis of the 
$\cO(N^0)$ terms in \eqref{naive_large_N_saddle_point_eq}.}
However, for the saddle point solution \eqref{rho_sol}, \eqref{y34_sol} that we have derived
for the chiral quiver dual to $Q^{1,1,1}/\bZ_{\rm k}$,
at least one of the bounds \eqref{free_energy_determination} is always saturated.
In such cases, equation \eqref{naive_large_N_saddle_point_eq} must be corrected. In a region where
\be
\label{beta_correction}
    2\pi\Delta_J\,+\,y_{b_J}(x_i)-y_{c_J}(x_i)\:=\:\cO(N^{-\beta})
\ee
for some $J$ and $\beta>0$, the $i=j=k$ terms in \eqref{saddle_point_eq_finite_N} would give a $\cO(N^\beta)$
contribution that corrects \eqref{naive_large_N_saddle_point_eq} when $a=b_J$ or $a=c_J$,
but cancels in the sum of these two corrected equations.%
\footnote{This is more evident by looking at equation \eqref{corrected_large_N_saddle_point_eq}.
    The corrections brought by the very small $z_{ii}^I$ vanish when summing over the $a=b_I$ and the $a=c_I$ equations
    due to the $(\delta_a^{b_I}-\delta_a^{c_I})$ term.}
In the previous sections, we determined the solution
to the saddle point by implicitly solving only for the sum of the $a=b_J$ and $a=c_J$
saddle point equations when \eqref{beta_correction} is satisfied, which allowed us to ignore the computation of the
right hand side of \eqref{beta_correction} and the resulting corrections to \eqref{naive_large_N_saddle_point_eq}.%
\footnote{Indeed, substituting the tail region ansatz \eqref{Q111_solution_ansatz} in the functional $\cF(\Delta,\rho,y,\nu)$
    and then extremizing with respect to the subset of $y_a(x)$ functions that is not constrained by \eqref{Q111_solution_ansatz}
    is equivalent to only imposing the linear combinations of the saddle point equations \eqref{saddle_point_eq_finite_N}
    for which the corrections vanish.}
This is a standard procedure even for non-chiral quivers \cite{Herzog:2010hf}, where the
$N^{-1}\delta\cF/\delta y_a(x_i)$ term in \eqref{naive_large_N_saddle_point_eq} is at most of order
$\cO(N^{1/2})$ and therefore the exponent $\beta$ in \eqref{beta_correction} should be $\beta=1/2$.
For the chiral quivers of interest in this paper, the $N^{-1}\delta\cF/\delta y_a(x_i)$ term can
be of order $\cO(N)$ due to the presence of long range forces, which in turn requires $\beta=1$.
We will now compute the $\cO(N^{-1})$ corrections in the right hand side of \eqref{beta_correction}
that are required for the cancellation of long range forces in the chiral quiver dual to $Q^{1,1,1}/\bZ_{\rm k}$.
For this purpose, we will ignore $\cO(N^{1/2})$ terms and only focus on the $\cO(N)$ long range forces.

Since $N^{1/2}(x_{i+1}-x_i)=N^{-1/2}\rho(x_i)^{-1}$ by definition of $\rho(x_i)$, only $i=j=k$ terms in
\eqref{saddle_point_eq_finite_N} are expected to give a $\cO(N)$ correction.
Before taking the large\,-$N$ limit of \eqref{saddle_point_eq_finite_N}, we isolate the $i=j=k$ term
in the sum in the first line of \eqref{saddle_point_eq_finite_N}, and we leave it as is.
The saddle point equation \eqref{saddle_point_eq_finite_N} is thus rewritten as follows:
\be
\label{corrected_large_N_saddle_point_eq}
    0\,=\,\frac1{\ii N\rho(x_i)}\,\frac{\delta\cF(\Delta,\rho,y,\nu)}{\delta y_a(x_i)}+\frac\ii2\sum_{I=1}^{|\cR|}
        (\delta_a^{b_I}-\delta_a^{c_I})\,z_{ii}^I\,\cot(\pi z_{ii}^I)+\cO(N^{1/2})\:.
\ee
In order to cancel the $\cO(N)$ long range forces coming from the first piece, we take the following ansatz for
the subleading corrections to the tail region equations:%
\footnote{We are keeping $\lambda_{a,\,i}=N^{1/2}\,x_i+\ii\,y_a(x_i)$ and allowing $y_a(x)$ to have corrections beyond its leading $\cO(N^0)$ behavior. In particular, $\,y_{b_J}(x_i)-y_{c_J}(x_i)=\big(\,y_{b_J}(x_i)-y_{c_J}(x_i)\big)_{\text{leading order}}+\big(\ii N\,Y_J(x_i)\big)^{-1}+\cO(N^{-3/2})$, where the leading order was determined in section \ref{sec: Saddle point of Q111} for the quiver dual to $Q^{1,1,1}/\bZ_{\rm k}$. Since this correction term is purely imaginary, $y_a(x_i)$ is no longer the imaginary part of $\lambda_{a,\,i}$, and we have that $Y_J(x_i)=\big(N\,\text{Re}(\lambda_{b_J,\,i}-\lambda_{c_J,\,i})\big)^{-1}$.}
\be
\label{Y_def}
    2\pi\Delta_J\,+\,y_{b_J}(x_i)-y_{c_J}(x_i)\,=\big(\ii N\,Y_J(x_i)\big)^{-1}+\cO(N^{-3/2})\:,\qquad\forall\:J\in T_{x_i}\:.
\ee
Here we denote with $T_x$ the set of chiral multiplet indices for which $ 2\pi\Delta_J\,+\,y_{b_J}(x)-y_{c_J}(x)$
is of order $\cO(N^{-1})$.
Then the saddle point equation \eqref{corrected_large_N_saddle_point_eq} is rewritten as
\be
    0\,=\,\frac1{\ii N\rho(x_i)}\,\frac{\delta\cF(\Delta,\rho,y,\nu)}{\delta y_a(x_i)}
        +N\sum_{J\,\in\,T_{x_i}}(\delta_a^{b_J}-\delta_a^{c_J})\:Y_J(x_i)+\cO(N^{1/2})\:,
\ee
and by explicitly computing the functional derivatives in the expression \eqref{free_energy_functional} for
$\cF(\Delta,\rho,y,\nu)$ we get
\be
\label{Y_equation}
    \sum_{J\,\in\,T_x}(\delta_a^{b_J}-\delta_a^{c_J})\:Y_J(x)=\frac12\int\dd\wt x\:\rho(\,\wt x\,)\:\text{sign}(x-\wt x)
        \Bigg[\cD_a-\frac1{2\pi}\sum_{b=1}^{|\cG|}n_{a,b}\:y_b(\,\wt x\,)\Bigg]\:.
\ee
For the chiral quiver dual to $Q^{1,1,1}/\bZ_{\rm k}$, we can evaluate the integrals on the right hand side of the
above equation on the saddle point solution \eqref{rho_sol}, \eqref{y34_sol}, which gives us the subleading corrections
to the tail region equations \eqref{Q111_solution_ansatz} required for the cancellation of long range forces:
\begin{equation}
\label{Q111_tail_corrections}
\begin{cases}
	Y_{A_1}(x)=-Y_{C_2}(x)=\displaystyle\frac{x\,(\pi\,{\rm k}\,\Delta_{122}\,x+\mu)}{(2\pi)^3\,(\Delta_{B_1}-\Delta_{B_2})\,\Delta_{121}}+c_L\:,&\makebox[15mm][r]{$\xmin<\:$}x<\xL\:,\\[7mm]
	Y_{A_1}(x)=-Y_{C_2}(x)=\displaystyle\frac{x\,(\pi\,{\rm k}\,(\,\Delta_{A_1}-\Delta_{A_2})\,x+\mu)}{(2\pi)^3\,\Delta_{111}\,\Delta_{121}}+c_0\:,&\makebox[15mm][r]{$\xL<\:$}x<0\:,\\[7mm]
	Y_{A_2}(x)=-Y_{C_1}(x)=\displaystyle\frac{x\,(\pi\,{\rm k}\,(\,\Delta_{A_1}-\Delta_{A_2})\,x+\mu)}{(2\pi)^3\,\Delta_{111}\,\Delta_{121}}+c_0\:,&\makebox[15mm][r]{$0<\:$}x<\xR\:,\\[7mm]
	Y_{A_2}(x)=-Y_{C_1}(x)=\displaystyle\frac{x\,(-\pi\,{\rm k}\,\Delta_{221}\,x+\mu)}{(2\pi)^3\,(\Delta_{B_1}-\Delta_{B_2})\,\Delta_{121}}+c_R\:,&\makebox[15mm][r]{$\xR<\:$}x<\xmax\:,
\end{cases}
\end{equation}
where we have defined the constants
\begin{align}
    {}&c_0=\frac{\mu^2(\Delta_{C_1}-\Delta_{C_2})(\Delta_{121}\sum_I\Delta_I+\Delta_{112}\,\Delta_{211})}
        {2\,(2\pi)^4\,{\rm k}\,\Delta_{111}\,\Delta_{112}\,\Delta_{121}\,\Delta_{122}\,\Delta_{211}\,\Delta_{221}}\:,\\[2mm]\nn
    {}&c_L=c_0+\frac{\mu^2}{2\,(2\pi)^4\,{\rm k}\,(\Delta_{B_1}-\Delta_{B_2})\,\Delta_{111}\,\Delta_{112}}\:,\quad
        c_R=c_0-\frac{\mu^2}{2\,(2\pi)^4\,\,{\rm k}\,(\Delta_{B_1}-\Delta_{B_2})\,\Delta_{111}\,\Delta_{211}}\:.
\end{align}
Notice the absence of $Y_{B_2}(x)$ for $\xmin<x<\xL$ and $\xR<x<\xmax$.
If we assumed that $B_2\in T_x$, or in other words that $2\pi\Delta_{B_2}+y_3(x)-y_4(x)=\cO(N^{-1})$,
from \eqref{Y_equation} we would find $Y_{B_2}(x)=0$, from which we would conclude $B_2\notin T_x$,
since it would not be possible to have $2\pi\Delta_{B_2}+y_3(x)-y_4(x)=\cO(N^{-\beta})$ with $\beta>1$.
Therefore $\beta=1/2$ and these corrections contribute to the $\cO(N^{1/2})$
terms in \eqref{corrected_large_N_saddle_point_eq}, but play no role in the cancellation of long range forces.

\begin{figure}[h]
     \centering
     \begin{subfigure}[b]{0.49\textwidth}
         \centering
         \includegraphics[width=\linewidth]{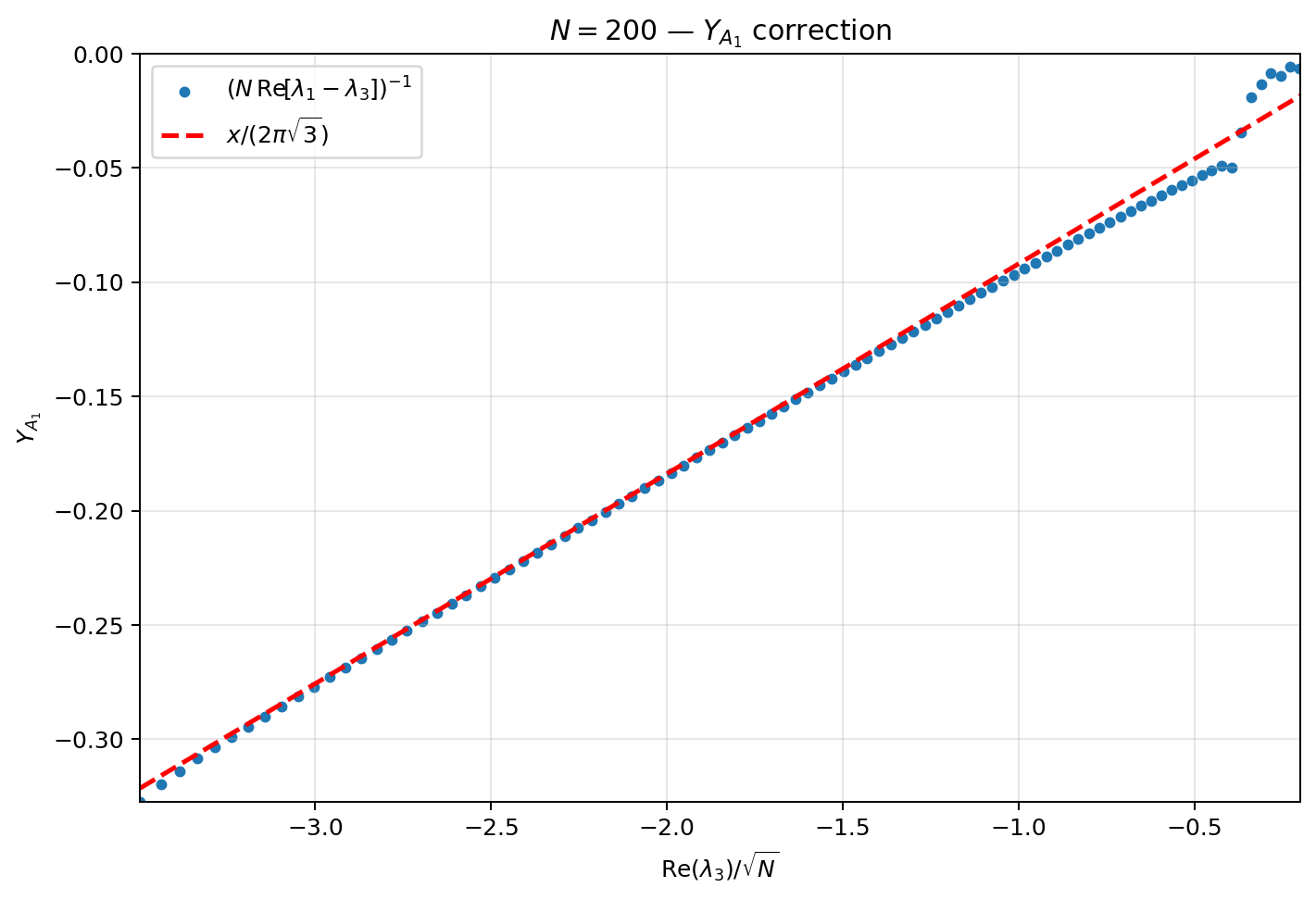}
         \caption{$\Delta_I=1/3\:$ solution.}
         \label{fig:YA1_canonical}
     \end{subfigure}
     \hfill
     \begin{subfigure}[b]{0.49\textwidth}
         \centering
         \includegraphics[width=\linewidth]{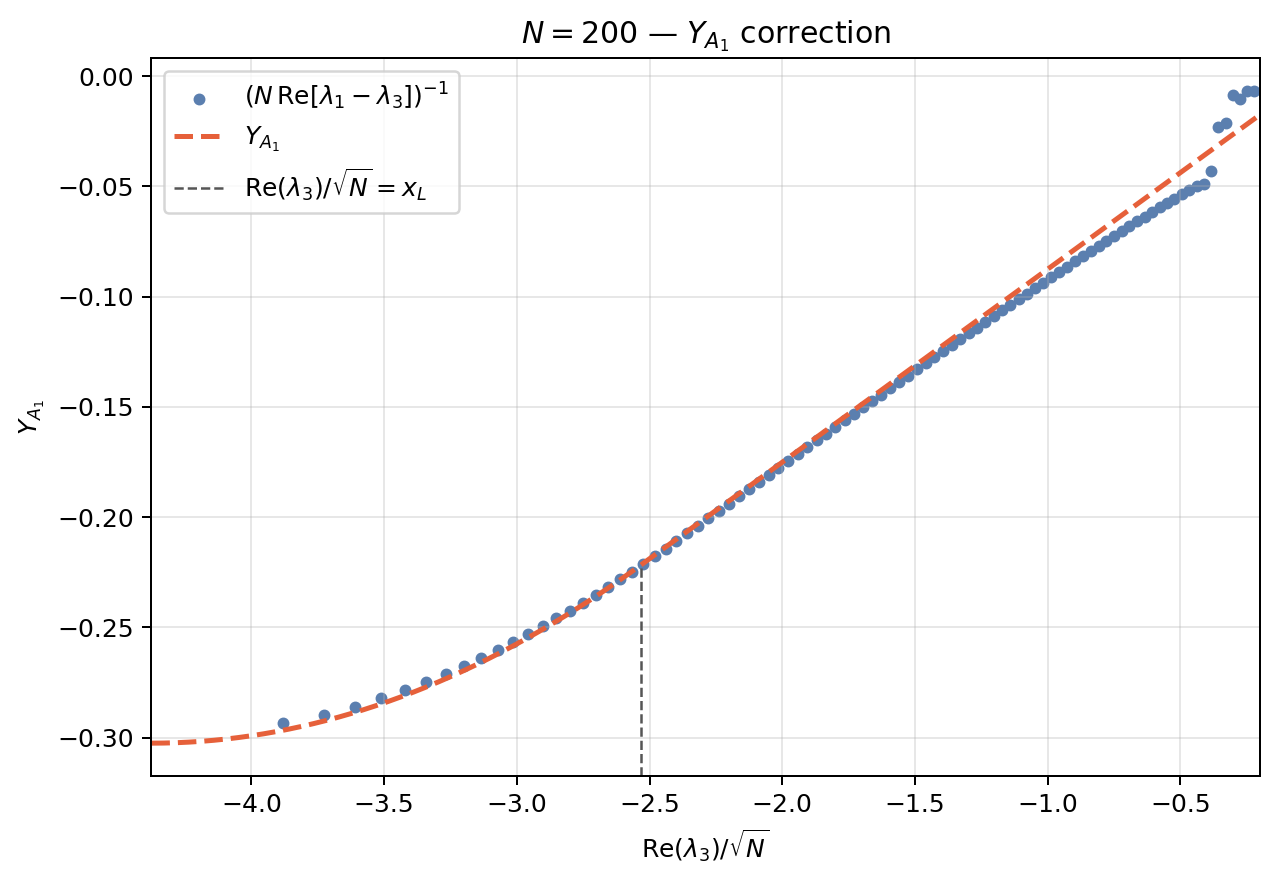}
         \caption{$\Delta_{A_i}=\Delta_{B_1}=0.6,\,\Delta_{C_i}=\Delta_{B_2}\:$ solution.}
         \label{fig:YA1_non_canonical}
     \end{subfigure}
     
     \caption{Comparison between $\left(N\,\text{Re}[\lambda_{1,i\,}-\lambda_{3,\,i}]\right)^{-1}$ from the $N=200$
            numeric solutions and the analytic prediction for $Y_{A_1}(x_i)$, with $x_i\,\equiv\,\text{Re}(\lambda_{3,\,i})/\sqrt{N}$.}
     \label{fig:YA1}
\end{figure}

Let us restrict \eqref{Q111_tail_corrections} to ${\rm k}=1$, $\Delta_{A_i}=\Delta_{B_1}=\Delta_A$, \:$\Delta_{C_i}=\Delta_{B_2}=2/3-\Delta_A$ and compare the value of the $Y_J(x)$ that we
have computed analytically with the numerical solutions at $N=200$. First, \eqref{Q111_tail_corrections} becomes
\begin{equation}
\begin{aligned}
{}&Y_{A_1}(x)=-Y_{C_2}(x)=
\begin{cases}
	\displaystyle\frac{2\pi^2\,x^2+\frac{2\pi\,\mu}{\frac43-\Delta_A}\,x+\frac{\mu^2}{2(\Delta_A+\frac23)^2}}{(2\pi)^4(2\Delta_A-\frac23)}\:,&\makebox[15mm][r]{$\xmin<\:$}x<\xL\:,\\[7mm]
	\displaystyle\frac{\mu\,x}{(2\pi)^3\,(\Delta_A+\frac23)\,(\frac43-\Delta_A)}\:,&\makebox[15mm][r]{$\xL<\:$}x<0\:,
\end{cases}\\[4mm]
{}&Y_{A_2}(x)=-Y_{C_1}(x)=-Y_{A_1}(-x)\:,\hspace{53mm} x>0\:.
\end{aligned}
\end{equation}
where $\xL=-\mu/(2\pi(\Delta_A+\frac23))$ and $\xmin=\mu/(2\pi(\Delta_A-\frac43))$ and $\mu$ is given by \eqref{mu_Delta_A}.
If we further restrict $\Delta_A=1/3$ and $\mu=(2\pi)^2/\sqrt3$, we obtain the subleading corrections for the canonical R-charge assignment $\Delta_I=1/3$:
\begin{equation}
\begin{cases}
	Y_{A_1}(x)=-Y_{C_2}(x)=\displaystyle\frac{x}{2\pi\sqrt3}\:,&\makebox[15mm][r]{$-\frac{2\pi}{\sqrt3}<\:$}x<0\:,\\[7mm]
	Y_{A_2}(x)=-Y_{C_1}(x)=\displaystyle\frac{x}{2\pi\sqrt3}\:,&\makebox[15mm][r]{$0<\:$}x<\frac{2\pi}{\sqrt3}\:.
\end{cases}
\end{equation}
Given the definition \eqref{Y_def} of $Y_J(x)$, we can compare $Y_{A_1}(x_i)$ with
$\left(N\,\text{Re}[\lambda_{1,i\,}-\lambda_{3,\,i}]\right)^{-1}$ as computed from the $N=200$ numerical solutions for $x_i<0$.
The comparison is shown in Figure \ref{fig:YA1}, where the blue dots correspond to the numerical solution and the red dashed line indicates the analytic prediction, which for $\Delta_{A}\ne1/3$ transitions from a parabola to a line at $x=\xL$, marked on the plot with a vertical line. The match between the analytic prediction and the numerical results is good.
By virtue of symmetry, there is no need to repeat the comparison for $Y_{C_2}$, $Y_{A_2}$, $Y_{C_1}$. 

For the superconformal and topologically twisted indices on $S^2\times S^1$, the analogue of formula
\eqref{corrected_large_N_saddle_point_eq} is
\be
\label{indices_large_N_saddle_point_eq}
    0\,=\,\frac{\pm\pi\sigma_\pm}{\epsilon N\rho^\pm(x^\pm_i)}\,\frac{\delta\cF(\Delta^\pm,\rho^\pm,y^\pm,\nu^\pm)}
    {\delta y^\pm_a(x^\pm_i)}+\sum_{I=1}^{|\cR|}\partial_{u^\pm_{a;\,i}}\log\diag^\pm_I+\cO(N^{1/2})\:,
\ee
where $\diag^\pm_I$ is defined in \eqref{blocks}, it gives no leading order contributions to
$\cF(\Delta^\pm,\rho^\pm,y^\pm,\nu^\pm)$, but it can give $\cO(N)$ corrections to the above saddle point equation.
Indeed, if we take the ansatz
\be
\label{Y_pm_def}
    2\pi\Delta^\pm_J\,+\,y^\pm_{b_J}(x)-y^\pm_{c_J}(x)\,=\pm\,2\,\epsilon\,\big(\ii N\,Y^\pm_J(x)\big)^{-1}+\cO(N^{-3/2})\:,\qquad\forall\:J\in T^\pm_{x_i}\:,
\ee
we then find that
\be
    \partial_{u^\pm_{a;\,i}}\log\diag^\pm_J=\frac{\pi\ii N}\epsilon(\delta_a^{b_J}-\delta_a^{c_J})Y^\pm_J(x_i^\pm)+\cO(N^{1/2})\:,
\ee
and the saddle point equation \eqref{indices_large_N_saddle_point_eq} is solved by
\be
    \sum_{J\,\in\,T^\pm_x}(\delta_a^{b_J}-\delta_a^{c_J})\:Y^\pm_J(x)=\frac12\int\dd\wt x\:\rho^\pm(\,\wt x\,)\:\text{sign}(x-\wt x)
        \Bigg[\cD^\pm_a-\frac1{2\pi}\sum_{b=1}^{|\cG|}n_{a,b}\:y^\pm_b(\,\wt x\,)\Bigg]\:,
\ee
which is perfectly analogous to \eqref{Y_equation}. The $Y^\pm_J(x)$ corrections for the quiver dual to
$Q^{1,1,1}/\bZ_{\rm k}$ are then exactly the same as \eqref{Q111_tail_corrections} with the addition of $\pm$ superscripts.
We note that the ansatz \eqref{Y_pm_def} is only valid for $\epsilon\ne0$.
The topologically twisted index is also well defined for $\epsilon=0$, and in that case the corrections to
the tail equation are of the form $\cO(e^{-N^\beta})$ \cite{Hosseini:2016tor}.

\subsection{A large\,-$N$ saddle for general $\Delta$ values}

\label{sec: A proposal for general Delta}

In section \ref{sec: Saddle point of Q111} we have found an analytic saddle point solution to the chiral quiver dual to $Q^{1,1,1}/\bZ_{\rm k}$
under the assumption that the parameters satisfy the condition $\Delta_{A_1}+\Delta_{C_1}=\Delta_{A_2}+\Delta_{C_2}$.
In this section we explore the possibility of relaxing this condition, and propose a solution for general values of $\Delta_I$, under the assumption that the large\,-$N$ saddle is still subdivided into four
tail regions with a discontinuity at $x=0$ in the same manner as \eqref{Q111_solution_ansatz}.
In order to deal with the nonzero long range forces \eqref{Q111_long_range_forces} that remain after \eqref{Q111_solution_ansatz} is substituted
but no condition on $\Delta_I$ is imposed, we modify the ansatz \eqref{eigenvalue_Ansatz} for the distribution of the eigenvalues
of the $S^3$ free energy as follows:
\be
\label{eigenvalue_Ansatz_corrected}
    \lambda_{a,\,i}=\ii\,N\,w(x_i)+N^{1/2}\,x_i+\ii\,y_a(x_i)\:.
\ee
First, let us notice that since $w(x_i)$ has no dependence on the node $a$, so its addition does not change the long range forces coming from the 1-loop
determinants. The $\cO(N^{3/2})$ contribution of the 1-loop determinants on the other hand may be influenced, as can be seen for example by the fact that
the polylogarithm asymptotic formula \eqref{Li_asymptotic} is generalized to
\be
    \int_0^{\bar t>0}\dd t\,f(t)\Li_s\left(e^{\ii N \wt w(t)-N^{1/2} t\,+\,\ii y(t)}\right)=
		\frac{f(0)}{N^{1/2}-\ii\,N\,\wt w'(0)}\Li_{s+1}\left(e^{\ii y(0)}\right)+\cO(N^{-1})\:,
\ee
for $\wt w(t)$ such that $\wt w(0)=0$.%
\footnote{In order to apply this identity to \eqref{chiral_contribution_to_F} we would need to take $t=|x-\wt x|$ and $\wt w(t)$ to be
        $w(x-t)-w(x)$ or $w(x)-w(x+t)$ depending on the sign of $x-\wt x$. Either way $\wt w(0)=0$.}
From the above relation it would follow that for $w'(x)\ne0$ the 1-loop determinant would not bring any $N^{3/2}$ contribution;
to avoid this, we will take $w'(x)=0$, so that $w$ is a piecewise constant function.
Since we allow discontinuities at $x=0$ in our ansatz, we will assume that $w$ takes a constant value in $x>0$ and another constant value in $x<0$:
\be
    w(x<0)={w_\text L}\:,\qquad w(x>0)=w_{\text R}\:.
\ee
With these assumptions, the value of the 1-loop determinants at large\,-$N$ is unaltered by the addition of $w(x)$.
On the other hand the classical terms \eqref{F_classical_terms} would receive a correction,
and the large\,-$N$ functional \eqref{free_energy_functional} would change as follows:
\be
    \frac{\cF(\Delta,\rho,y,w,\nu)}{N^{3/2}}\,=\,\frac{\cF(\Delta,\rho,y,\nu)}{N^{3/2}}+\frac\nu\pi\int\dd x\,\rho(x)\cdot w(x)\cdot
        \sum_{a=1}^{|\cG|}\left(\frac{{\rm k}_a}{2\pi}\,y_a(x)+\Delta_{m;\,a}\right)\:,
\ee
where we recall that $\nu=\pi\ii\,N^{1/2}$. There are now new $N^2$ terms in $\cF$, which compete with the long range forces.
For the chiral quiver dual to $Q^{1,1,1}/\bZ_{\rm k}$, after substituting the ansatz \eqref{Q111_solution_ansatz} the total $N^2$ terms would be
\begin{align}
\label{total_N^2_terms}
	\frac{\cF-\cF\,\big|_{\nu\,=\,0}}{\nu N^{3/2}}{}&=\frac{\rm k}\pi\,w_{\text L}\int_{\xmin}^0\dd x\,\rho(x)\cdot\Big(\tfrac{\delta y_{34}(x)}{2\pi}-\Delta_{A_1}-\Delta_{C_2}\Big)-\\\nn
    {}&\:-\frac{\rm k}\pi\,w_{\text R}\int^{\xmax}_0\dd x\,\rho(x)\cdot\Big(\tfrac{\delta y_{34}(x)}{2\pi}-\Delta_{A_2}-\Delta_{C_1}\Big)+\big(\Delta_{A_1}\!+\!\Delta_{C_1}\!-\!\Delta_{A_2}\!-\!\Delta_{C_2}\big)\cdot\\\nn
	{}&\cdot\int_{\xmin}^0\!\dd x\int_0^{\xmax}\!\dd\wt x\:\rho(x)\rho(\,\wt x\,)\Big[\tfrac{\delta y_{34}(x)+\delta y_{34}(\,\wt x\,)}{2\pi}-
    \big(\Delta_{A_1}+\Delta_{C_1}+\Delta_{A_2}+\Delta_{C_2}\big)\Big],
\end{align}
where as usual we have taken $\Delta_{m}=0$ without loss of generality, see footnote \ref{Delta_m_reintroduction}.
Imposing that the above terms are extremized with respect to variations in $\delta y_{34}(x)$, we are led to an equation for the constants $w_{\text L}$
and $w_{\text R}$ that is solved by
\be
\begin{aligned}
    {}&w_{\text L}\,=\,\frac\pi{\rm k}\,\big(\Delta_{A_2}+\Delta_{C_2}-\Delta_{A_1}-\Delta_{C_1}\big)\int_0^{\xmax}\dd x\,\rho(x)\:,\\
    {}&w_{\text R}\,=\,\frac\pi{\rm k}\,\big(\Delta_{A_1}+\Delta_{C_1}-\Delta_{A_2}-\Delta_{C_2}\big)\int_{\xmin}^0\dd x\,\rho(x)\:.
\end{aligned}
\ee
With these values, the $N^2$ terms \eqref{total_N^2_terms} vanish. Extremization of the $N^{3/2}$ terms would then lead to
\begin{equation}
\label{general_analytic_rho}
	\rho(x)=\begin{cases}
		\displaystyle\frac{2\pi\,{\rm k}\,x\,\Delta_{122}+\mu}{(2\pi)^3(\Delta_{B_1}-\Delta_{B_2})\,\Delta_{121}\,\Delta_{222}}\:,
			&\makebox[15mm][r]{$\xmin<\:$}x<\xL\:,\\[7mm]
		\displaystyle\frac{2\pi\,{\rm k}\,x\left(\Delta_{122}\DeltaS-\Delta_{121}\,\Delta_{222}\right)+\mu\,\DeltaS}
			{(2\pi)^3\,\Delta_{111}\,\Delta_{121}\,\Delta_{212}\,\Delta_{222}}\:,&\makebox[15mm][r]{$\xL<\:$}x<0\:,\\[7mm]
		\displaystyle\frac{-2\pi\,{\rm k}\,x\left(\Delta_{221}\DeltaS-\Delta_{121}\Delta_{222}\right)+\mu\,\DeltaS}
			{(2\pi)^3\,\Delta_{111}\,\Delta_{121}\,\Delta_{212}\,\Delta_{222}}\:,&\makebox[15mm][r]{$0<\:$}x<\xR\:,\\[7mm]
		\displaystyle\frac{-2\pi\,{\rm k}\,x\,\Delta_{221}+\mu}{(2\pi)^3(\Delta_{B_1}-\Delta_{B_2})\,\Delta_{121}\,\Delta_{222}}\:,
			&\makebox[15mm][r]{$\xR<\:$}x<\xmax\:,\end{cases}
\end{equation}
and also
\begin{equation}
\label{general_analytic_y34}
	\frac{\delta y_{34}(x)}{2\pi}=\begin{cases}
		-\Delta_{B_2}\:,&\makebox[7mm][r]{$\xmin<\:$}x<\xL\:,\\[5mm]
		\Delta_{A_2}+\Delta_{C_2}+\displaystyle\frac{\left(2\pi\,{\rm k}\,x\,(\Delta_{C_1}-\Delta_{C_2})-\mu\right)\,\Delta_{212}\,\Delta_{222}}
			{2\pi\,{\rm k}\,x\left(\Delta_{122}\DeltaS-\Delta_{121}\,\Delta_{222}\right)+\mu\,\DeltaS}\:,&\makebox[7mm][r]{$\xL<\:$}x<0\:,\\[7mm]
		\Delta_{A_1}+\Delta_{C_1}+\displaystyle\frac{\left(2\pi\,{\rm k}\,x\,(\Delta_{C_1}-\Delta_{C_2})-\mu\right)\,\Delta_{111}\,\Delta_{121}}
			{-2\pi\,{\rm k}\,x\left(\Delta_{221}\DeltaS-\Delta_{121}\,\Delta_{222}\right)+\mu\,\DeltaS}\:,&\makebox[7mm][r]{$0<\:$}x<\xR\:,\\[7mm]
		-\Delta_{B_2}\:,&\makebox[7mm][r]{$\xR<\:$}x<\xmax\:,\end{cases}
\end{equation}
where $\xmin$, $\xL$, $\xR$ and $\xmax$ are given by the same expression as \eqref{x_bounds}.
All the other eigenvalue differences $y_a(x)-y_b(x)$ can be obtained from $\delta y_{34}(x)$ and \eqref{Q111_solution_ansatz}.
This solution would be valid in a region where
the $\Delta_I$ satisfy the inequalities \eqref{x_bounds_inequalities} together with
\be
    \Delta_{121},\:\Delta_{222}\in(0,2)\:,\qquad\Delta_{111},\:\Delta_{212}>0\:,\qquad\Delta_{111}\Delta_{212}<2\textstyle\sum_I\Delta_I\:,
\ee
which are obtained by imposing that both inequalities in \eqref{free_energy_determination} are satisfied.
The large\,-$N$ free energy would then be given by
\begin{align}
\label{free_energy_with_general_Delta}
\nonumber
	\frac{\cF_{\text{crit.}}({\rm k},\Delta)}{N^{3/2}}=\frac{\mu}{3\pi}=\frac{4\pi}3\,\sqrt{2{\rm k}\,\Delta_{121}}\,\Big[{}&
		\Delta_{111}^{-1}\,\Delta_{212}^{-1}\,\big(\Delta_{112}^{-1}+\Delta_{211}^{-1}\big)
		+\Delta_{112}^{-1}\,\Delta_{222}^{-1}\,\big(\Delta_{122}^{-1}+\Delta_{212}^{-1}\big)\,+\\
	{}&+\Delta_{211}^{-1}\,\Delta_{222}^{-1}\,\big(\Delta_{212}^{-1}+\Delta_{221}^{-1}\big)\Big]^{-1/2}\:.
\end{align}

Notice that the eigenvalue density $\rho$ and the eigenvalue differences $y_a(x)-y_b(x)$ of the analytic solution that we found in this section match perfectly with the operator counting prediction of section \ref{sec:operator_counting}.%
\footnote{Here, unlike in section \ref{sec:operator_counting}, we have set $\Delta_m=0$ for simplicity, but as already explained in footnote \ref{Delta_m_reintroduction} we can use the flat directions \eqref{flat_directions} to reintroduce the parameter $\Delta_m$ here by sending $\Delta_{A_1}\longrightarrow\Delta_{A_1}-\frac1{\rm k}\Delta_{m}\:$,
$\:\Delta_{C_1}\longrightarrow\Delta_{C_1}+\frac1{\rm k}\Delta_{m}\:$
and $\:y_1(x)\longrightarrow y_1(x)+\frac{2\pi}{\rm k}\Delta_{m}\:$. The match with the operator counting prediction is then perfect.}

\subsection{Verifying the general analytic saddle with a numerical analysis}
\label{sec:Delta_1_numeric_analysis}

\begin{figure}[h]
  \centering
  \includegraphics[width=\textwidth]{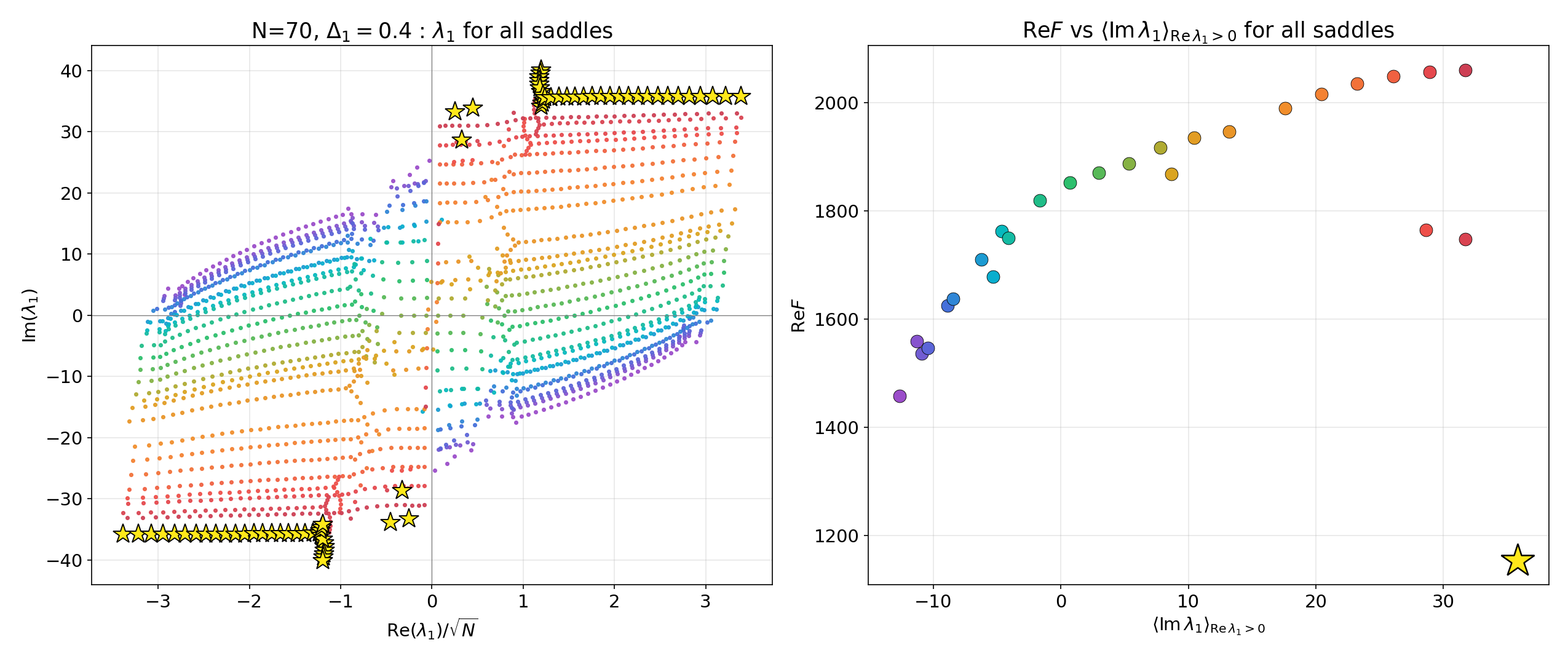}
  \caption{On the left: the eigenvalue distribution for the node 1 of 27 different $N=70$ numerical solutions for R-charges $\Delta_{A_1}=\Delta_{B_1}=\Delta_{C_1}=0.4,\:\Delta_{A_2}=\Delta_{B_2}=\Delta_{C_2}$. Each solution has a different color and the one with lowest $\text{Re}(\cF)$ is marked with stars.\\
  On the right: the value of $\text{Re}(\cF)$ for all 27 solutions is plotted on the y-axis, while the x-axis displays the average value of $\text{Im}(\lambda_1)$ among eigenvalues with $\text{Re}(\lambda_1)>0$. Each solution is marked with the same color as in the plot on the left, and the solution with lowest $\text{Re}(\cF)$ is marked with a star (bottom right corner).}
  \label{fig:N70_all_saddles}
\end{figure}

In this section we will try to verify numerically the analytic solution presented in the previous section, characterized by a large imaginary part $\ii\,N\,w(x)$ equal for all nodes in the eigenvalue ansatz \eqref{eigenvalue_Ansatz_corrected}.
For the numerical analysis we will fix ${\rm k}=1$ and $\Delta_m=0$,
and consider R-charge assignments of the form
\be
    \Delta_{A_i}=\Delta_{B_i}=\Delta_{C_i}\,\equiv\,\Delta_i\:,\qquad i=1,2\:,
\ee
with $\Delta_1+\Delta_2=2/3$.
When $\Delta_1=1/3$ these R-charges correspond to the canonical ones, for which a numerical solution is presented in section \ref{sec:numerical_solution}. Similarly to what we did in section \ref{sec:noncanonical}, we will use the solution for canonical R-charges as a starting point, and gradually increase the value of $\Delta_1$ in small steps, using the solution in the previous step to construct an initial guess for the numerical solver.
We stop at the value $\Delta_1=0.4$.
Rather than using the algorithm described in section \ref{sec:noncanonical}, we will make a modification in order to account for the fact that the expected saddle has large imaginary part $\ii\,N\,w(x)$, with the analytic solution having $w(x)=2\pi(\Delta_1-1/3)\,\text{sign}(x)$ for the R-charges that we are considering. At each step where the value of $\Delta_1$ is increased, we will therefore build the initial guess as $\lambda_{a,\,i}^{\text{new guess}}=\lambda_{a,\,i}^{\text{old solution}}+\ii\,N\,\delta\,\text{sign(Re}(\lambda_{a,\,i}))$
for some choice of $\delta$. After a few test runs of this algorithm, it becomes clear that adjusting the value of $\delta$ makes the algorithm converge to a different solution each time.
Therefore the matrix model possesses, at these values of R-charges, a large number of saddles, whose eigenvalue distributions often follow very similar profiles. All these saddles compete for dominance, and the one with lowest $\text{Re}(\cF)$ is reasonably expected to be the dominant one.%
\footnote{Picard–Lefschetz theory may exclude certain saddles, but such an analysis is beyond the scope of this paper.}
For concreteness, we set $N=70$, and search for as many numerical solutions to the saddle point equations as we can find at $\Delta_1=0.4$. We start from one such solution, then apply the $\delta$-deformation to the eigenvalues before running the Newton numerical solver to see if it converges to a different saddle. By scanning a wide range of values for $\delta$, we find a total of 27 numerical solutions. In Figure \ref{fig:N70_all_saddles} all these distinct solutions are compared. We cannot claim that these 27 solutions that we have found form an exhaustive list; we can nonetheless focus on the saddle with lowest $\text{Re}(\cF)$ among the ones that we have found, which we might reasonably expect to be the dominant one, and we can compare it to the analytic solution that we described in the previous section.

 \begin{figure}[h]
  \centering
  \includegraphics[width=0.7\textwidth]{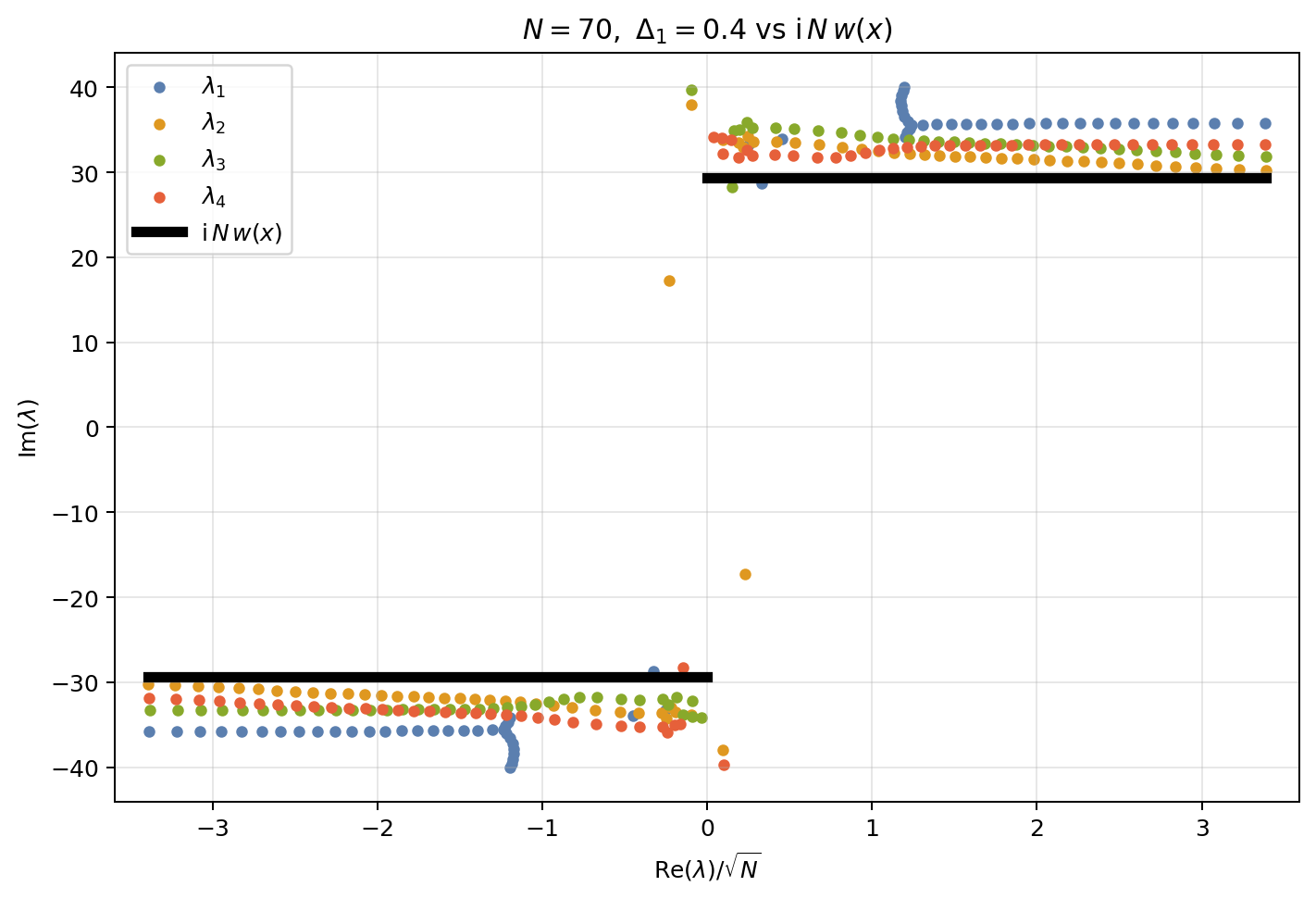}
  \caption{The eigenvalue distributions of all four nodes at $N=70$ for the dominant solution with $\Delta_{A_1}=\Delta_{B_1}=\Delta_{C_1}=0.4,\:\Delta_{A_2}=\Delta_{B_2}=\Delta_{C_2}$, compared with the analytic prediction for the large imaginary part $\ii\,N\,w(x)$.}
  \label{fig:N=70_Delta_1_eigenvalues}
\end{figure}

\begin{figure}[htbp]
     \centering
     \begin{subfigure}[b]{0.48\textwidth}
         \centering
         \includegraphics[width=\linewidth]{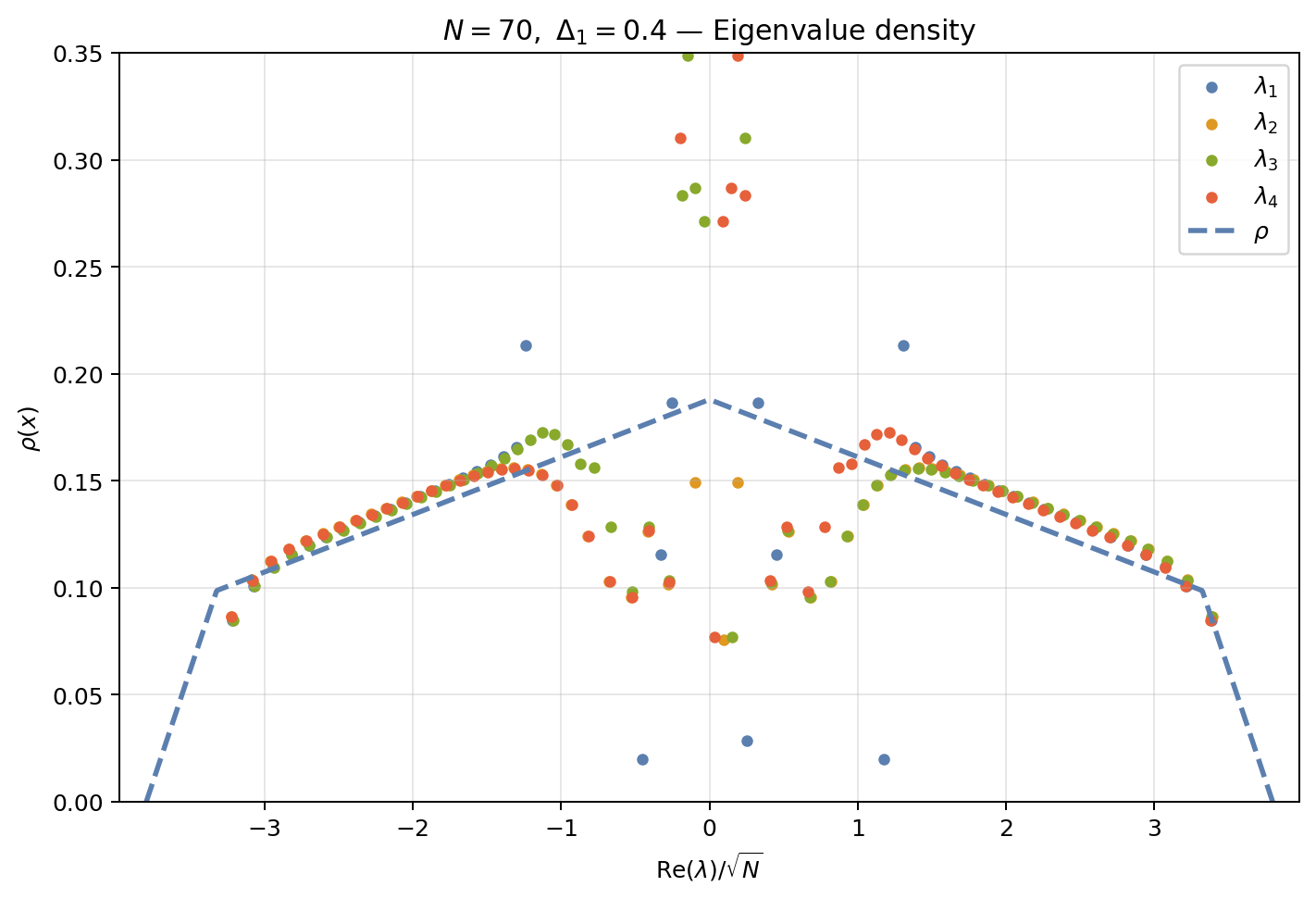}
         \caption{Eigenvalue density $\rho(x)$}
         \label{fig:rho_Delta_1}
     \end{subfigure}
     \hfill
     \begin{subfigure}[b]{0.48\textwidth}
         \centering
         \includegraphics[width=\linewidth]{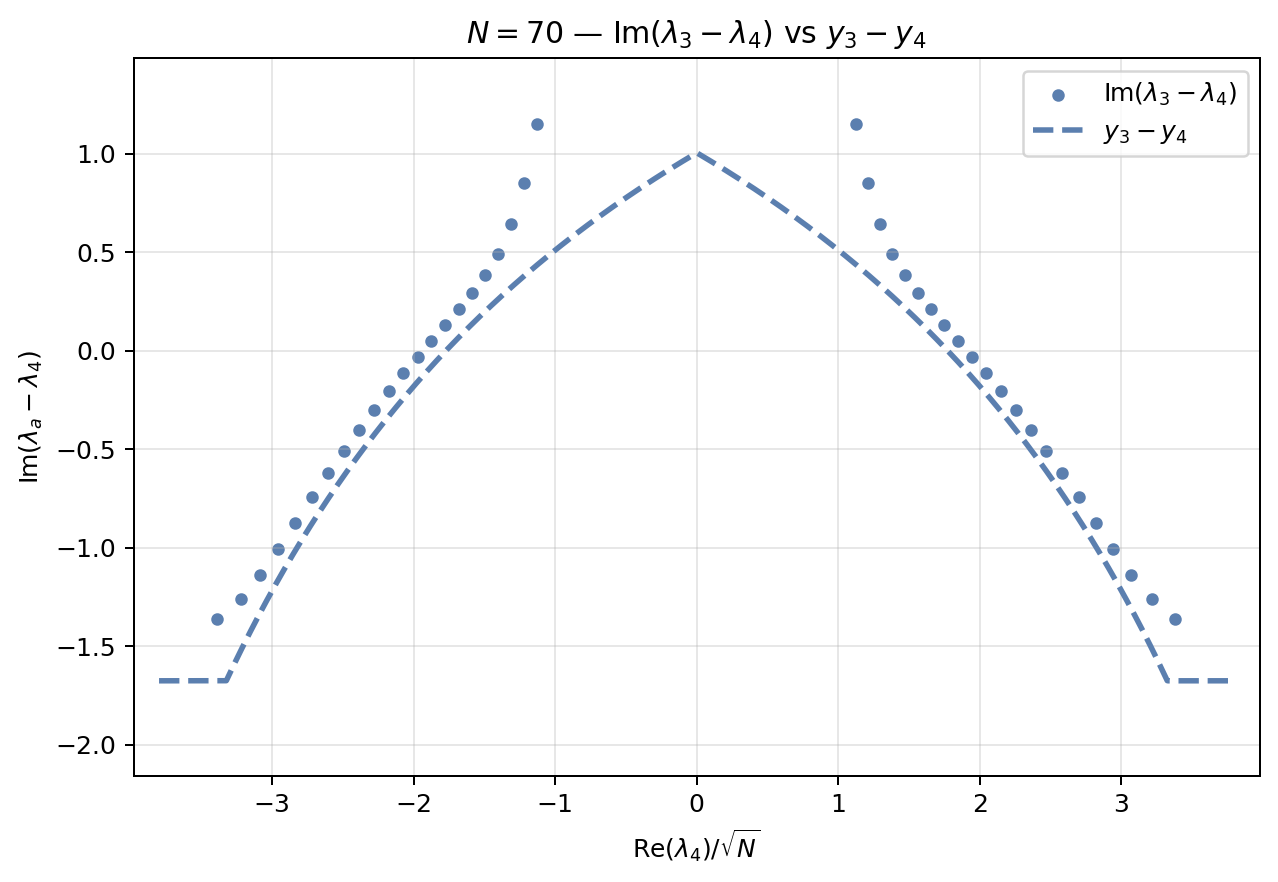}
         \caption{Eigenvalue difference $y_3(x)-y_4(x)$}
         \label{fig:y34_Delta_1}
     \end{subfigure}
     
     \vspace{3mm}
     \begin{subfigure}[b]{0.48\textwidth}
         \centering
         \includegraphics[width=\linewidth]{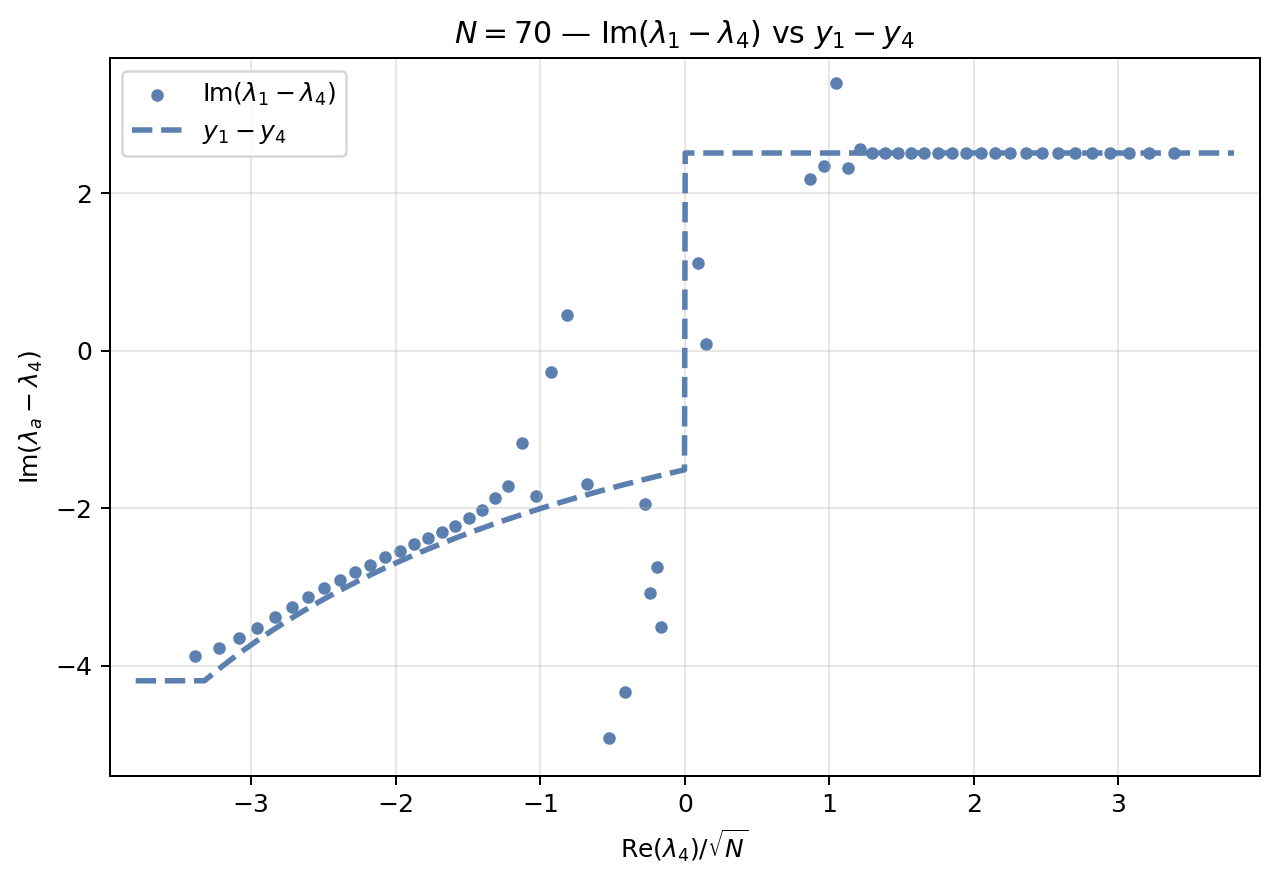}
         \caption{Eigenvalue difference $y_1(x)-y_4(x)$}
         \label{fig:y14_Delta_1}
     \end{subfigure}
     \hfill
     \begin{subfigure}[b]{0.48\textwidth}
         \centering
         \includegraphics[width=\linewidth]{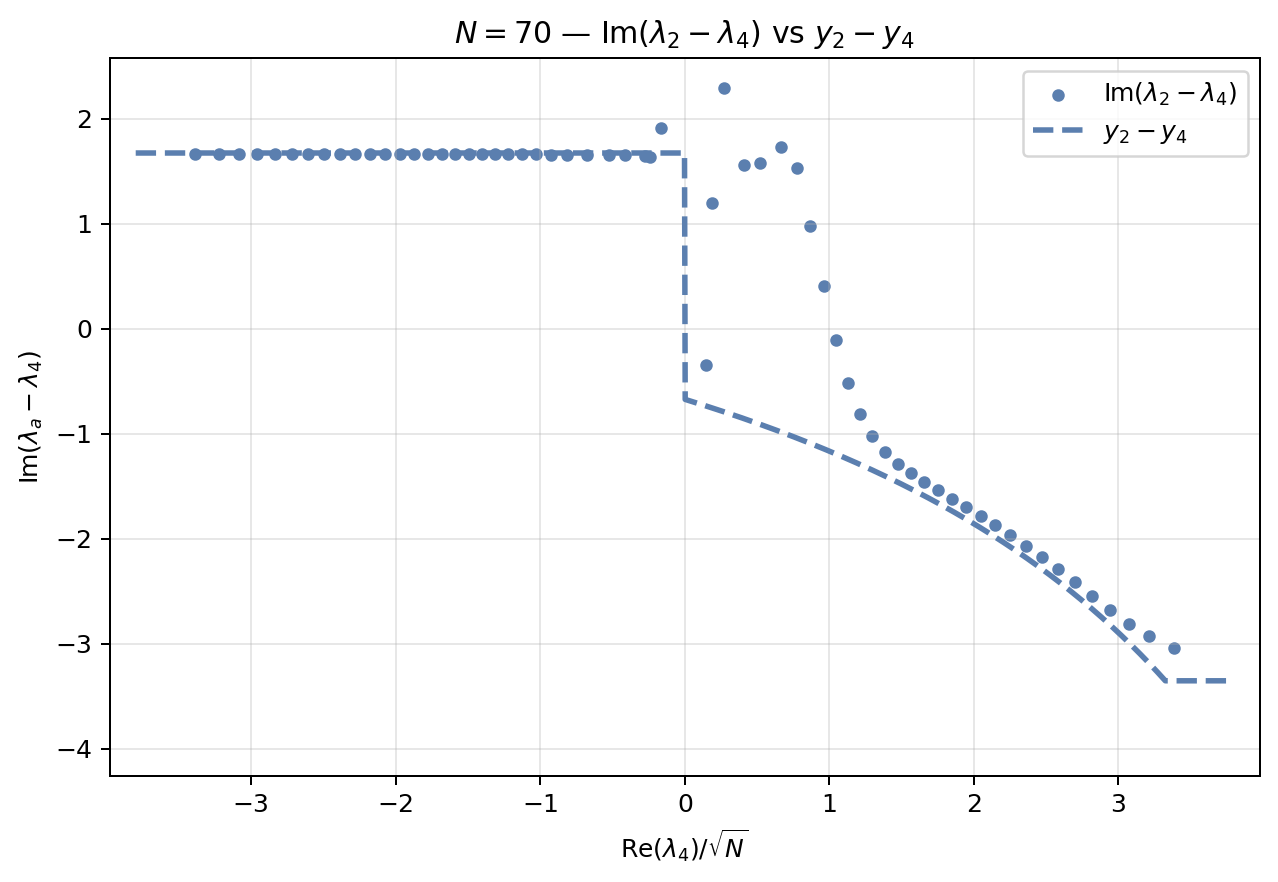}
         \caption{Eigenvalue difference $y_2(x)-y_4(x)$}
         \label{fig:y24_Delta_1}
     \end{subfigure}     
     \caption{Comparison of the numeric solution with lowest $\text{Re}(\cF)$ for $\Delta_{A_i}=\Delta_{B_i}=\Delta_{C_i}=\Delta_i$, $\Delta_1=0.4$, with the analytic solution of section \ref{sec: A proposal for general Delta} / operator counting result of section \ref{sec:operator_counting}.}
     \label{fig:Delta_1_comparison_plots}
\end{figure}

In Figure \ref{fig:N70_all_saddles} the solution with lowest $\text{Re}(\cF)$ is marked with stars. In Figure \ref{fig:N=70_Delta_1_eigenvalues} we have plotted the eigenvalue distribution for all the nodes of this solution, and as a comparison we have plotted the analytic prediction for the large imaginary part $\ii\,N\,w(x)$ with a black line. The numerical solution adheres quite closely to the analytic prediction. Then in Figure \ref{fig:Delta_1_comparison_plots} we compare the eigenvalue density and eigenvalue differences with the analytic predictions given by \eqref{general_analytic_rho}, \eqref{general_analytic_y34}, \eqref{Q111_solution_ansatz}, or equivalently with the same expressions obtained from operator counting in  section \ref{sec:operator_counting}. With the exception of the bulk region around $x=0$, the match is reasonably good. We expect that at bigger values of $N$, the bulk region would shrink and the match would improve, just as we observed in sections \ref{sec:numerical_solution} and \ref{sec:noncanonical}. However, we will not try to repeat such an analysis for the case studied in this section, since it is much more computationally demanding compared to the numerical solutions discussed previously.

\FloatBarrier

\subsection{Comparison with the Sasakian volume}
\label{Comparison with the Sasakian volume}

In this section we consider the Chern-Simons level assignment ${\rm k}_a=(1,-1,0,0)$ and compare the result for the free energy
of the chiral quiver with the Sasakian volume of $Q^{1,1,1}$.%
\footnote{Since the volume of $Q^{1,1,1}/\bZ_{\rm k}$ is $1/{\rm k}$ times the volume of $Q^{1,1,1}$, the matching for the more general ${\rm k}_a=({\rm k},-{\rm k},0,0)$ assignment follows trivially.}

The toric data for the cone over $Q^{1,1,1}$ can be read off 
equation (4.4) of \cite{Franco:2008um}. The toric fan of $C(Q^{1,1,1})$ is comprised of the cone generated by the
lattice vectors
\be
\begin{alignedat}{3}
    &V^{A_1}=(1,1,0,-1)\:,\qquad &&V^{B_1}=(1,1,-1,0)\:,\qquad &&V^{C_1}=(0,1,0,0)\:,\\
    &V^{A_2}=(0,0,0,1)\:,\qquad &&V^{B_2}=(0,0,1,0)\:,\qquad &&V^{C_2}=(1,0,0,0)\:,
\end{alignedat}
\ee
together with its faces. Notice that the vector $\nu_{\text{CY}}=(1,1,1,1)$ has dot product with the above vectors equal to one,
which ensures that there exists a Calabi-Yau metric on the cone $C(Q^{1,1,1})$ that restricts to the unique Sasaki-Einstein
metric on $Q^{1,1,1}$ \cite{Futaki:2006cc}.
Among all possible toric Sasakian metrics, the Sasaki-Einstein metric has the property of minimizing the
volume, which is called Sasakian volume minimization \cite{Martelli:2005tp,Martelli:2006yb}.
In order to compute the Sasakian volume associated to a generic toric Sasakian metric,
we need to resolve the conical singularity of $C(Q^{1,1,1})$ first. The resolution is not unique,
one possible choice is to consider the toric fan whose four-dimensional cones are
\be
\begin{alignedat}{2}
    &\sigma_1=\text{Span}_{\,\bR_+}(V^{A_1},V^{A_2},V^{B_1},V^{C_1})\:,\qquad
        &&\sigma_2=\text{Span}_{\,\bR_+}(V^{A_1},V^{A_2},V^{B_1},V^{C_2})\:,\\
    &\sigma_3=\text{Span}_{\,\bR_+}(V^{A_1},V^{A_2},V^{B_2},V^{C_1})\:,\qquad
        &&\sigma_4=\text{Span}_{\,\bR_+}(V^{A_1},V^{A_2},V^{B_2},V^{C_2})\:.
\end{alignedat}
\ee
The orbifold singularity numbers for the above resolved fan are all equal to one,
\be
    d_{\sigma_k}=\big|\,{\det}_{\,V^{I_j}\,\in\,\sigma_k}(V^{I_1},V^{I_2},V^{I_3},V^{I_4})\,\big|=1\:,
\ee
meaning that the resolution is smooth. 

A convenient way to derive the Sasakian volume and the volumes of the divisors is to first compute the equivariant volume
\cite{Duistermaat,Martelli:2023oqk}. We mostly follow the conventions of \cite{Hosseini:2025mgf}.
The equivariant volume of the resolved cone is given by
\begin{align}
\nn
    \bV(\lambda_I,b_i)={}&\sum_{\sigma_k}\:\frac1{d_{\sigma_k}}\:{\prod}_{\,V^{I}\,\in\,\sigma_k}
        \frac{e^{-\,\epsilon_{I}^{\sigma_k}\,\lambda_{I}}}{\epsilon_{I}^{\sigma_k}}=\frac{e^{-(b_1+b_3)\lambda_{A_1}-(b_1+b_3+b_4)\lambda_{A_2}+b_3\lambda_{B_1}+(b_1-b_2)\lambda_{C_1}}}{(b_1-b_2)b_3(b_1+b_3)(b_1+b_3+b_4)}-\\[2mm]\nn
    {}&-\frac{e^{-b_1\lambda_{A_1}-(b_1+b_4)\lambda_{A_2}-b_3\lambda_{B_2}+(b_1-b_2)\lambda_{C_1}}}{(b_1-b_2)b_3b_1(b_1+b_4)}
        +\frac{e^{-b_2\lambda_{A_1}-(b_2+b_4)\lambda_{A_2}-b_3\lambda_{B_2}-(b_1-b_2)\lambda_{C_2}}}{(b_1-b_2)b_3b_2(b_2+b_4)}-\\[2mm]
    {}&-\frac{e^{-(b_2+b_3)\lambda_{A_1}-(b_2+b_3+b_4)\lambda_{A_2}+b_3\lambda_{B_1}-(b_1-b_2)\lambda_{C_2}}}{(b_1-b_2)b_3(b_2+b_3)(b_2+b_3+b_4)}\:,
\end{align}
where the $\lambda_I$ are the K\"ahler moduli, the $b_i$ are the components of the Reeb vector $\sum_ib_i\,\partial_{\phi_i}$
with respect to the generators $\partial_{\phi_i}$ of the U(1)$^4$ isometry of $C(Q^{1,1,1})$,
and we have denoted with $\epsilon_{I}^{\sigma_k}$ the weights of the action of the Reeb vector at the fixed points,
which are given by
\be
    \epsilon_{I}^{\sigma_k}=\frac{\det(b,V^{I_{2}},V^{I_{3}},V^{I_{4}})}{\det(V^{I},V^{I_{2}},V^{I_{3}},V^{I_{4}})}\:,
        \qquad V^{I_{j}}\,\in\,\sigma_k\:.
\ee
Then the Sasakian volume and the volumes of the toric divisors are easily derived from the equivariant volume as follows:
\be
    \text{Vol}\big(Y_7[b]\big)=\frac{\pi^4}3\,\bV(\lambda_I,b_i)\,\big|_{\lambda_I\,=\,0}\:,\qquad
    \text{Vol}\big(D_I[b]\big)=-\,\pi^3\,\big(\partial_{\lambda_I}\bV(\lambda_I,b_i)\big)\big|_{\lambda_I\,=\,0}\:.
\ee
Even though the equivariant volume is dependent on the specific choice of resolution, the above expressions are not.

The free energy result \eqref{free_energy_with_general_Delta} can be matched with the Sasakian volume as follows
\cite{Herzog:2010hf,Martelli:2011qj,Jafferis:2011zi}.
First, the R-charges are parametrized in terms of the Reeb vector $b$ as a ratio of volumes \cite{Gubser:1998fp,Fabbri:1999hw},
\be
\label{mesonic_parametrization}
    \Delta_I[b]=\frac{2\pi}{3\,\nu_{\text{CY}}\cdot b}\:\frac{\text{Vol}\big(D_I[b]\big)}{\text{Vol}\big(Y_7[b]\big)}\:,
\ee
where as before $\nu_{\text{CY}}=(1,1,1,1)$ is the vector such that $\nu_{\text{CY}}\cdot V^I=1$ for all $I$.
This parametrization is purely mesonic; the large\,-$N$ free energy does not depend on baryonic variables due to the flat directions.
When the above parametrization for the R-charges is substituted in the free energy \eqref{free_energy_with_general_Delta}
(with ${\rm k}=1$),
we recover precisely the expected \emph{off-shell} relation \cite{Martelli:2011qj,Jafferis:2011zi} between
the large-\,$N$ limit of the free energy and the Sasakian volume:%
\footnote{If we wanted to compare the result \eqref{free_energy_with_Delta_constraint} for the free energy in the special case
    where $\Delta_{A_1}+\Delta_{C_1}=\Delta_{A_2}+\Delta_{C_2}$ with the Sasakian volume,
    the computation would work out the same but we would need to be careful that this restriction on the $\Delta_I$
    together with the parametrization \eqref{mesonic_parametrization} would correspond to a restriction of the values of the
    Reeb vector, namely $b_1 - b_2 + b_4=0$.}
\be
\label{F_vs_Sasakian}
    F_{S^3}(\Delta_I=\Delta_I[b]\,)\:=\frac{16\,N^{3/2}}{(\nu_{\text{CY}}\cdot b)^2}\:\sqrt{\frac{2\pi^6}{27\,\text{Vol}\big(Y_7[b]\big)}}+o(N^{3/2})\:.
\ee
Toric Sasakian geometry fixes $\nu_{\text{CY}}\cdot b=4$; among all the remaining possible values of $b$,%
\footnote{$b$ must also lie in the interior of the cone spanned by the $V^I$.}
the one that minimizes 
$\text{Vol}\big(Y_7[b]\big)$ gives the value of the components of the Reeb vector for the Sasaki-Einstein metric of $Q^{1,1,1}$
\cite{Martelli:2005tp,Martelli:2006yb}.
As the above equation suggests, the field theory dual of the Sasakian volume minimization is $F$-extremization \cite{Jafferis:2010un},
which singles out the values of the superconformal R-charges among all possible R-charge assignments. 
At the extremum, or in other words \emph{on-shell}, the relation \eqref{F_vs_Sasakian} reduces to
\be
    F_{S^3}(\Delta_I=1/3)\:=\:N^{3/2}\,\sqrt{\frac{2\pi^6}{27\,\text{Vol}\big(Q^{1,1,1}\big)}}+o(N^{3/2})
            =\frac{4\pi}{3\sqrt 3}\,N^{3/2}+o(N^{3/2})\:,
\ee
which we also verified numerically in section \ref{sec:numerical_solution}, \eqref{ReF0_numerical_fit}.
Indeed, it is easy to verify that the Sasakian volume is minimized at $b_1=b_2=2$ and $b_3=b_4=0$,
and the volume of the Sasaki-Einstein metric is $\text{Vol}\big(Q^{1,1,1}\big)=\text{Vol}\big(Y_7[b_1=b_2=2,b_3=b_4=0]\big)=\pi^4/8$.

\section{Discussion}
\label{sec:Discussion}

In this paper we have studied the large\,-$N$ limit of the supersymmetric partition functions of chiral $\cN=2$ Chern-Simons-matter theories, focusing on the chiral quiver dual to M-theory on AdS$_4 \times Q^{1,1,1}/{\mathbb{Z}_{\rm k}}$.
We have provided an analytic large\,-$N$ solution to the saddle point equations, which has been a long-standing problem for 3d chiral theories with an M-theory dual. Our results apply to the large\,-$N$ limit of the free energy on the round three-sphere and the superconformal and topologically twisted indices on $S^2\times S^1$ in the Cardy limit, and they agree with the holographic predictions from the gravity side off-shell.
In order to find a good ansatz for the analytic solution and test its correctness, we have carried out a finite\,-$N$ numerical analysis that extends the one of \cite{Hosseini:2025jxb} to larger values of $N$ and assignments of trial R-charges different from the superconformal ones.
We also considered the operator counting method conjectured in \cite{Gulotta:2011si, Gulotta:2011aa}, which had already been applied to the chiral quiver dual to AdS$_4 \times Q^{1,1,1}/{\mathbb{Z}_{\rm k}}$ in \cite{Kim:2012vza}; we extended the results of the latter and compared them with the numerical and analytical solutions to the saddle point equations, finding a good match.
Our results therefore provide further evidence for the validity of the operator counting method.

The saddles that we find have several features that have not been described before in the literature.
At finite $N$ they have a bulk region that shrinks to zero at large $N$, they can have large $\cO(N)$ imaginary parts in the eigenvalues and the long range forces cancel through a combination of these $\cO(N)$ imaginary parts and $\cO(N^{-1})$ subleading corrections in the tail regions. It is reasonable to think that these phenomena also occur in the large\,-$N$ saddle of more general chiral quiver models, which is an interesting direction for future work. A simple candidate would be the D3 quiver \cite{Franco:2008um} for which numeric results have already been obtained in \cite{Hosseini:2025jxb}. It would especially be interesting to consider chiral quiver theories for which the method of \cite{Amariti:2026djf}, based on integral identities \cite{vandeBult:2007} related to Giveon-Kutasov dualities \cite{Giveon:2008zn}, does not apply, such as the chiral quiver dual to AdS$_4 \times M^{1,1,1}/{\mathbb{Z}_{\rm k}}$ \cite{Martelli:2008si}.

The lessons learned in this paper may also be useful to find new saddles that carry dependence on baryonic charges. None of the explicit saddles found in the literature, including the ones of this paper, depend on them, despite the fact that black hole solutions with baryonic charges are known \cite{Halmagyi:2013sla,Azzurli:2017kxo,Hong:2019wyi,Kim:2020qec}.
New saddles with baryonic dependence would provide a microstate counting for these black holes and fill the gap.
The large\,-$N$ limit of the corresponding partition functions evaluated on these saddles is expected to reproduce the recent proposal of \cite{Hosseini:2025ggq,Hosseini:2025mgf}, which extends the results of \cite{Hosseini:2019ddy,Gauntlett:2019roi,Kim:2019umc} and is based on the master volume \cite{Gauntlett:2018dpc}, a generalization of the Sasakian volume that can also be derived from the equivariant volume \cite{Martelli:2023oqk}.

In section \ref{sec:Delta_1_numeric_analysis} we have found a large number of competing numerical saddles. It would be nice to gain analytic understanding on all of them and pinpoint their gravitational duals. In the absence of explicit solutions on the gravitational side, equivariant localization methods could help identify them, using for example the results of \cite{BenettiGenolini:2026cdw} or by making contact with the equivariant-volume-based approach of \cite{Martelli:2023oqk,Colombo:2023fhu}.

Beyond searching for new saddles, interesting future directions include studying the saddles that we have found in this paper beyond the leading order in $N$ using methods similar to those of \cite{Bobev:2026lvl}, and also extending our results to include other supersymmetric partition functions such as the spindle index \cite{Inglese:2023wky,Inglese:2023tyc,Colombo:2024mts}.

\acknowledgments
We would like to thank Junho Hong, Seyed Morteza Hosseini, Seok Kim, Noppadol Mekareeya and Alberto Zaffaroni for useful discussions. This work is supported by the National Research Foundation of Korea under the grants: RS-2023-00243491 (H.K.), RS-2025-00518906 (E.C., N.K., H.S.), RS-2025-25441170 (E.C., N.K.), RS-2025-25457100 (N.K.), and RS-2026-25477643 (H.K.). An earlier version of this work was presented at ``Black hole thermodynamics -- from semiclassical to quantum'' held in Kyung Hee University July 13--17 2026.
All analytic derivations in this paper were carried out by the authors without the use of large language models; these were only used to re-check them for typos and sign errors. The Python code for the numerical analysis, including the plots and fits, was generated with the assistance of Anthropic's Claude Fable~5.1 and Opus~5.1--5.5, and Google's Gemini.

\bibliography{chiral}{}
\end{document}